\documentclass[
prd,
aps,
superscriptaddress,
groupedaddress,
showpacs,
preprintnumbers,
amsmath,
floatfix,
manuscript
]{revtex4}

\usepackage{epsfig}
\usepackage{latexsym}
\usepackage[psamsfonts]{amssymb}
\usepackage{graphicx}
\usepackage{ulem}
\usepackage{epstopdf}
\usepackage{bm}
\usepackage{color}

\begin{document}
\bibliographystyle{apsrev}
\preprint{Edinburgh 2008/51, KEK-TH-1282, RBRC-762, TKYNT-09-05, YITP-08-100}

\bibliographystyle{apsrev}

\title{Nucleon form factors with 2+1 flavor dynamical domain-wall fermions}

\author{Takeshi Yamazaki\footnote{present address: Center for Computational Sciences, University of Tsukuba, Tsukuba, 305-8577, Japan}}
\affiliation{Physics Department, University of Connecticut, Storrs, CT 06269-3046}
\affiliation{Yukawa Institute for Theoretical Physics, Kyoto University, Kyoto 606-8502, Japan}

\author{Yasumichi Aoki} 
\affiliation{RIKEN-BNL Research Center, Brookhaven National Laboratory, Upton, NY 11973}

\author{Tom Blum} 
\affiliation{Physics Department, University of Connecticut, Storrs, CT 06269-3046}
\affiliation{RIKEN-BNL Research Center, Brookhaven National Laboratory, Upton, NY 11973}

\author{Huey-Wen Lin}
\affiliation{Thomas Jefferson National Accelerator Facility, Newport News, VA 23606, USA}

\author{Shigemi Ohta
} 
\affiliation{Institute of Particle and Nuclear Studies, KEK, Tsukuba, 305-0801, Japan}
\affiliation{Physics Department, Sokendai Graduate U.\ Adv.\ Studies, Hayama, Kanagawa 240-0193, Japan}
\affiliation{RIKEN-BNL Research Center, Brookhaven National Laboratory, Upton, NY 11973}

\author{Shoichi Sasaki}
\affiliation{Department of Physics, University of Tokyo, Hongo 7-3-1, Tokyo 113-0033, Japan}

\author{Robert Tweedie}
\affiliation{School of Physics \& Astronomy, The University of Edinburgh, Edinburgh EH9 3JZ, UK}

\author{James Zanotti}
\affiliation{School of Physics \& Astronomy, The University of Edinburgh, Edinburgh EH9 3JZ, UK}


\collaboration{RBC and UKQCD Collaborations}


\bibliographystyle{apsrev}

\date{
\today
}

\begin{abstract}
We report our numerical lattice QCD calculations of the isovector nucleon form factors for the vector and axialvector currents: the vector, induced tensor, axialvector, and induced pseudoscalar form factors.
The calculation is carried out with the gauge configurations generated with \(N_f = 2 + 1\) dynamical domain wall fermions and Iwasaki gauge actions at \(\beta = 2.13\), corresponding to a cutoff \(a^{-1}\) = 1.73 GeV, and a spatial volume of \((2.7\;{\rm fm})^3\).
The up and down quark masses are varied so the pion mass lies between 0.33 and 0.67 GeV while 
the strange quark mass is about 12 \% heavier than the physical one.
We calculate the form factors in the range of momentum transfers, \(0.2 < q^2 < 0.75\) GeV$^2$.
The vector and induced tensor form factors are well described by the 
conventional dipole forms and result in significant underestimation 
of the Dirac and Pauli mean-squared radii and the anomalous magnetic 
moment compared to the respective experimental values.
We show that the axialvector form factor is significantly affected by the finite spatial volume of the lattice.
In particular in the axial charge, \(g_A/g_V\), the finite volume effect scales with a single dimensionless quantity, \(m_\pi L\), the product of the calculated pion mass and the spatial lattice extent.
Our results indicate that for this quantity, \(m_\pi L>6\) is required to ensure that finite volume effects are below 1\%.
\end{abstract}

\maketitle

\section{Introduction}
\label{sec:Introduction}
The isovector nucleon form factors are probes for nucleon structure associated with the isovector vector and axialvector currents, $V^+_\mu = \overline{u}\gamma_\mu d$ and $A^+_\mu = \overline{u}\gamma_\mu \gamma_5 d$, with up- and down-quark spinors \(u\) and \(d\).
From these currents, four isovector form factors arise in neutron $\beta$ decay: the vector $(F_V)$ and induced tensor $(F_T)$ form factors from the vector current,
\begin{equation}\label{eq:cont_vector}
\langle p| V^+_\mu(x) | n \rangle = \bar{u}_p \left[\gamma_\mu
F_V(q^2) + \sigma_{\mu \lambda}q_{\lambda} {F_T(q^2)} \right]
u_n e^{iq\cdot x},
\end{equation}
where \(F_V\) is equivalent to \(F_1\) and \(F_T\) to \(F_2/(2M_N)\) in the isovector part of
electromagnetic form factors under the isospin symmetry, and the axial $(F_A)$ and induced
pseudoscalar $(F_P)$ form factors from the axialvector current,
\begin{equation}\label{eq:cont_axial}
\langle p| A^+_\mu(x) | n \rangle = \bar{u}_p
            \left[\gamma_\mu  \gamma_5 F_{A}(q^2)
             +i q_\mu \gamma_5 {F_{P}(q^2)} \right]  u_n e^{iq\cdot x}.
\end{equation}
We use the Euclidean metric convention as in the recent RBC works
~\cite{{Sasaki:2007gw},{Lin:2008uz}}. Thus $q^2$ stands for Euclidean four-momentum squared, and corresponds to the time-like momentum squared since $q_M^2=-q^2<0$ in 
Minkowski space.
Here $q=p_n-p_p$ is the momentum transfer between the proton ($p$) and neutron ($n$).

The vector-current form factors have been studied experimentally with high accuracy at both small ($<$ 1 GeV$^2$) and large ($>$ 1 GeV$^2$) momentum transfers, through electron elastic scattering 
off proton and nuclei~\cite{Arrington:2006zm}.
Early experiments revealed that the proton is a composite particle
\cite{{Hofstadter:1955ae,Bumiller:1960zz,Bumiller:1961zz,Janssens:1965kd}}:
{\it i.e.} non-zero Pauli and Dirac mean-squared radii and anomalous magnetic moments were measured among other observables.
Recent experiments have improved the accuracy of these form factors
and deviations from earlier perturbative QCD predictions have been
observed \cite{{Jones:1999rz,Gayou:2001qd}}.

As is well-known, the isovector axialvector current is strongly affected by the spontaneous chiral symmetry breaking in the strong interaction \cite{Nambu:1961tp,Nambu:1961fr}.
A consequence for  the nucleon is that the isovector axial charge \(g_A\) deviates from the corresponding vector charge $g_V$.
These isovector vector and axialvector charges, respectively the vector and axialvector form factors at the zero momentum transfer, are most accurately measured in neutron beta decay experiments: $g_A/g_V = F_A(0)/F_V(0) = 1.2695(29)$~\cite{Amsler:2008zz}.
Whether lattice QCD calculations can accurately reproduce this ratio, \(g_A/g_V\), is an important test of lattice QCD.

The spontaneous breaking of chiral symmetry also means that the corresponding form factors of
the axialvector current are strongly coupled with the Nambu-Goldstone particles, {\it i.e.}\ the pions.
Using the axial Ward-Takahashi identity and the pion-pole dominance assumption on the induced pseudoscalar, one can derive the Goldberger-Treiman relation~\cite{Goldberger:1958tr}, which relates the nucleon mass $(M_N)$, the axial charge $(g_A)$, the pion decay constant $(F_{\pi})$ and the pion-nucleon coupling $(g_{\pi NN})$: $M_N g_A = F_{\pi}g_{\pi NN}$.
It is an interesting challenge for lattice QCD if it can reproduce this relation.

The $q^2$ dependence of the axialvector form factor has also been studied in experiments \cite{Bernard:2001rs}.
It again provides a stringent test of QCD through a comparison of lattice QCD calculations with such experiments.
While recent experiments report the induced pseudoscalar form factors~\cite{{Choi:1993vt,Andreev:2007wg}}, it is less well known than the other form factors.
Hence this provides an excellent opportunity for lattice QCD to play a
leading role and guide future experiments.

In the past years, many lattice QCD studies have been made for these isovector form factors in the above-mentioned contexts~\cite{{Hagler:2007hu,Zanotti:2008zm}}.
Many earlier
works~\cite{{Gockeler:2003ay,Sasaki:2003jh,Tang:2003jh,Sasaki:2007gw,Boinepalli:2006xd}}
were performed either in the quenched approximation,
neglecting dynamical sea-quark effects or were either limited to two 
dynamical flavors of Wilson fermion quarks that explicitly violate 
chiral symmetry~\cite{{Alexandrou:2006ru,Khan:2006de,Gockeler:2007hj}}, 
limited to non-unitary combination of valence and sea 
quarks~\cite{{Edwards:2005ym,Alexandrou:2007xj,Hagler:2007xi,Bratt:2008uf}}, 
or just two dynamical flavors of domain-wall fermions (DWFs)~\cite{Lin:2008uz}.
There has also been an increasing amount of interest in the form
factors of other baryons~\cite{{Alexandrou:2007xj,Alexandrou:2007dt,Guadagnoli:2006gj,Sasaki:2008ha,Lin:2008mr}}.

In this paper we present our results with more realistic ``2+1 flavor'' dynamical quarks: reasonably light and degenerate up and down quarks and strange quark with a realistic mass are all described by the DWF scheme~\cite{{Ginsparg:1981bj,Kaplan:1992bt,Kaplan:1992sg,Shamir:1993zy,Furman:1994ky}} that preserves the flavor and chiral symmetries sufficiently.
Earlier studies were often performed on small spatial volumes 
($\sim(2\,{\rm fm})^3$) which are now widely regarded to be too small to
accommodate a nucleon at light quark masses that yield
realistic axial charge~\cite{Khan:2006de,Yamazaki:2008py}.
We use larger spatial lattice volume, as large as 2.7 fm across, 
to better address the finite-size question.

The rest of the paper is organized as follows:
We explain our method of calculation in Sec.~\ref{sec:Method}.
In Sec.~\ref{sec:Ensembles} we first summarize the numerical lattice QCD ensembles used for this work.
Then we discuss in detail the known systematic errors in the relevant form factors calculated on these ensembles.
The numerical results are presented in section~\ref{sec:Results}.
Finally, we give the conclusions in Sec.~\ref{Conclusions}.

Since we vary only light quark mass in our simulation while the strange quark mass is fixed, in the following we call the light up and down quark mass as quark mass, $m_f$, in the lattice unit, unless explicitly stated otherwise.
We note that some preliminary results from this study were presented in Refs.~\cite{{Yamazaki:2007mk,Ohta:2008kd,Yamazaki:2008py}}.

\section{Method}
\label{sec:Method}
\subsection{Two- and three-point functions}

Following earlier studies with quenched and two 
dynamical flavors~\cite{Sasaki:2007gw,Lin:2008uz}, 
we define the two-point function of proton
\begin{equation}
C_S(t-t_{\rm src},p) = \frac{1}{4}\sum_{\vec{x}}e^{i\vec{p}\cdot\vec{x}}
\mathrm{Tr}\left[ \mathcal{P}_4
\langle
0 | \chi_S(\vec{x},t) \overline{\chi}_G(\vec{0},t_{\rm src})| 0 
\rangle
\right],
\label{eq:2pt}
\end{equation}
where $S$ is the index of the smearing of the quark operator and 
$t_{\rm src}$ is the time location of
the source operator.
The projection operator $\mathcal{P}_4 = (1+\gamma_4)/2$ 
eliminates the contributions from the opposite-parity
state for $p^2=0$~\cite{{Sasaki:2001nf},{Sasaki:2005ug}}.
We use the standard proton operator, 
\begin{equation}
\chi_S(x) = \epsilon_{abc} ([u^S_a(x)]^T C \gamma_5d^S_b(x))u^S_c(x),
\end{equation}
where $C$ is the matrix of the charge conjugation,
and $a,b,c$ are color indices, 
to create and annihilate proton states. 
In order to improve the overlap with the ground state, we apply
Gaussian smearing~\cite{Alexandrou:1992ti} at the source, while at
the sink we employ both local and Gaussian-smeared operators, $S=L$ or $G$.

In this paper we measure the nucleon isovector matrix elements for 
the vector and axialvector currents,
\begin{eqnarray}
\langle p | V_\mu^3(x) | p \rangle &=& 
\langle p | \overline{u}(x)\gamma_\mu u(x) - 
\overline{d}(x)\gamma_\mu d(x) | p \rangle,\\
\langle p | A_\mu^3(x) | p \rangle &=& 
\langle p | \overline{u}(x)\gamma_5\gamma_\mu u(x) - 
\overline{d}(x)\gamma_5\gamma_\mu d(x) | p \rangle.
\end{eqnarray}
While we employ the local currents in most of the calculations,
the point-split conserved vector current~\cite{Furman:1994ky} 
is used for the
vector charge at the lightest quark mass which will be described later.

In order to obtain the matrix elements, we define the three-point function
with the current $J$ and the projector $\mathcal{P_\alpha}$
\begin{eqnarray}
C^{\mathcal{P}_\alpha}_{J_\mu}(\vec{q},t) &=& 
\frac{1}{4} \sum_{\vec{x},\vec{z}}e^{i\vec{q}\cdot\vec{z}}
\mathrm{Tr}\left[\mathcal{P}_\alpha
\langle 0 | \chi_G(\vec{x},t_{\rm snk}) J_\mu(\vec{z},t) 
\overline{\chi}_G(\vec{0},t_{\rm src}) | 0 \rangle
\right]
\label{eq:3pt_0}
\\
&=&
\Lambda^{J}(q)\times f(t_{\rm src},t_{\rm snk},t,M_N,E(q),q) + \cdots,
\label{eq:3pt}
\end{eqnarray}
where $t_{\rm snk}$ is the sink time slice fixed as 
$t_{\rm snk} - t_{\rm src} = 12$, and $E(q) = \sqrt{M_N^2+\vec{q}^2}$.
The ellipsis denotes the higher excited state contributions, which can be ignored for long time separations $t_{\rm snk} \gg t\gg  t_{\rm src} $.
The time independent part of $\Lambda^{J}(q)$ is a matrix element, which is a linear combination of the form factors we seek. 
The time dependent part of $f(t_{\rm src},t_{\rm snk},t,M_N,E(q),q)$
includes the kinematical factor and the normalization of the proton
operator which we Gaussian smear at both the source and sink.
We employ the sequential source method to reduce statistical fluctuations, 
as in Ref.~\cite{Sasaki:2003jh,Wilcox:1991cq}.
In the three-point function, initial and final proton states carry $\vec{q}$ and zero momenta, respectively. This is because the 
spatial momentum should be conserved in the function as in the two-point
function.

The time dependence of $f(t_{\rm src},t_{\rm
  snk},t,M_N,E(q),q)$ is removed by taking an appropriate ratio of the three- and two-point functions~\cite{Hagler:2003jd}
\begin{eqnarray}
R^{\mathcal{P}_\alpha}_{J_\mu}(q,t) & = &
K\cdot
\frac{C^{\mathcal{P}_\alpha}_{J_\mu}(\vec{q},t)}{C_G(t_{\rm snk}-t_{\rm src},0)}
\left[
\frac{C_L(t_{\rm snk}-t,q)C_G(t-t_{\rm src},0)C_L(t_{\rm snk}-t_{\rm src},0)}
{C_L(t_{\rm snk}-t,0)C_G(t-t_{\rm src},q)C_L(t_{\rm snk}-t_{\rm src},q)}
\right]^{1/2},
\label{eq:2-3_ratio}
\end{eqnarray}
where $K=M_N \sqrt{2 E(q) ( M_N + E(q) )}$.
The ratio $R^{\mathcal{P}_\alpha}_{J_\mu}$ should display a plateau from which the matrix element
we seek is extracted.

For each of the vector or axialvector currents, we first obtain $\Lambda^J(q)$ in eq.(\ref{eq:3pt}) which is a linear combination of the form factors.
For convenience, using the ratio $R$ we define 
\begin{eqnarray}
\Lambda_4^V(q,t) &=& 
\frac{R^{\mathcal{P}_4}_{V_4}(q,t)}{M_N(M_N+E(q))}, 
\label{eq:lambda_v0_t}\\
\Lambda_T^V(q,t) &=& -\frac{1}{2}\left(
\frac{R^{\mathcal{P}_{53}}_{V_1}(q,t)}{iq_2 M_N} -
\frac{R^{\mathcal{P}_{53}}_{V_2}(q,t)}{iq_1 M_N} \right),
\label{eq:lambda_vt_t}
\end{eqnarray}
for the vector current, and
\begin{eqnarray}
\Lambda_L^A(q,t) &=& 
\frac{R^{\mathcal{P}_{53}}_{A_3}(q,t)}{M_N(M_N+E(q))}, 
\label{eq:lambda_al_t}\\
\Lambda_T^A(q,t) &=& -\frac{1}{2}\left(
\frac{R^{\mathcal{P}_{53}}_{A_1}(q,t)}{q_2 q_3} +
\frac{R^{\mathcal{P}_{53}}_{A_2}(q,t)}{q_1 q_3} \right),
\label{eq:lambda_at_t}
\end{eqnarray}
for the axialvector current.
Here we also define $q^2 = 2 M_N ( E(q) - M_N )$, and $\mathcal{P}_{53} = (1+\gamma_4)\gamma_5\gamma_3/2$ implies the z-direction is chosen as the polarization direction in our calculation.
In the plateau region of $\Lambda^J(q,t)$ we determine the matrix elements of each current, $\Lambda^J(q)$ which has the following relation to the form factors: 
\begin{eqnarray}
\Lambda_4^V(q) &=& 
F_1(q^2) - \frac{q^2}{4 M_N^2} F_2(q^2),
\label{eq:lambda_v0}\\
\Lambda_T^V(q) &=& 
F_1(q^2) + F_2(q^2),
\label{eq:lambda_vt}
\end{eqnarray}
for the vector current, and
\begin{eqnarray}
\Lambda_L^A(q) &=& 
F_A(q^2) - \frac{q_3^2}{M_N + E(q)} F_P(q^2),
\label{eq:lambda_al}\\
\Lambda_T^A(q) &=& 
M_N F_P(q^2),
\label{eq:lambda_at}
\end{eqnarray}
for the axialvector current.
In the following we use the isovector part of 
the Dirac and Pauli form factors, $F_{1,2}$,
rather than the vector and induced tensor form factors.
They are identical through the isospin symmetry except the normalization
of the Pauli form factor, $F_2 = 2 M_N F_T$.
We will see that the signal of these combinations is reasonable 
in Sec.~\ref{sec:vector}.
Finally respective form factors are obtained by solving the sets of linear equations,
(\ref{eq:lambda_v0}) and (\ref{eq:lambda_vt}), or (\ref{eq:lambda_al}) and (\ref{eq:lambda_at}), at fixed $q^2$.

\subsection{Double source method}
\label{sec:double_source}

We find the ensemble with the lightest quark mass of \(m_f=0.005\) is much noisier than the ones with heavier mass values: it is insufficient and takes enormous amount of calculation time to obtain reasonable statistical error if we used only a single nucleon source/sink combination per configuration.

\begin{figure}[!h]
\begin{center}
\includegraphics[width=.6\textwidth,clip]{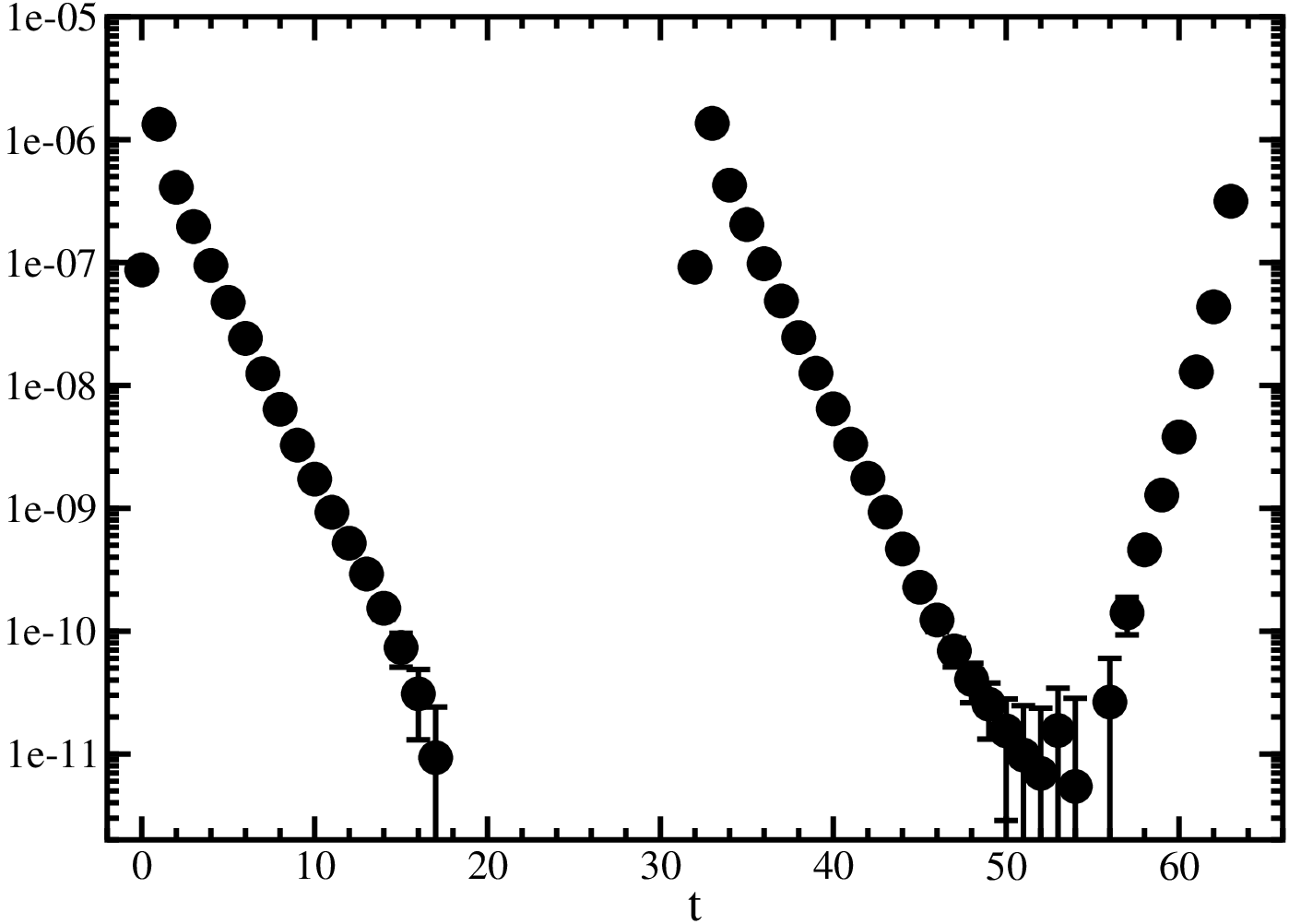}
\end{center}
\caption{Two-point function with the double source at $t=(0,32)$
for $m_f = 0.005$.
\label{fig:2pt}
}
\end{figure}

Fortunately, the time extent of the lattice, \(64 \times a = 7.3\) fm, is very large compared to the inverse of the nucleon mass, \(M_N^{-1} = (1.15 {\rm \ GeV})^{-1} = 0.17\) fm.
Hence, we can easily accommodate a pair of source/sink combinations on each configuration without letting them interfere with each other if the sources are separated by 32 units, as shown in Fig.~\ref{fig:2pt}.
We call this the double-source method.

The three-point functions are calculated with the sequential source method, and 
the sink operators are placed 12 time slices from their respective sources.
The number of the measurements is effectively doubled in this calculation while the cost remains the same as one single source measurement.

\section{Ensembles}
\label{sec:Ensembles}
\subsection{Statistics}

The RBC-UKQCD joint (2+1)-flavor dynamical DWF coarse ensembles \cite{Allton:2008pn} are used for the calculations.
These ensembles are generated with Iwasaki gauge action~\cite{Iwasaki:1983yi} at the coupling \(\beta=2.13\) which corresponds to
the lattice cutoff of \(a^{-1}=1.73(3)\) GeV, determined
from the $\Omega^{-}$ baryon mass~\cite{Allton:2008pn}.

The dynamical strange and up and down quarks are described by DWF actions with the fifth-dimensional extent of \(L_s=16\) and the domain-wall height of \(M_5=1.8\).
The strange quark mass is set at 0.04 in lattice units and turned out
to be about 12\% heavier than the physical
strange quark, after taking into account the additive correction of the residual mass, \(m_{\rm res}=0.003\).
The degenerate light quark masses in lattice units, 0.005, 0.01, 0.02 and 0.03, correspond to pion masses of about 0.33, 0.42, 0.56 and 0.67 GeV and nucleon masses, 1.15, 1.22, 1.39 and 1.55 GeV.

Two lattice volumes used are \(16^3\times 32\) and \(24^3\times 64\), corresponding to linear spatial extent of approximately 1.8 and 2.7 fm, respectively.
The smaller volume ensembles, calculated only with the heavier three light
quark masses, are used for a finite volume study of the axial charge and form factors
discussed in Section~\ref{sec:Results}.
On the $16^3$ ensembles we use 
3500 trajectories separated by 
5 trajectories at $m_f = 0.01$ and 0.02, and by 10 at 0.03.
The main results are obtained from the larger volume ensembles
with the number of the configurations summarized in table~\ref{table:statistics}.

\begin{table}[!t]
\begin{center}
\begin{tabular}{lccccc} \hline\hline
\multicolumn{1}{c}{$m_f$} & 
\multicolumn{1}{c}{$N_{\rm conf}$} &
\multicolumn{1}{c}{$N_{\rm sep}$} &
\multicolumn{1}{c}{$N_{\rm meas}$} &
\multicolumn{1}{c}{$m_\pi$ [GeV]} &
\multicolumn{1}{c}{$M_N$ [GeV]}\\ \hline
0.005 & 932$^a$
& 10 & 4$^b$
& 0.3294(13) &1.154(7)\\ 
0.01  & 356 & 10 & 4 & 0.4164(12) &1.216(7)\\
0.02  &  98 & 20 & 4 & 0.5550(12) &1.381(12)\\
0.03  & 106 & 20 & 4 & 0.6681(15) &1.546(12)\\ \hline\hline
\end{tabular}
\end{center}
\caption{$N_{\rm conf}$, $N_{\rm sep}$ and $N_{\rm meas}$ denote
number of gauge configurations, trajectory separation between each 
measured configuration, and number of measurements on each configuration,
respectively, on (2.7 fm)$^3$ volume. Table also contains
the pion and nucleon mass for each ensemble.\\
$^a$
Total number of configurations is actually 646. We
carry out extra measurements on a subset of these (286 configurations) to improve the statistics using different source positions.\\
$^b$
Two measurements with the double-source method gives effectively four measurements.
}
\label{table:statistics}
\end{table}

On the larger volume at the heavier three quark masses, we make four measurements on each configuration with the conventional single source method using $t_{\rm src} = 0,$ 16, 32, 48, or 8, 19, 40, 51.
At the lightest mass the double-source method is used,
and two measurements on each configuration are carried out using the source pairs of (0, 32) and (16, 48), or (8, 40) and (19, 51).
We make an additional two measurements on roughly half of the configurations with another source pairs. This means that 
we make four, double-source measurements on almost half of the
configurations, while two, double-source measurements are carried
out on the remaining configurations.
We have checked the independence of these measurements from each other by
changing the block size in the jackknife analysis, {\it e.g.,} treating each source/sink measurement as independent.
None of these resulted in significantly different error estimate.
Thus in the following we treat the two double-source measurements
performed on a single configuration, one with the source pairs of (0, 32) 
and (16, 48), and the other with the source pairs of (8, 40) and
(19, 51), as being independent of each other.

In the following, in order to reduce possible auto-correlations
at the larger volume the measurements are blocked into bins of 40 trajectories each, while 20 trajectories at the smaller volume.
The statistical errors are estimated by the jackknife method.

\subsection{Correlation functions}

The quark propagator is calculated with an
anti-periodic boundary condition in the temporal direction, and
periodic boundary conditions for the spatial directions.
We employ gauge-invariant Gaussian smearing at the source with
smearing parameters $(N,\omega) = (100,7)$, which were chosen after a
series of pilot calculations, as described in Ref.~\cite{Berruto:2005hg}.
For the calculation of the three-point functions,
we use a time separation of 12 timeslices between the source and sink operators
to reduce effects from excited state contributions as much as possible.

To obtain the form factors at non-zero $q^2$, we evaluate the two-  
and three-point functions, eqs.(\ref{eq:2pt}) and (\ref{eq:3pt_0}),  
with the four lowest nonzero momenta: $\vec{p} = 2\pi/L \times$  
(0,0,1), (0,1,1), (1,1,1), and (0,0,2), corresponding to a $q^2$ range  
from about 0.2 to 0.75 GeV$^2$ on the large volume, while on the small  
volume we use only the smallest two momentum transfers, corresponding  
to $q^2\approx 0.4$ and $0.8\,{\rm GeV}^2$.
All possible permutations of the momentum including the positive and  
negative directions are taken into account.

\begin{figure}[!t]
\begin{center}
\includegraphics[width=.6\textwidth,clip]{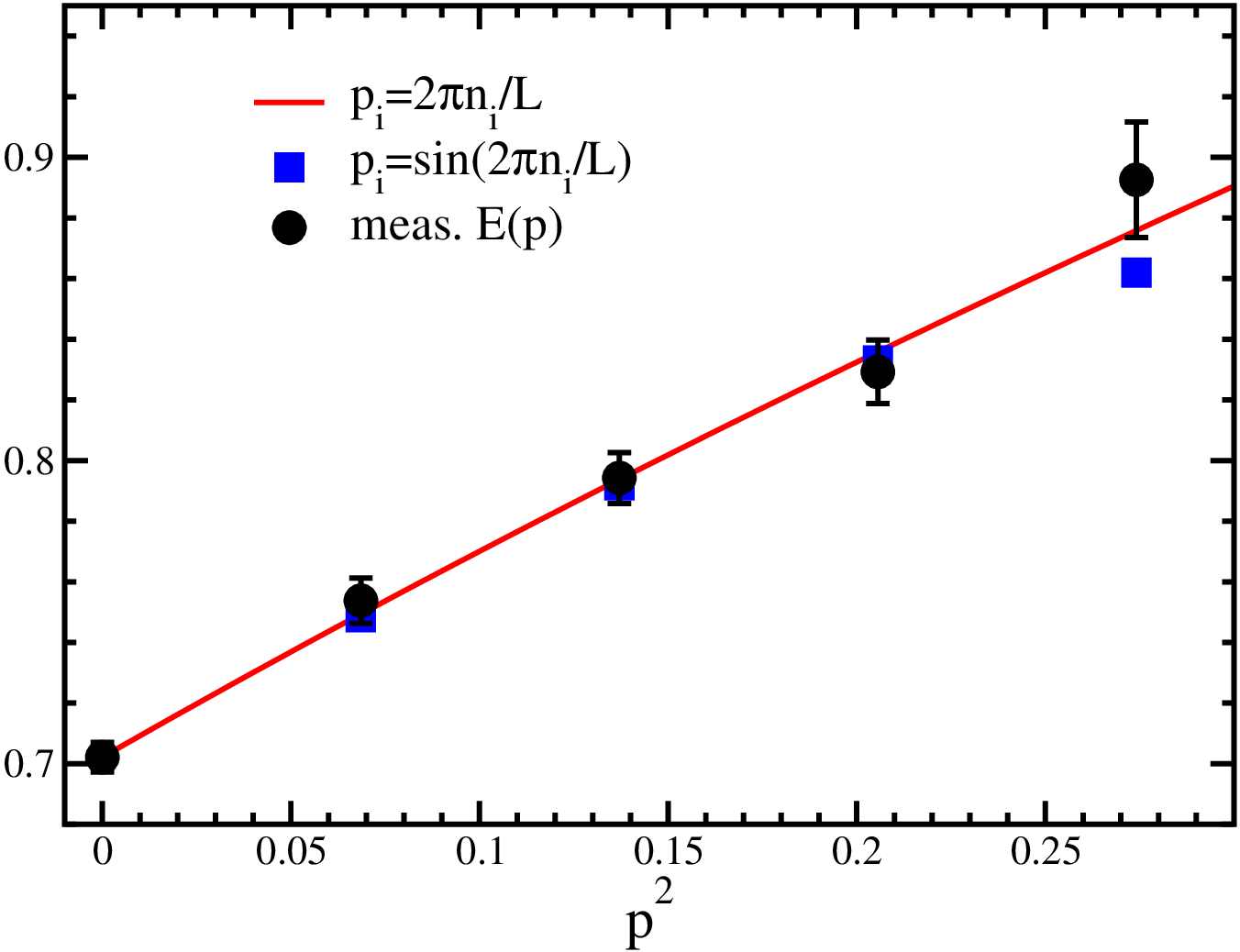}
\end{center}
\caption{Measured nucleon energies in lattice unit at $m_f = 0.01$.
Estimated energies by continuum and lattice momenta are also plotted.
\label{fig:ene}
}
\end{figure}

There are several choices for the definition of the momentum
in the lattice calculation, 
{\it e.g.}, $p_i = 2\pi/L\cdot n_i$, $\sin(2\pi a/L\cdot n_i)/a$, 
or one determined from the measured energy in the two-point function.
Figure~\ref{fig:ene} shows that the three energies with the different momentum definitions reasonably agree with each other.
In the following we choose the continuum momentum definition $p_i = 2\pi/L\cdot n_i$, since this simple definition gives smaller statistical error for the energy than the measured one.

\subsection{Systematic errors}

There are two important sources of systematic error: finite spatial size of the lattice and excited state contamination.
Chiral-perturbation-theory-inspired analysis of the former for meson
observables suggests the dimensionless product, \(m_\pi L\), of the
calculated pion mass \(m_\pi\) and lattice linear spatial extent
\(L\), should be set greater than 4 to ensure that the finite-volume correction is negligible below one percent, and the available lattice calculations seem to support this.
While our present parameters satisfy this condition, it should be emphasized that such a practical criterion is not known sufficiently for baryon observables.
It is important to check this through the present calculations, and it is indeed an important purpose of this work.

On the other hand, one should adjust the time separation between the nucleon source and sink appropriately so the resultant nucleon observables are free of contamination from excited states.
The separation has to be made longer as the quark masses decrease.
In  our previous study with two dynamical flavors of DWF 
quarks~\cite{Lin:2008uz} with a similar lattice cutoff of about 1.7 GeV, 
we saw systematic differences between observables calculated with 
the shorter time separation of 10, or about 1.16 fm, and longer 12, 
or 1.39 fm: the differences amount to about 20 \%, or two standard deviations.
This would suggest that at the shorter time separation of about 1.2 fm, the excited-state contamination has not decayed sufficiently to guarantee correct calculations for the ground-state observables~\cite{Ohta:2008kd}.
There is, however, a price to pay for the larger time separation as
the nucleon correlation function suffers from large statistical noise
at large times, especially with light quark masses.
Since the hadron masses are much lighter in the present work than we
considered previously (the lightest pion mass is 0.33 GeV and nucleon 1.15 GeV) we decided to use the separation of 12 lattice units, or about 1.4 fm.

\begin{figure}[!th]
\begin{center}
\includegraphics[width=.6\textwidth,clip]{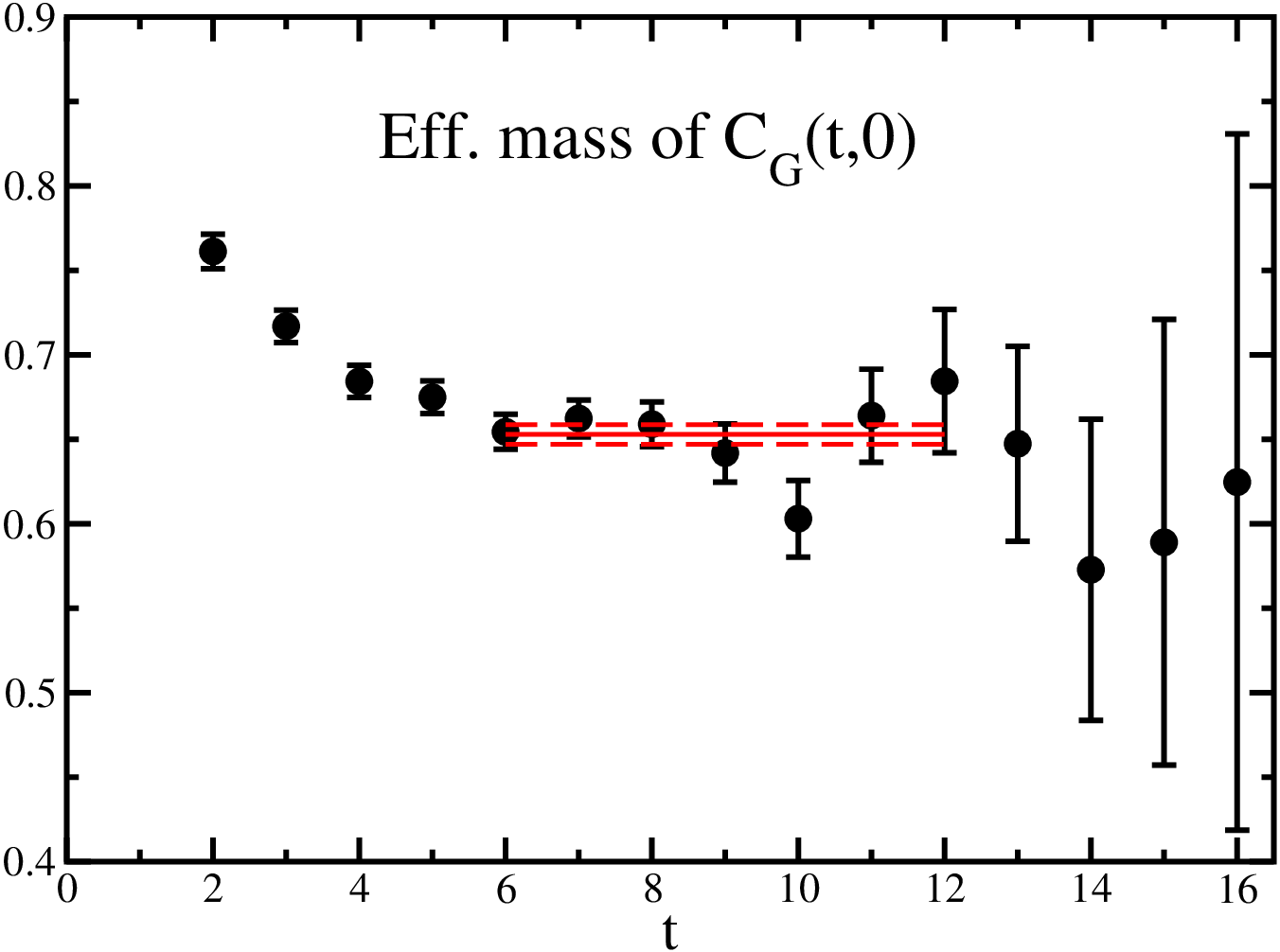}
\end{center}
\caption{Effective mass of nucleon correlator with Gaussian smearing
applied at both source and sink, for quark mass $m_f=0.005$.
\label{fig:Nemass}
}
\end{figure}

While it is desirable to use a longer separation, it cannot be made too long in practice without losing control of statistical errors.
In Fig. \ref{fig:Nemass} we present the nucleon effective mass at the lightest quark mass, \(m_f=0.005\).
The nucleon signal begins to decay at \(t=12\), or about 1.4 fm: this is about longest distance we can choose without losing the signal, and hence about as free of excited-state contamination as we can achieve with the present statistics. As will be shown in detail in this paper, the bare three-point function signals for the form factors for this source-sink separation of \(t=12\) are acceptable.
Whether this is sufficiently long separation between the source and sink to guarantee correct calculations of ground-state observables remains a future problem.

\section{Results}
\label{sec:Results}
%
%
\subsection{Vector and axial charges}
\label{sec:charges}
%
%

Much of the results and discussion in this subsection have appeared in Ref.~\cite{Yamazaki:2008py}. We repeat them here for convenience, and to lay some of the ground work necessary for discussion of the form-factor results that follow.

\begin{figure}[!t]
\centering
\includegraphics*[width=.6\textwidth,clip]{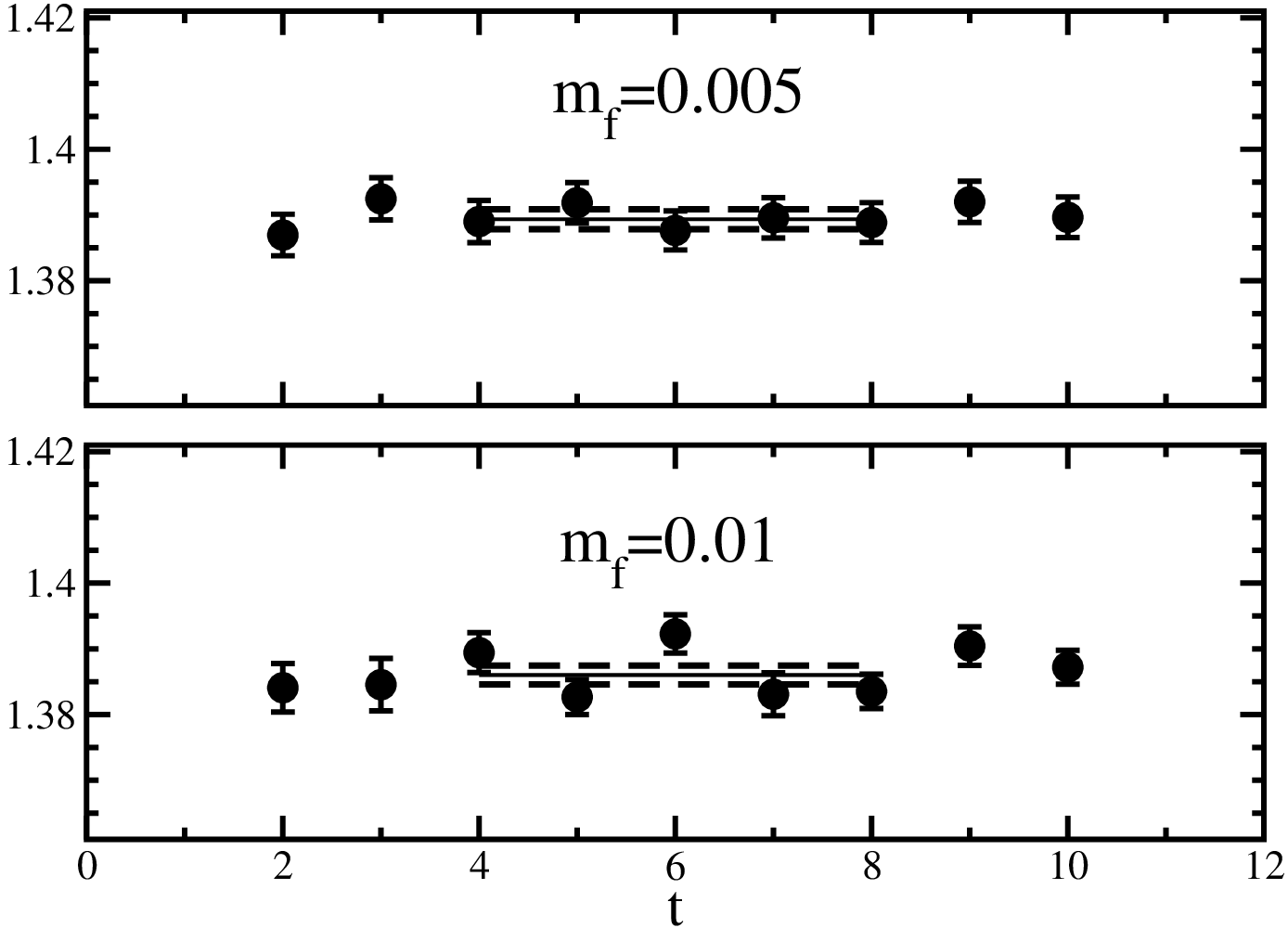}
\caption{
Plateaus of $g_V^{\mathrm{lat}}$ for $m_f = 0.005$ (top) and 0.01 (bottom).
Statistical errors are of comparable sizes for the two \(m_f\) values 
despite difference in the methods.
Solid lines denote fit results with one standard deviation.
}
\label{fig:plateau_gv}
\end{figure}

At zero momentum transfer the time component of the vector form factor gives the vector charge, \(g_V = F_1(0)\).
For our calculations at the heaviest three quark masses, 
we use the 4-dimensional local current. As a result,
the value of \(g_V^{\mathrm{lat}}\), measured from the bare $F_1(0)$,
deviates from unity, and gives the inverse of the renormalization,
\(Z_V\), for the local current. At the lightest quark mass,
$m_f = 0.005$, we evaluate the vector charge using the point
split conserved vector current~\cite{Furman:1994ky}, $\mathcal{V}_4$ as well.
This is to alleviate a problem that arises from the double source method described in Sec.~\ref{sec:double_source}:
Conventionally the vector charge is calculated from the ratio of the three-point function with the local vector current to the two-point function with zero momentum, as in eq.(\ref{eq:2-3_ratio}); a strong correlation between the denominator and numerator suppresses the statistical error associated with such calculations.
This correlation is lost in the double-source calculation and results in larger statistical errors.
Fortunately, the three-point functions of the local and conserved currents are highly correlated, even in this method.
Therefore we evaluate the vector charge from the ratio of 
the three-point functions 
$g_V^{\mathrm{lat}} = C^{\mathcal{P}_4}_{V_4}(\vec{0},t) 
/ C^{\mathcal{P}_4}_{\mathcal{V}_4}(\vec{0},t)$ at $m_f = 0.005$.
Figure~\ref{fig:plateau_gv} shows that the error
in this ratio is as small as that coming from the single source 
calculation at $m_f = 0.01$.

\begin{figure}[!t]
\centering
\includegraphics*[width=.6\textwidth,clip]{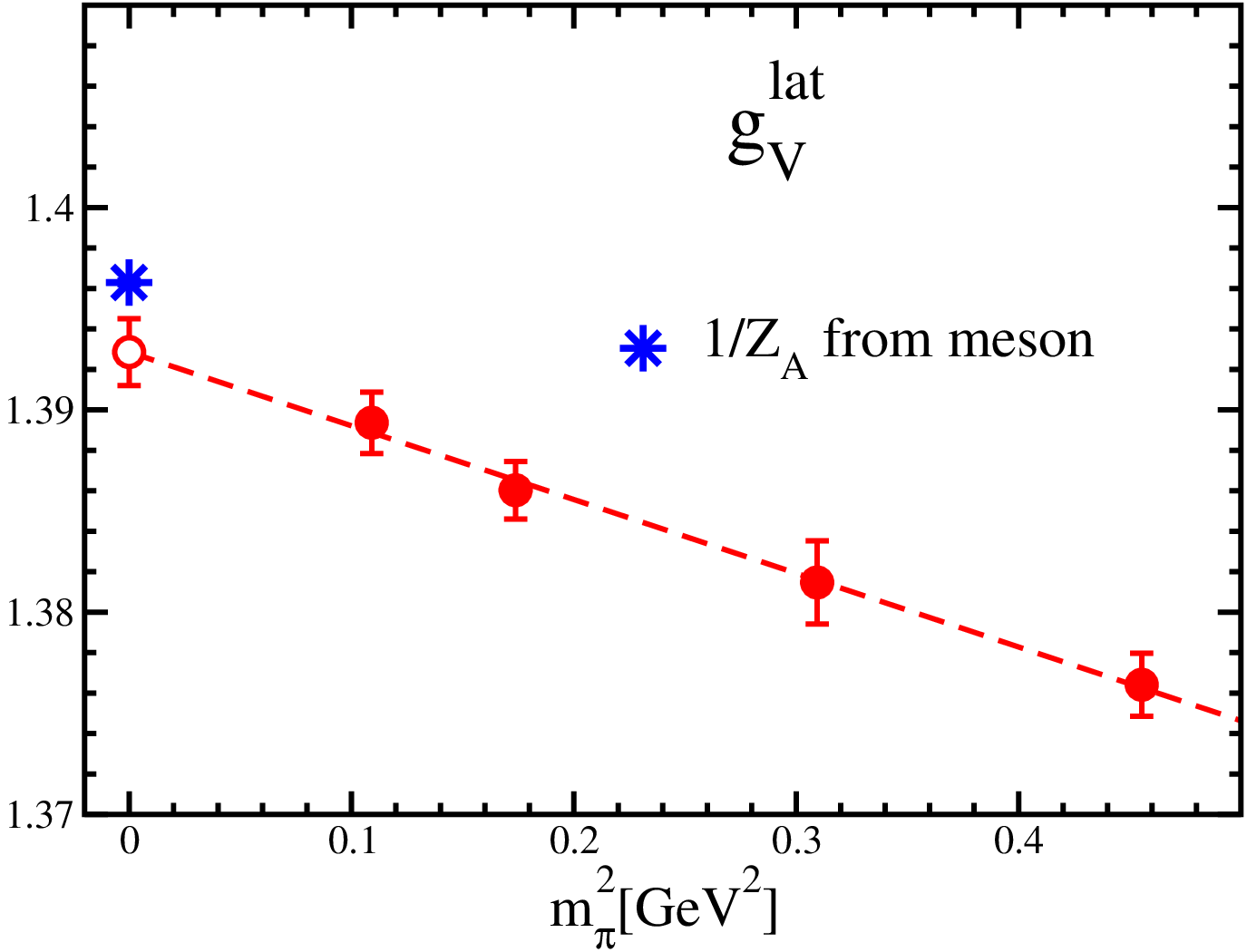}
\caption{
$g_V^{\mathrm{lat}}$ and $1/Z_A$ obtained from the pion-to-vacuum matrix element of the conserved axialvector current~\cite{Allton:2008pn}.
}
\label{fig:gv}
\end{figure}

A linear extrapolation to the chiral limit yields an accurate estimate of
\(g_V^{\mathrm{lat}}=1.3929(17)\), as shown in Fig.~\ref{fig:gv}.
This corresponds to \(Z_V=0.7179(9)\) in the chiral limit and agrees well with an independent calculation in the meson sector~\cite{Allton:2008pn}, \(Z_A=0.7161(1)\), up to the discretization error.

%
%
\label{sec:axial_charge}

The axial charge is calculated from the ratio of the vector and axialvector form factors
$\displaystyle{
g_A = F_A(0) / F_1(0).
}$
This ratio gives the renormalized axial charge since
the vector and axial currents, $V_\mu$ and $A_\mu$, share a common
renormalization thanks to the good chiral symmetry properties of DWF, up to small discretization error of \(O(a^2)\).

\begin{figure}[!t]
\centering
\includegraphics*[width=.6\textwidth,clip]{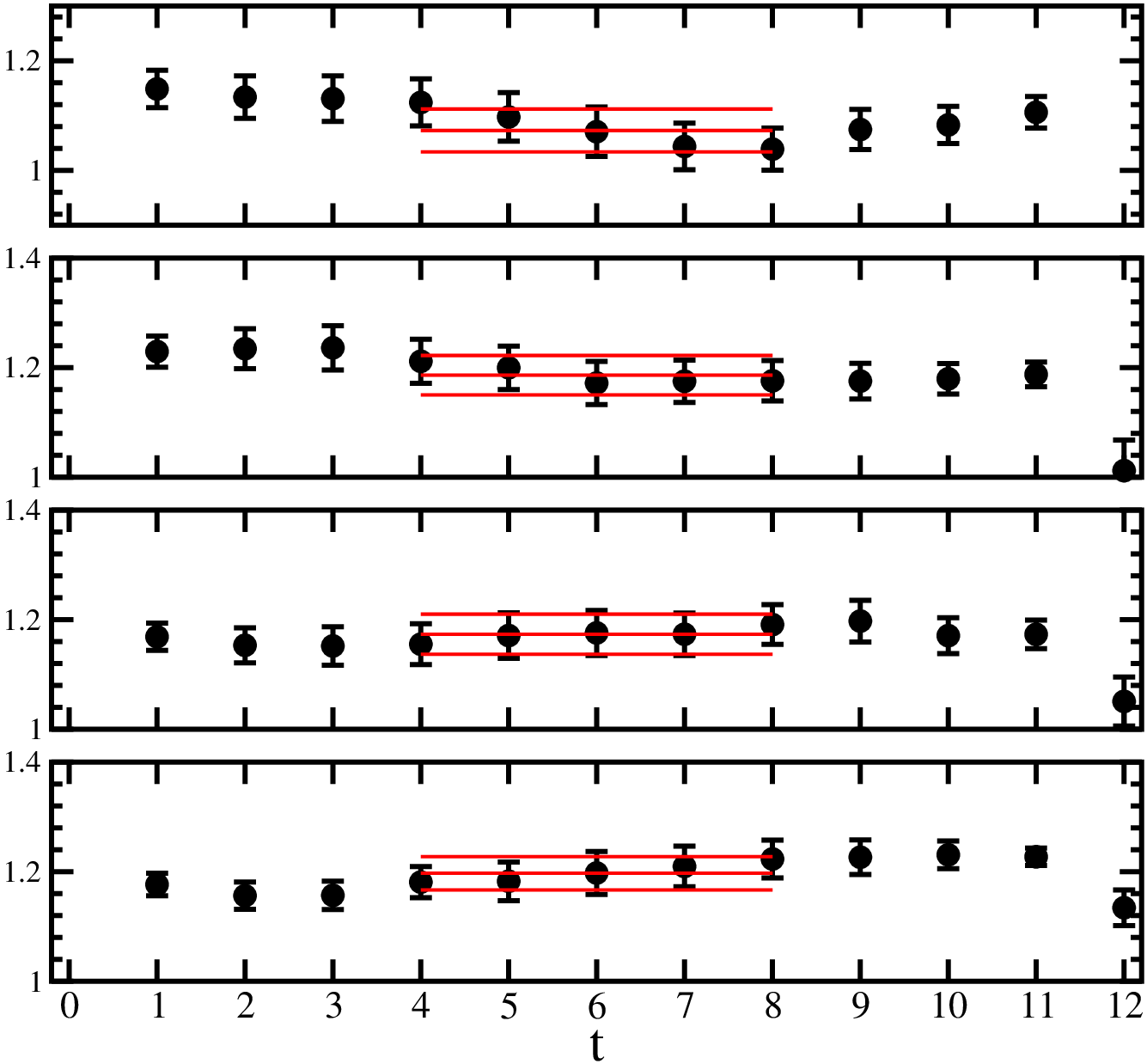}
\caption{
Plateaus of $g_A$. $V=(2.7$ fm$)^3$ and $m_f = 0.005$, 0.01, 0.02, and 0.03,
from top to bottom.
}
\label{fig:plateau_ga}
\end{figure}

The plateaus of $g_A$ computed on volume $V=(2.7$ fm$)^3$ are shown in Fig.~\ref{fig:plateau_ga}.
We checked that consistent results are obtained 
by either fitting or averaging over appropriate time slices,
$t=4$--8, and also by fitting the data symmetrized about $t=6$.
The data can be symmetrized
because the source and sink operators are identical in the limit of large statistics.
We note that the statistics at our lightest mass is the largest we 
know of for comparable simulation parameters in the literatures.
Results obtained from the fit using the
unsymmetrized data, presented in the figure with
one standard deviation, are employed in the analysis.
These results are compiled in table~\ref{tab:g_a}.
\begin{table}[!t]
\begin{tabular}{ccccc}\hline\hline
$m_f$ & 0.005 & 0.01 & 0.02 & 0.03 \\\hline
(2.7 fm)$^3$ & 1.073(39) & 1.186(36) & 1.173(36) & 1.197(30) \\
(1.8 fm)$^3$ & N/A       & 1.066(72) & 1.115(58) & 1.149(32) \\
\hline\hline
\end{tabular}
\caption{\label{tab:g_a} Summary of axial charge, $g_A$, for both volumes. }
\end{table}

\begin{figure}[!t]
\centering
\includegraphics*[width=.6\textwidth,clip]{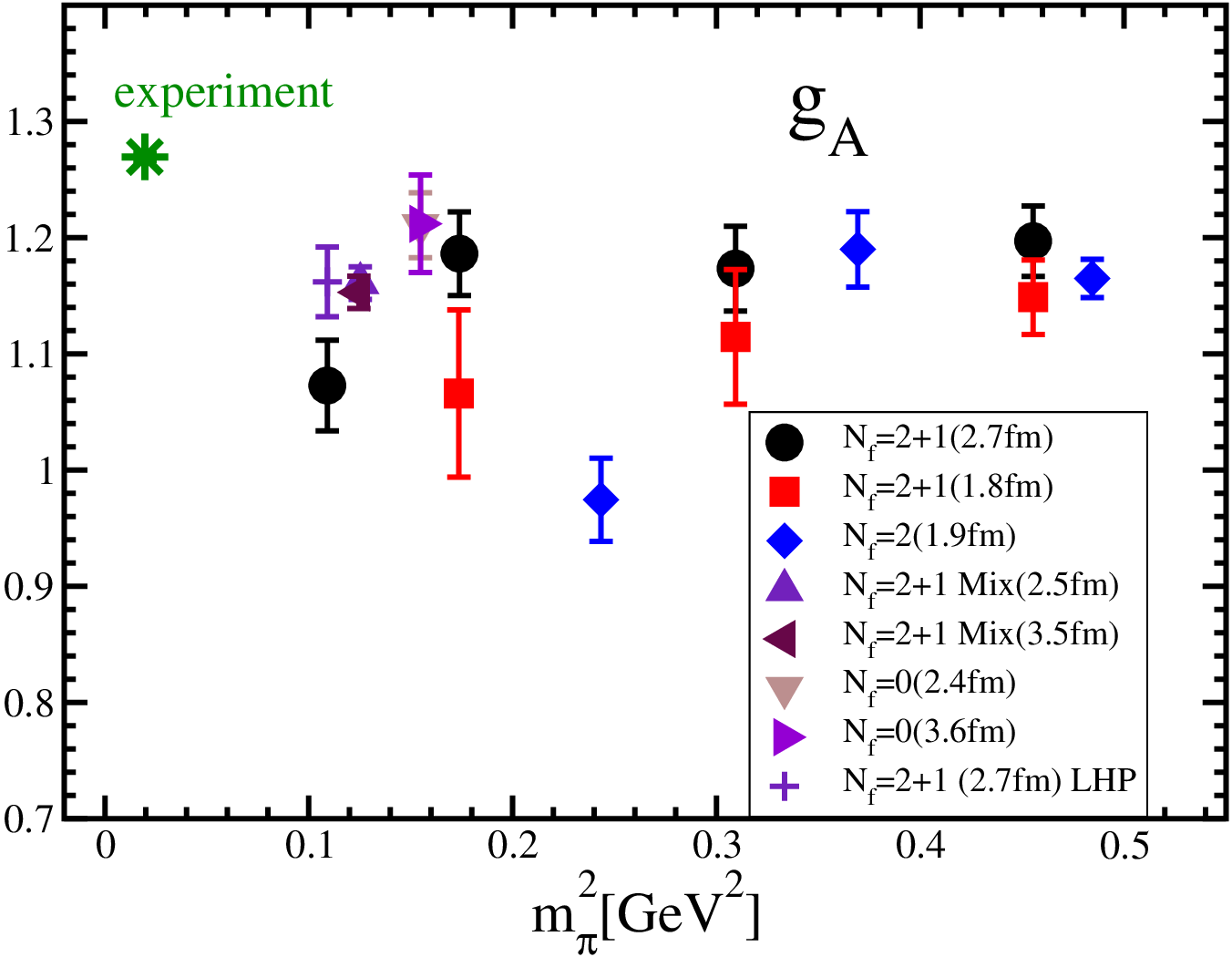}
\caption{
Axial charge $g_A$ together with two-flavor~\cite{Lin:2008uz} and quenched~\cite{Sasaki:2003jh,Sasaki:2007gw} DWF, and mixed action~\cite{Edwards:2005ym,Bratt:2008uf}
calculations. Recent $N_f = 2+1$ DWF by LHP~\cite{Bratt:2008uf} is also plotted.
}
\label{fig:ga_mpi}
\end{figure}

Figure~\ref{fig:ga_mpi} shows that the (2.7 fm)$^3$ data are almost independent of the pion mass (squared) except for the lightest point which is about 9\% smaller than the others.
A set of the results obtained with a smaller volume, (1.8 fm)$^3$
shows a similar downward behavior,
albeit with relatively larger statistical uncertainties.
An earlier two flavor calculation by 
RBC~\cite{Lin:2008uz} with spatial volume (1.9 fm)$^3$ 
and $1/a=1.7$ GeV showed a clear downward behavior, 
but it sets in at heavier pion mass.

We suspect that this pion mass dependence driving $g_A$ away from
the experimental value is caused by the finite 
volume of our calculation.
Similar behavior was observed in
quenched DWF studies~\cite{Sasaki:2003jh,Sasaki:2007gw} and was
predicted in a model calculation~\cite{Thomas:2005qm}.
However, for pion masses close to our lightest point
such a sizable shift
is not observed when $V$ is larger than about (2.4 fm)$^3$,
not only in the quenched case,
but also the 2+1 flavor, mixed action calculation in~\cite{Edwards:2005ym}
and their updated results~\cite{Bratt:2008uf}. 
Both the results of quenched~\cite{Sasaki:2003jh,Sasaki:2007gw} 
and mixed action~\cite{Bratt:2008uf} calculations
on larger volumes are presented in Fig.~\ref{fig:ga_mpi}.
On the other hand,
our results suggest that a volume of $V=(2.7$ fm)$^3$ is not
large enough to
avoid a significant finite volume effect on $g_A$
when $m_\pi \le 0.33$ GeV in dynamical fermion calculations.
It is worth noting that the bending of the axial charge
comes from only the axialvector part $F_A(0)$, since
the vector part $F_1(0)$ does not have such a pion mass dependence
(see Fig.~\ref{fig:gv}).

\begin{figure}[!t]
\centering
\includegraphics*[width=.5\textwidth,clip]{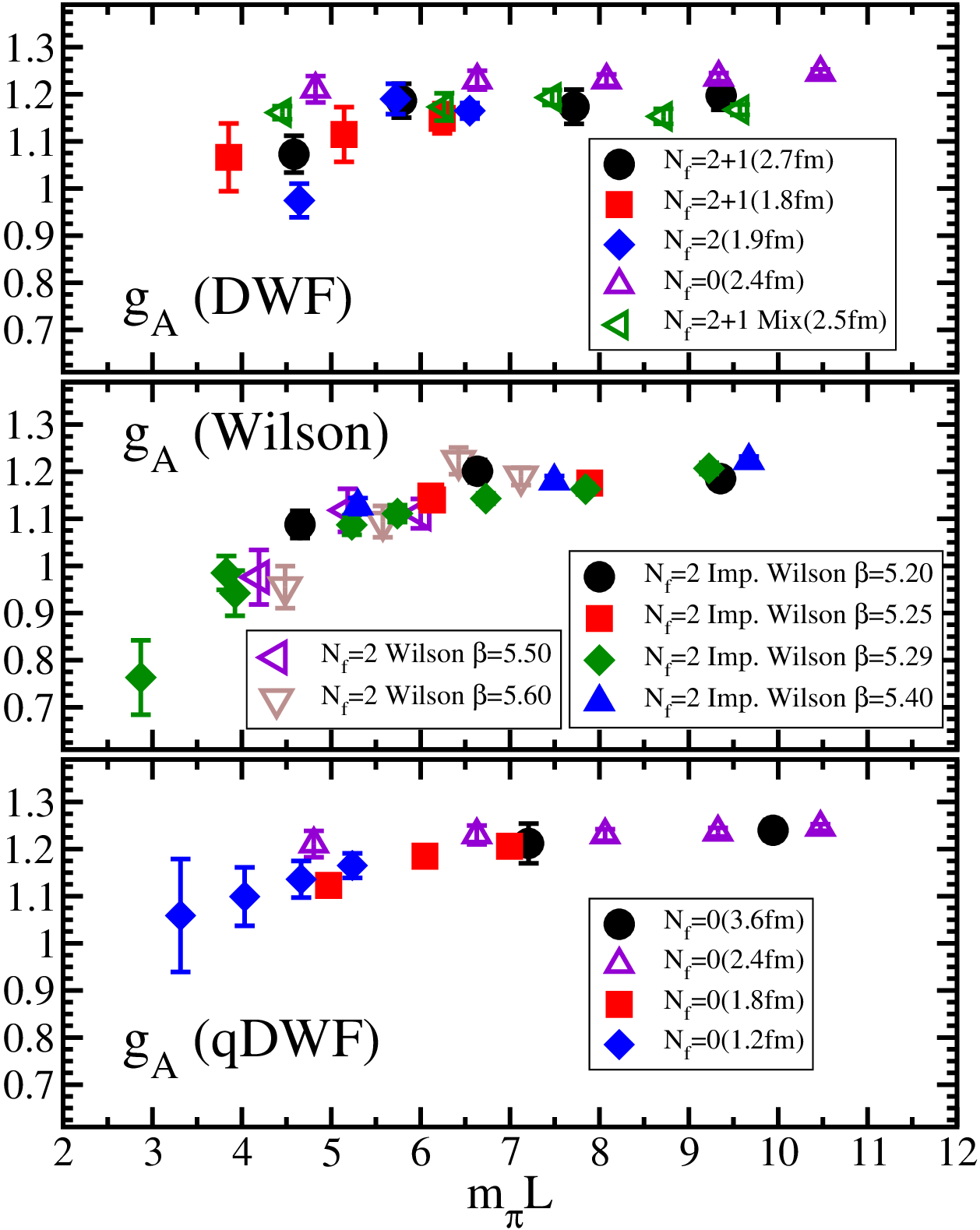}
\caption{
$m_\pi L$ scaling of the axial charge.
Top, middle, and bottom panels are dynamical DWF, dynamical Wilson,
and quenched DWF results~\cite{Sasaki:2003jh,Sasaki:2007gw}, respectively.
In the top panel circle, square, and diamond denote 2+1 flavor
larger, smaller volumes, and 2 flavor data~\cite{Lin:2008uz}, respectively,
and open up and left triangles denote quenched~\cite{Sasaki:2003jh}
and mixed action~\cite{Edwards:2005ym,Bratt:2008uf} data, respectively.
In the middle panel closed symbols denote
dynamical improved-Wilson fermions~\cite{Khan:2006de}, and
open symbols are dynamical Wilson 
fermion~\cite{Dolgov:2002zm,Alexandrou:2007xj}.
In the bottom panel, the open triangle is the same as in top panel.
}
\label{fig:ga_mpiL}
\end{figure}

In order to more directly compare the various results, we plot $g_A$ against the dimensionless quantity, $m_\pi L$, in the top panel of Fig.~\ref{fig:ga_mpiL}.
We find that the 2+1 flavor results on both volumes
reasonably collapse onto a single curve that monotonically increases
with $m_\pi L$;  in other words, they exhibit scaling in this variable.
The two flavor results~\cite{Lin:2008uz} display a similar behavior
which is also evident in dynamical two flavor 
(improved) Wilson fermion calculations
as shown in the middle panel~\cite{Khan:2006de,Dolgov:2002zm,Alexandrou:2007xj} for
the unitary points $\kappa_{\rm sea} = \kappa_{\rm val}$,
with various volumes (0.95--2.0 fm)$^3$,
pion masses 0.38--1.18 GeV, and gauge couplings.
While the trend is similar in the quenched DWF case~\cite{Sasaki:2003jh,Sasaki:2007gw} with pion masses in the range 0.39--0.86 GeV
and $1/a= 1.3$ GeV (see bottom panel),
the scaling is violated for the point with smallest $m_\pi L$
on $V=(2.4$ fm)$^3$.
The lightest point does not follow the (1.8 fm)$^3$ 
data: they differ by 2.5 standard deviations ($\sigma$)
at $m_\pi L \sim 5$, suggesting that there are non-universal terms that
depend separately on $m_\pi$ and $V$. In particular, this effect may be
due to the presence of a quenched chiral log~\cite{Kim:1996bz}.
From Ref.~\cite{Kim:1996bz}, the size of the effect at this mass can readily explain
the discrepancy observed with the dynamical $m_\pi L$ scaling.
Note, at this mass,
but going to $V=(3.6$ fm)$^3$, no finite volume effect is detected in 
the quenched case as can be seen in Fig.~\ref{fig:ga_mpi}.

The mixed action, 2+1 flavor result with a similar 
volume~\cite{Edwards:2005ym,Bratt:2008uf},
is denoted by the left triangle in the top panel.
We plot their recent result at our lightest point~\cite{Bratt:2008uf}.
At heavy pion masses the results are statistically
consistent with our larger volume data and essentially independent of $m_\pi L$.
At $m_\pi L \sim 4.5$ the mixed action result, however, is larger than
ours by (a combined) 2.1$\sigma$, and lies between
our lightest result and the quenched DWF result with (2.4 fm)$^3$ 
volume~\cite{Sasaki:2003jh} (the up triangle in the figure).

A possible explanation of the differences is that it is 
simply a dynamical fermion effect as discussed in Ref.~\cite{Yamazaki:2008py}.
While the mixed action result at $m_\pi L \sim 4.5$ has come down from higher value with larger error
(the previous result was consistent with the quenched result 
at the similar $m_\pi L$),
the explanation using the systematic error~\cite{Bar:2005tu,Prelovsek:2005rf} 
of the partially quenched effect of the mixed action results might 
be valid in the present data.
If the sea quark is effectively heavy,
a mixed action calculation will be closer to the quenched case.
Mixed action chiral perturbation
theory reveals the presence of partially-quenched logs whose size is
consistent with the observed effect~\cite{Jiang:2007sn,Chen:2007ug},
as in the quenched theory.
We should note that the preliminary result obtained by LHP~\cite{Bratt:2008uf}
at the same simulation parameter as our lightest point
appears inconsistent with our result (see Fig.~\ref{fig:ga_mpi}). 
We will discuss the difference later in this section.

For the chiral extrapolation of $g_A$,
we attempt to include the finite volume effect in our data.
While the pion mass dependence of $g_A$, including the finite volume effect, 
has been investigated in the small scale expansion
(SSE) scheme of heavy baryon chiral perturbation theory (HBChPT)~\cite{Khan:2006de}, 
the size of the finite volume effect
on $V=(2.7$ fm)$^3$ predicted in SSE
is less than 1\% in our pion mass region. The correction is much too small
to account for the observed finite volume effect in our data.
This suggests that the finite volume effect in HBChPT, which
is estimated by replacing all loop integrals by summations,
is not the leading finite volume effect in $g_A$, as in the $\varepsilon$ 
regime~\cite{Smigielski:2007pe}.
We also note that our attempts
to fit the mass dependence of the data to HBChPT failed, which is likely due
to the heavier quark mass points being beyond the radius of convergence of 
ChPT~\cite{Lin:2008uz,Bernard:2006te,Allton:2008pn}.

Instead of the SSE formula, we assume the following simple fit form, including 
the finite volume effect in a way that respects the scaling observed in the data,
\begin{equation}
A + B m_\pi^2 + C f_V(m_\pi L),
\label{eq:fit_g_A}
\end{equation}
with $f_V(x) = \mathrm{e}^{-x}$, and
where $A, B,$ and $C$ are fit parameters.
The third term corresponds to the observed finite volume effect,
taken as a function of $m_\pi L$ only, and vanishes rapidly 
towards the infinite volume limit, $L\to\infty$, at fixed pion mass.
The same $m_\pi L$ dependence appears in
one of the finite volume effect contributions in Ref.~\cite{Jaffe:2001eb}.
We note that this simple form is used to estimate the finite volume effects
in the data but {\it not} the value of $g_A$ in the chiral limit
at fixed $L$. In the end, we choose this simplest form, in part, because
the fit result at the physical point is not sensitive to the particular choice of $f_V(x)$, as
discussed below.

\begin{figure}[!t]
\centering
\includegraphics*[width=.6\textwidth,clip]{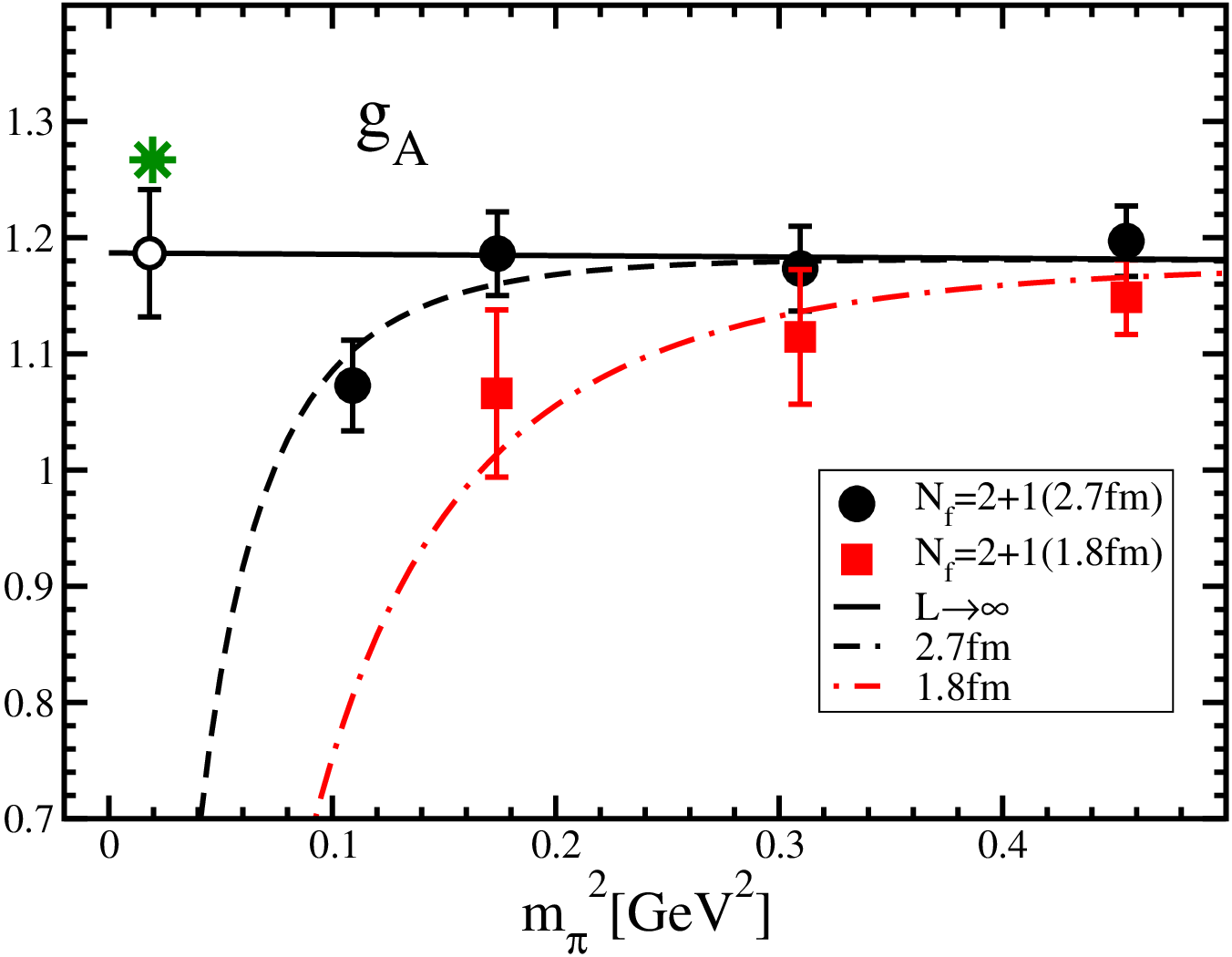}
\caption{
Chiral and infinite volume extrapolation of $g_A$ with 
finite volume effect $f_V = \mathrm{e}^{-x}$ in eq.~(\ref{eq:fit_g_A}). 
Fit is carried out with data on both the volumes, simultaneously.
}
\label{fig:ga_fit}
\end{figure}

In Fig.~\ref{fig:ga_fit} we see that the 2+1 flavor data are described very well by this simple fit ($\chi^2$/d.o.f$.=0.57$), using data computed on both volumes simultaneously.
The $L\to\infty$ extrapolation (solid line) in turn allows an extrapolation to the
physical pion mass ($m_\pi = 135$ MeV), $g_A = 1.19(6)(4)$,
where the first error is statistical.
The second error
is an estimate of the systematic error
determined by comparing this result with that from fits using different choices of $f_V(x)$, {\it e.g.},
the full form in~\cite{Jaffe:2001eb},
$x^{-3}$, and $m_\pi^2 \,\mathrm{e}^{-x} / x^{1/2}$. The latter is similar to HBChPT when 
$m_\pi L \gg 1$~\cite{Beane:2004rf,Detmold:2005pt,Khan:2006de}.
The results of some of the fit forms are summarized in table~\ref{tab:gagv_fit}.
\begin{table}[!t]
\begin{tabular}{cccccc}\hline\hline
$f_V$ & $A$ & $B$ & $C$ & $\chi^2/$d.o.f. & $m_\pi^{phys}$ \\\hline
$e^{-m_\pi L}$ & 
1.187(57) & $-$0.12(14) & $-$8.1(3.9) & 0.57 & 1.187(55) \\
$(m_\pi L)^{-3}$ & 
1.226(70) & $-$0.05(15) & $-$11.3(5.2) & 0.49 & 1.225(67) \\
$m_\pi^2 e^{-m_\pi L}/\sqrt{m_\pi L}$ &
1.148(46) & $-$0.09(12) & $-$75(41) & 0.80 & 1.150(44) \\
N/A &
1.172(58) & 0.05(1.7) & --- & 0.17 & 1.173(55) 
\\\hline\hline
\end{tabular}
\caption{\label{tab:gagv_fit} Fit results of $g_A$, together 
with the extrapolated result at $m_\pi^{phys}=135$ MeV. 
In the last row, the linear fit result
using only the three heavier points at $V=(2.7$ fm$)^3$ is presented.}
\end{table}
The extrapolated value is not
sensitive to the choice of $f_V$, and is also consistent
with a linear fit to the three heaviest points on the larger volume,
$g_A = 1.17(6)$.
The present data are insufficient  
to determine the detailed form of $f_V$, but do allow a reasonable estimate of the finite volume effect.

We also fit our data, with and without the $f_V$ term,
to the 2-loop formula from HBChPT~\cite{Bernard:2006te} and find
that the extrapolated result is less than 1 and  
that the fits are generally unstable. 
This is due to the many 
unknown low energy constants which cannot be determined accurately from only 
four data points, even if some of them are fixed. More importantly, 
though the 2-loop formula extends the range of applicability 
of the chiral expansion, 
it is still only large enough to include our lightest point, as 
demonstrated in Ref.~\cite{Bernard:2006te}.
The systematic error arising from 
the difference of the renormalization constants for $A_\mu$
and $V_\mu$ is much smaller than the quoted systematic error.
From the fit result with $f_V(x)=\mathrm{e}^{-x}$, 
we estimate that if one aims to keep finite volume effects
at or below 1\%, then for $m_\pi =0.33$ GeV, spatial sizes of 3.5--4.1 fm 
$(m_\pi L\approx 5.9$--$6.9$) are necessary.

As mentioned,  our lightest result on (2.7 fm)$^3$
differs from the preliminary findings from LHP~\cite{Bratt:2008uf} shown 
in  Fig.~\ref{fig:ga_mpi} by 1.8 $\sigma$. These calculations are carried out
with the same parameters except for the operator smearing and the 
time separation between the source and sink operators, 
$\Delta t=t_{\rm sink}-t_{\rm src}$.
So, while it is possible that this difference is simply due to
  the limited statistics in the preliminary result in
  \cite{Bratt:2008uf}, there is the possibility that this difference
  is due to a systematic error stemming from 
contaminations of higher excited states.
These contaminations will be negligible when the time separation of
the two nucleon operators in the three-point function, eq.~(\ref{eq:3pt_0}),
is large enough. The large separation, however, causes 
the statistical error of the three-point function to increase.
Thus, we employ a time separation of $\Delta t=12$, as 
described in Sec.~\ref{sec:Method}, while LHP uses $\Delta t=9$. 
While further investigation of this difference is desirable, it
is beyond the scope of this paper. 

Although there may be a systematic difference between our result and 
the result of LHP at the lightest quark mass on the (2.7 fm)$^3$ lattice,
all recent results (before chiral extrapolation) with dynamical quarks are about 10\% smaller than the experimental value.
In order to make a precise test of (lattice) QCD with the axial charge,
further study of the systematic errors as the
quark mass is decreased towards the physical point is required 
on large volumes.

%
%
\subsection{Form factors of the vector current}
\label{sec:vector}
\begin{figure}[!t]
\begin{center}
\includegraphics[width=.6\textwidth,clip]{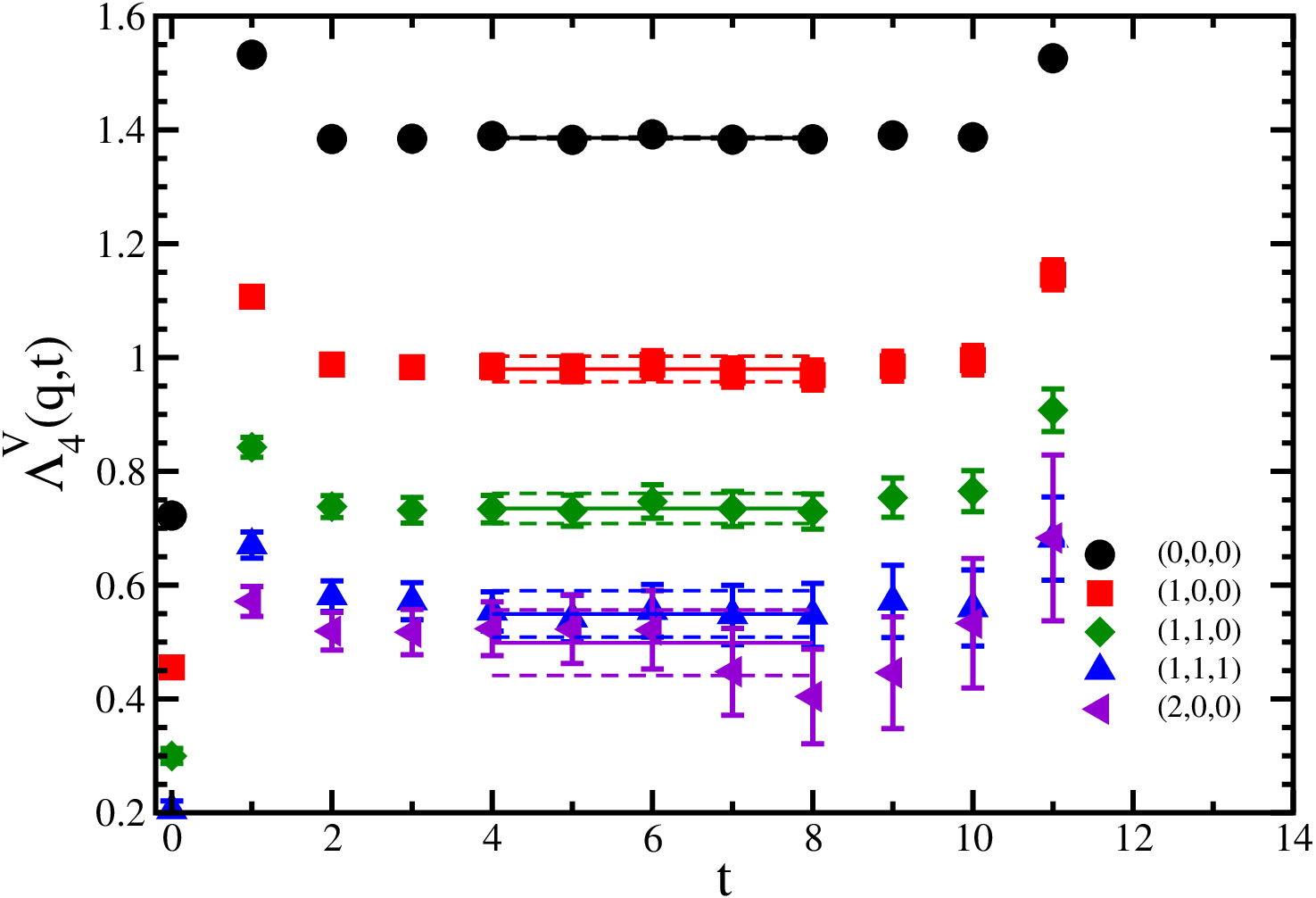}
\includegraphics[width=.6\textwidth,clip]{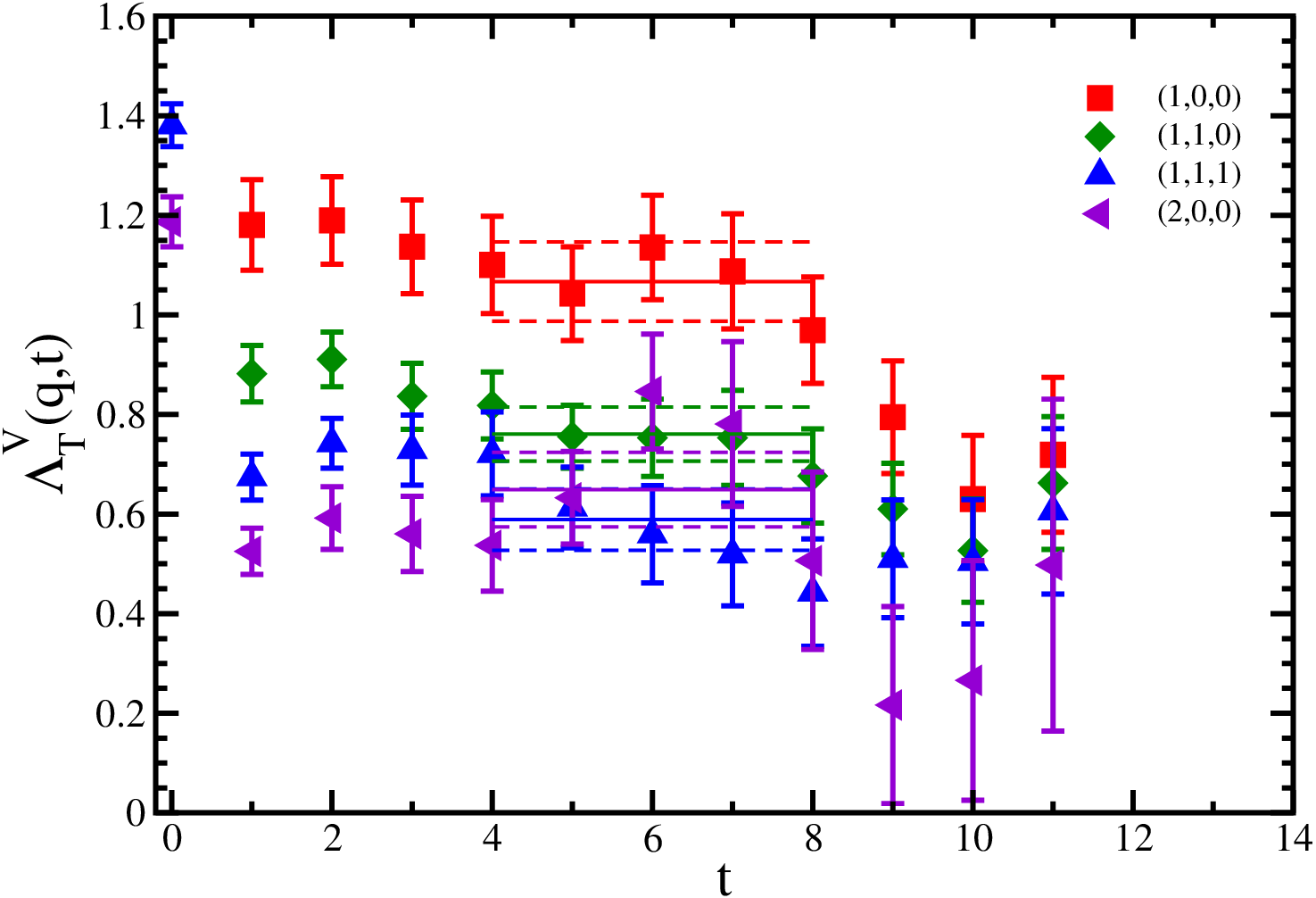}
\end{center}
\caption{
Ratios of 2- and 3-point functions of the vector current,
$\Lambda_4^V$ and $\Lambda_T^V$, at \(m_f=0.01\).
}
\label{fig:raw_vec_ff}
\end{figure}

In this subsection we discuss the isovector part of the Dirac and Pauli form factors,
$F_1(q^2)$ and $F_2(q^2)$.
In Fig.~\ref{fig:raw_vec_ff} we present the ratios of the three- and two-point functions, $\Lambda_4^V$ and $\Lambda_T^V$ defined in eqs.(\ref{eq:lambda_v0_t}) and (\ref{eq:lambda_vt_t}), at the quark mass \(m_f=0.01\) for each  
momentum transfer.
We find excellent plateaus in the middle time region between
the nucleon source and sink operators at $t=0$ and 12 for the smaller momenta,
while the plateau at $\vec{q} \propto (2,0,0)$ is not as well behaved 
and has a larger error. This is interpreted as simply a statistical
fluctuation. In order to remove this wiggle, we would need more statistics
at this momentum. 
To determine the values of the ratios, 
we perform a constant fit in the time interval, $t = 4$--8, for
all momentum combinations.

The form factors are obtained by solving
the linear equations (\ref{eq:lambda_v0}) and (\ref{eq:lambda_vt}),
\begin{eqnarray}
F_1(q^2) &=& \frac{\Lambda_4^V(q) + \tau \Lambda_T^V(q)}{1+\tau},
\ \mathrm{for\ all}\ q\\
F_2(q^2) &=& \frac{\Lambda_T^V(q) -   \Lambda_4^V(q)}{1+\tau},
\ \mathrm{for}\ q \ne 0
\end{eqnarray}
where $\tau = q^2/(4 M_N^2)$.
All the values of the two form factors are shown in table~\ref{tab:ffs}
in the appendix.
%

%
%
\subsubsection{Dirac form factor $F_1(q^2)$}
\begin{figure}[!t]
\begin{center}
\includegraphics[width=.6\textwidth,clip]{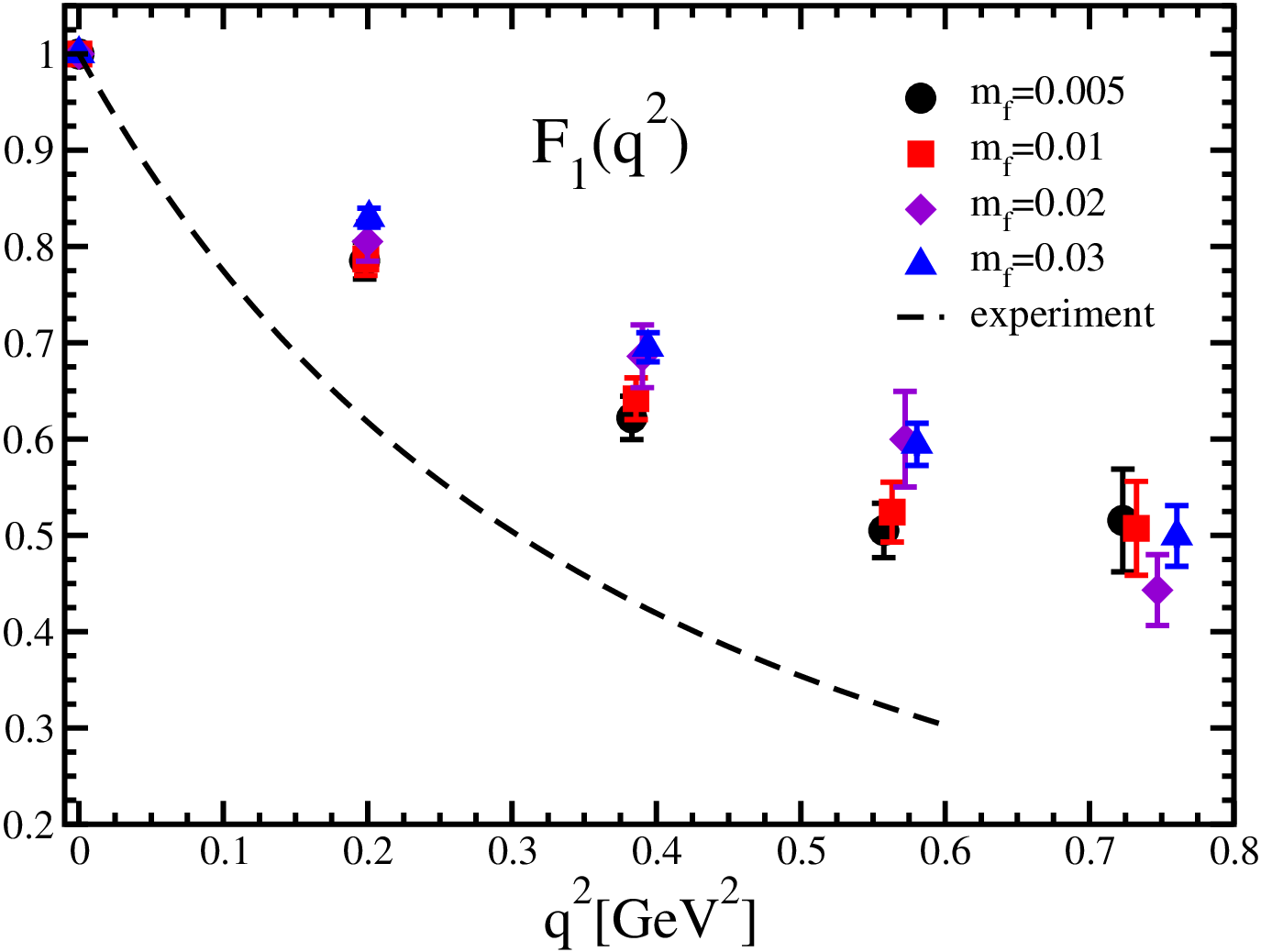}
\end{center}
\caption{The Dirac form factor, $F_1(q^2)$, normalized to unity at $q^2 = 0$.
}
\label{fig:f_1}
\end{figure}

Let us now turn our attention to the momentum dependence of the Dirac form factor.
In Fig.~\ref{fig:f_1} we present the form factor at each quark mass normalized by the respective values at zero momentum transfer.

\begin{figure}[!t]
\begin{center}
\includegraphics[width=.6\textwidth,clip]{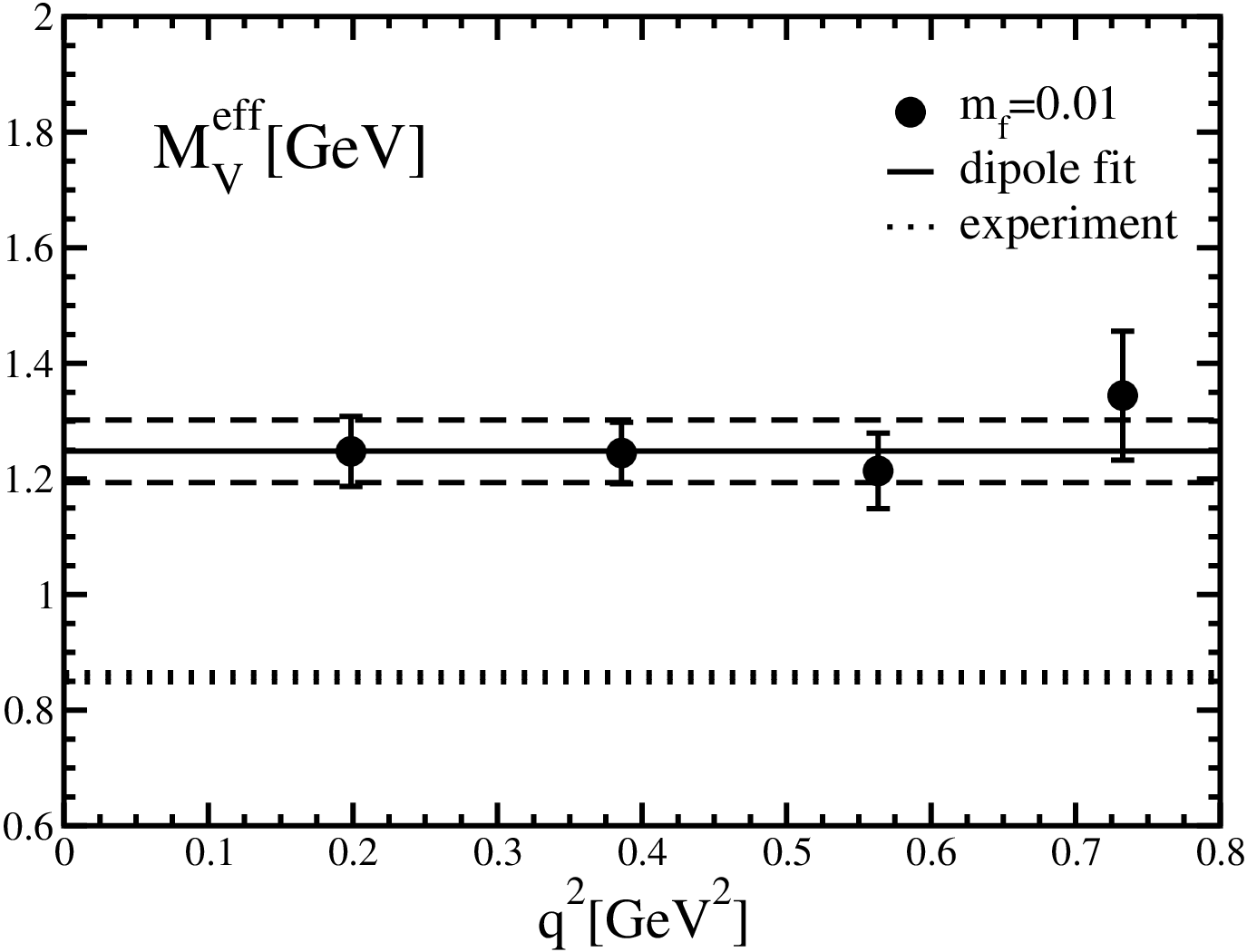}
\end{center}
\caption{
Effective dipole mass $M_1^{\mathrm{eff}}$ for $F_1(q^2)$ at $m_f = 0.01$ together with the
experimental result~\cite{Alexandrou:2006ru,Amsler:2008zz}.
Result of the dipole fit (solid line) with one standard deviation
(dashed line) is also presented.
}
\label{fig:eff_m_v}
\end{figure}

Phenomenologically the form factor is described by the conventional dipole form,  
\begin{equation}
F_1(q^2) = \frac{1}{(1+q^2/M_1^2)^2},
\label{eq:dipole_v}
\end{equation}
where $M_1$ is the dipole mass for this form factor, and fits to
experimental data give $M_1=0.857(8)$ GeV~\cite{Amsler:2008zz}.
In order to test the dipole form using our lattice results, 
for convenience we define an effective dipole mass
\begin{equation}
M_1^{\mathrm{eff}} = \sqrt{\frac{q^2}{\sqrt{1/F_1(q^2)}-1}}.
\end{equation}
Figure~\ref{fig:eff_m_v} shows that the effective dipole mass at $m_f = 0.01$ is almost flat against $q^2$. 
This means that
the form factor is well explained by the dipole form eq.(\ref{eq:dipole_v})
in the $q^2$ region where we measure.
The figure also shows that the effective mass is consistent with
the dipole fit result as expected.

\begin{figure}[!t]
\begin{center}
\includegraphics[width=.6\textwidth,clip]{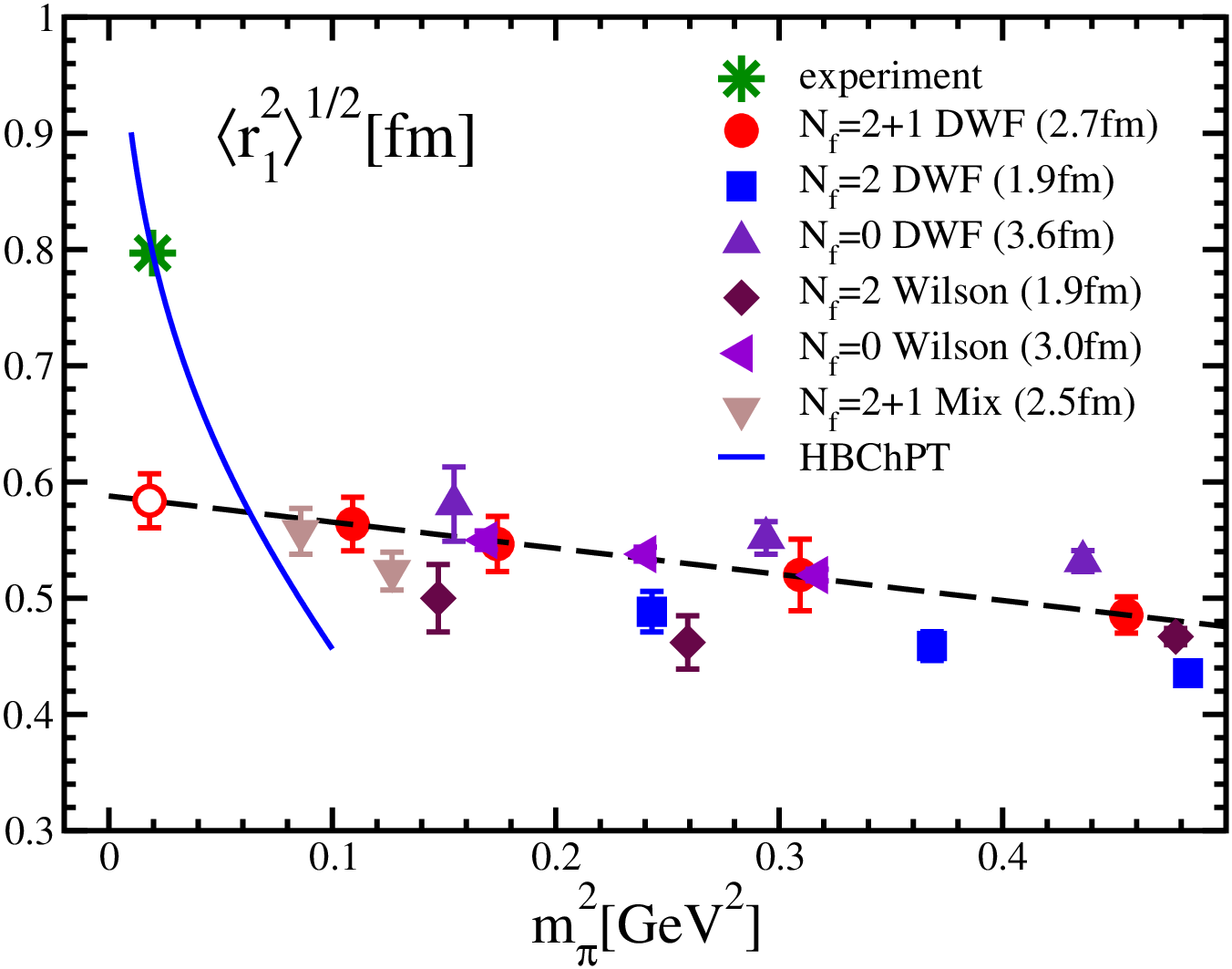}
\end{center}
\caption{
Dirac rms radius 
$\displaystyle{\langle r_1^2\rangle^{1/2}}$ determined from a dipole fit.
Dashed line represents a linear extrapolation of our results.
Square, up triangle, diamond, left triangle and down triangle denote 
two-flavor~\cite{Lin:2008uz} and quenched DWF~\cite{Sasaki:2007gw},
two-flavor and quenched Wilson~\cite{Alexandrou:2006ru},
and mixed action~\cite{Bratt:2008uf} calculations, 
respectively. A prediction from HBChPT with the
experimental result~\cite{Alexandrou:2006ru,Amsler:2008zz} 
is also plotted. 
}
\label{fig:r_1}
\end{figure}

We estimate the Dirac root mean-squared (rms) radius from the dipole
mass obtained by the fit as
\begin{equation}
\langle r_1^2\rangle^{1/2} = \frac{\sqrt{12}}{M_1},
\end{equation}
whose results are presented in table~\ref{tab:r1_r2_ft}.
Figure~\ref{fig:r_1} shows the pion mass dependence of our results for 
the rms radius. Here we also compare with other lattice calculations 
and the experimental value. Our results show a near-linear dependence 
in the pion mass squared which is quite different from the axial charge 
in Sec.~\ref{sec:axial_charge}. This suggests that the Dirac form
factor is less sensitive to finite volume effect than $g_A$, and
this is confirmed by an analysis of our results obtained on a
smaller volume $(1.8\,{\rm fm})^3$, shown in Fig.~\ref{fig:f_1_c}.
The smaller volume results are summarized in the appendix.
\begin{figure}[!t]
\begin{center}
\includegraphics[width=.6\textwidth,clip]{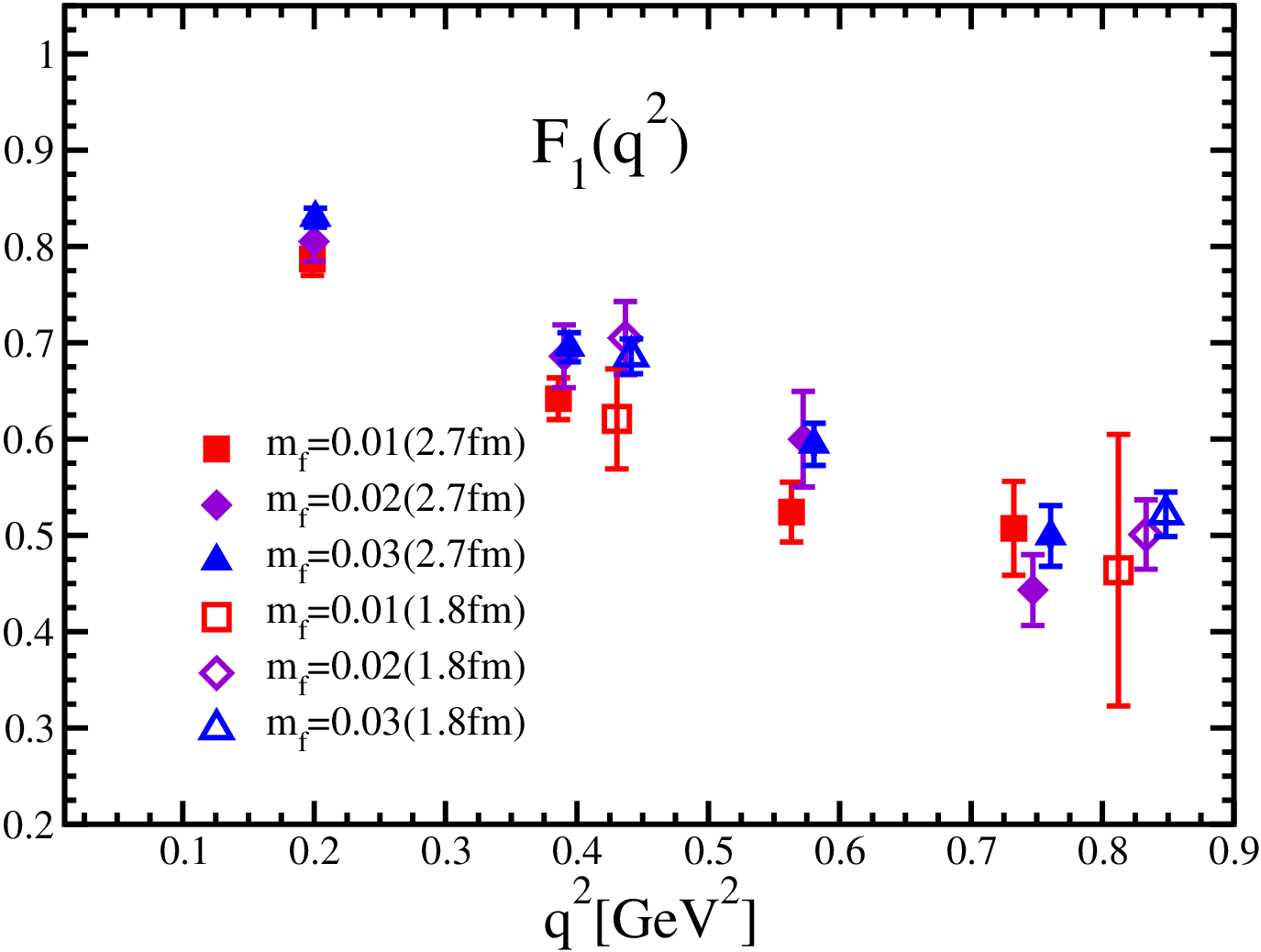}
\end{center}
\caption{
Comparison of $F_1$ with larger and smaller volumes denoted by 
closed and open symbols, respectively, at each quark mass.
}
\label{fig:f_1_c}
\end{figure}
Our results can be fit linearly and extrapolated to a value 
27\% smaller than experiment, 0.797(4) fm.
Other lattice calculations~\cite{Lin:2008uz,Sasaki:2007gw,Alexandrou:2006ru,Boinepalli:2006xd,Gockeler:2007hj} 
show similar trends.
The recent results of the mixed action calculation~\cite{Bratt:2008uf}
are also statistically consistent with our data and fit line.

This quantity is expected to logarithmically diverge 
in HBChPT~\cite{Beg:1973sc,Bernard:1998gv,Wang:2008vb}
at the chiral limit: such a behavior will help in bringing our present 
extrapolated results closer to experiment.
However, our results at $m_\pi>0.33$ GeV fail to reveal such a 
logarithmic divergence.
A naive determination of the HBChPT parameters at the physical point 
give the logarithmic contribution shown in Fig.~\ref{fig:r_1} 
by the solid line.
Future work will require simulations to be performed at lighter 
quark masses, {\it e.g.,} $m_\pi < 0.2$ GeV, if such logarithmic 
effects are to be seen in lattice results of the Dirac radius.

%
%
\subsubsection{Pauli form factor $F_2(q^2)$}
\begin{figure}[!t]
\begin{center}
\includegraphics[width=.6\textwidth,clip]{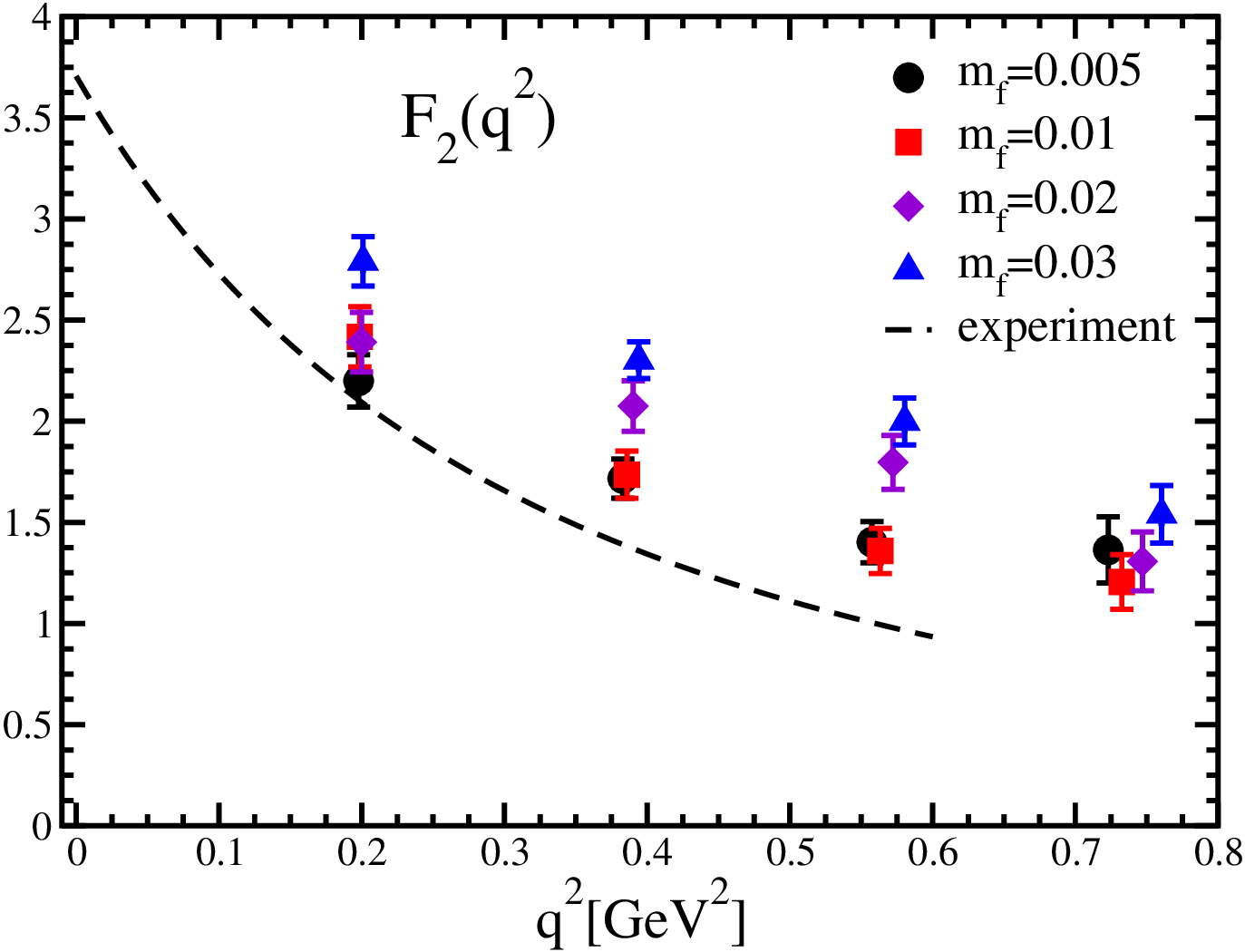}
\end{center}
\caption{The Pauli form factor, $F_2(q^2)$, renormalized by $Z_V = 1/F_1(0)$. The dashed curve is a fit to experimental data.
}
\label{fig:f_2}
\end{figure}

Figure~\ref{fig:f_2} shows the momentum-transfer dependence of our
results for the Pauli form factor at each quark mass.
These values are tabulated in table~\ref{tab:ffs}.
The form factor is renormalized by $F_1(0)$.

This form factor can also be described by the conventional dipole form,
\begin{equation}
F_2(q^2) = \frac{F_2(0)}{(1+q^2/M_2^2)^2},
\end{equation}
with $M_2 = 0.78(2)$ GeV and $F_2(0) = 3.70589$ extracted from fits to experimental data. 
In contrast to the Dirac form factor, there are two parameters, 
the over-all strength \(F_2(0)\) 
and the dipole mass $M_2$: the former gives the isovector part of 
the anomalous magnetic moment, \(\mu_p-\mu_n-1\), 
and the latter the Pauli mean-squared radius,
$\langle r_2^2 \rangle = 12/M_2^2$, as in the Dirac case.
We fit the form factor with these two parameters.

\begin{figure}[!t]
\begin{center}
\includegraphics[width=.6\textwidth,clip]{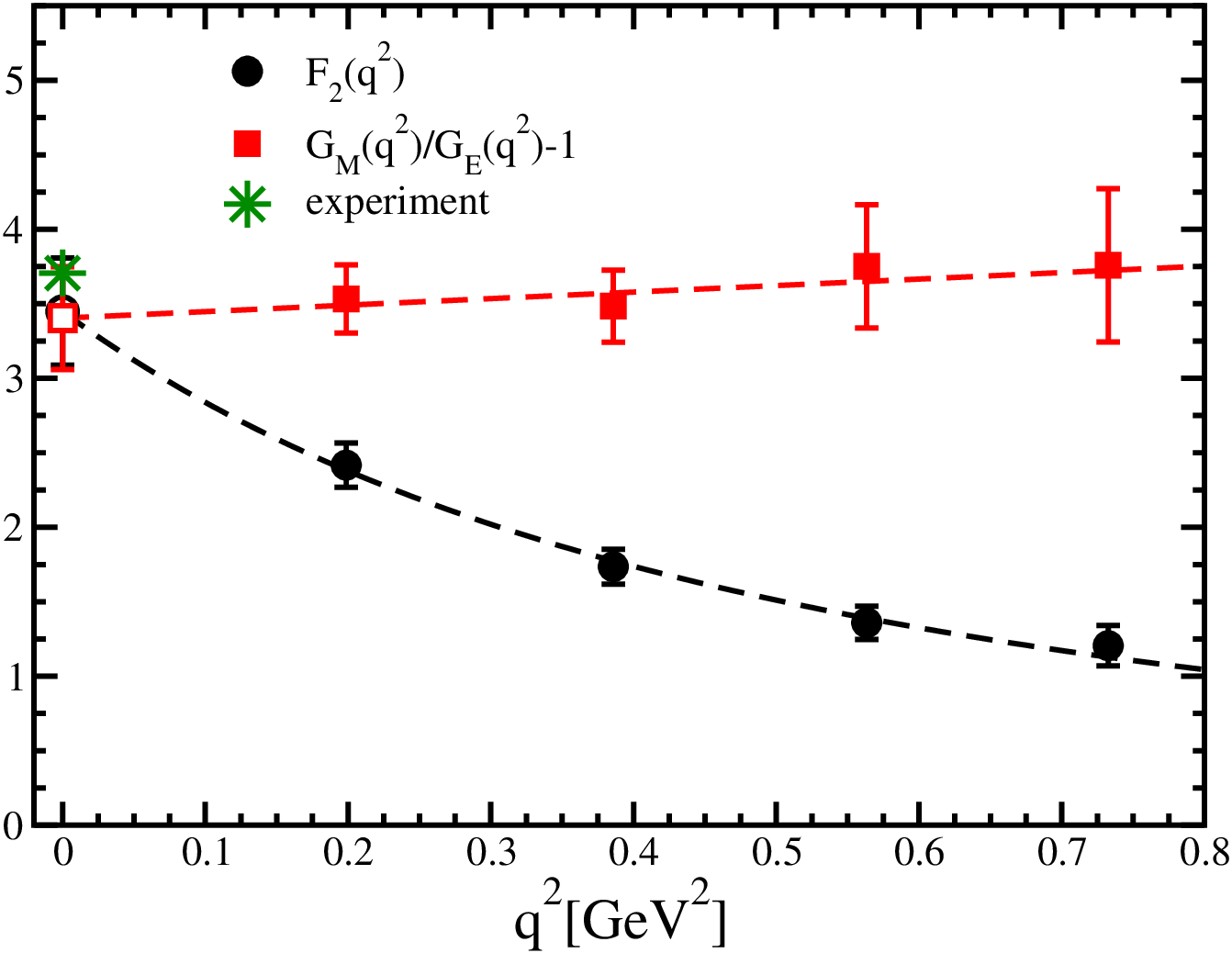}
\end{center}
\caption{
Dipole fit with $F_2(q^2)$
and linear fit with ratio of electric and magnetic form factors
$G_M(q^2)/G_E(q^2)-1$ at $m_f =0.01$.
}
\label{fig:f_me}
\end{figure}

To check reliability of the dipole fit, we measure the ratio of
the Sachs electric and magnetic form factors
\begin{equation}
\frac{G_M(q^2)}{G_E(q^2)} = \frac{\Lambda_T^V(q)}{\Lambda_4^V(q)}.
\end{equation}
At zero momentum transfer, we obtain $1+F_2(0)$ from the ratio.
Figure~\ref{fig:f_me} shows that the result for
$G_E(q^2)/G_M(q^2) - 1$ at $q^2 = 0$, obtained
via a linear fit in $q^2$, is consistent with the determination from a 
dipole fit of $F_2(q^2)$.

\begin{figure}[!t]
\begin{center}
\includegraphics[width=.6\textwidth,clip]{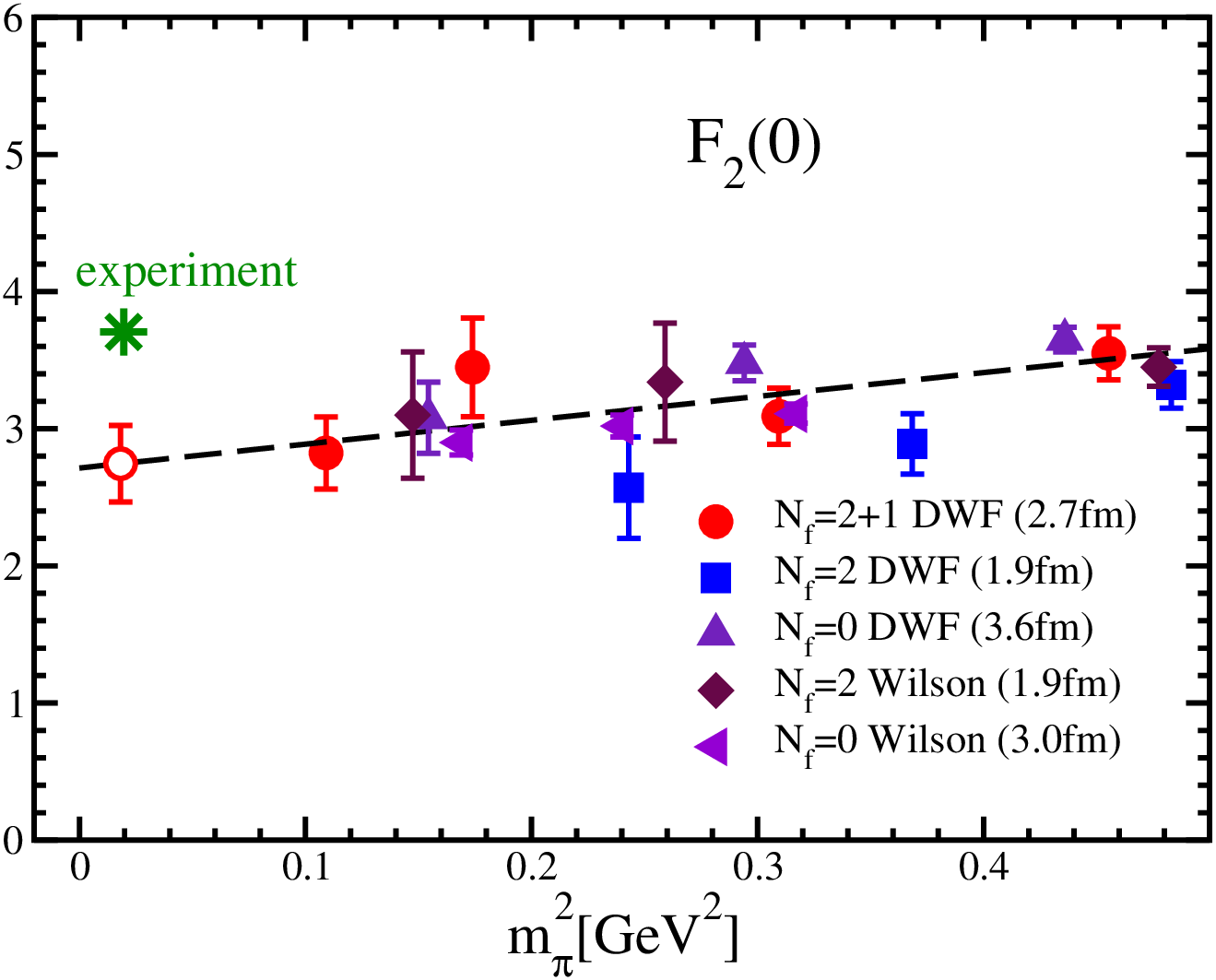}
\end{center}
\caption{
Same as Fig.~\ref{fig:r_1}
except anomalous magnetic moment, $F_2(0) = \mu_p - \mu_n - 1$, 
determined from dipole fit.
Experimental result~\cite{Amsler:2008zz} is also shown.
}
\label{fig:f_2_0}
\end{figure}

In Fig.~\ref{fig:f_2_0} we present the anomalous magnetic moment of
the nucleon, determined by the dipole fit presented in
table~\ref{tab:r1_r2_ft}, together with some other lattice QCD
calculations and the experimental value.
Our present results slightly decrease with the pion mass, in agreement with previous 
lattice calculations~\cite{Sasaki:2007gw,Alexandrou:2006ru}.
They extrapolate well linearly in the pion mass squared, and result in a value
26\% smaller than the experiment.
This result at the physical pion mass is consistent with those of
previous calculations~\cite{Sasaki:2007gw,Gockeler:2003ay}
using a linear fit.

\begin{figure}[!t]
\begin{center}
\includegraphics[width=.6\textwidth,clip]{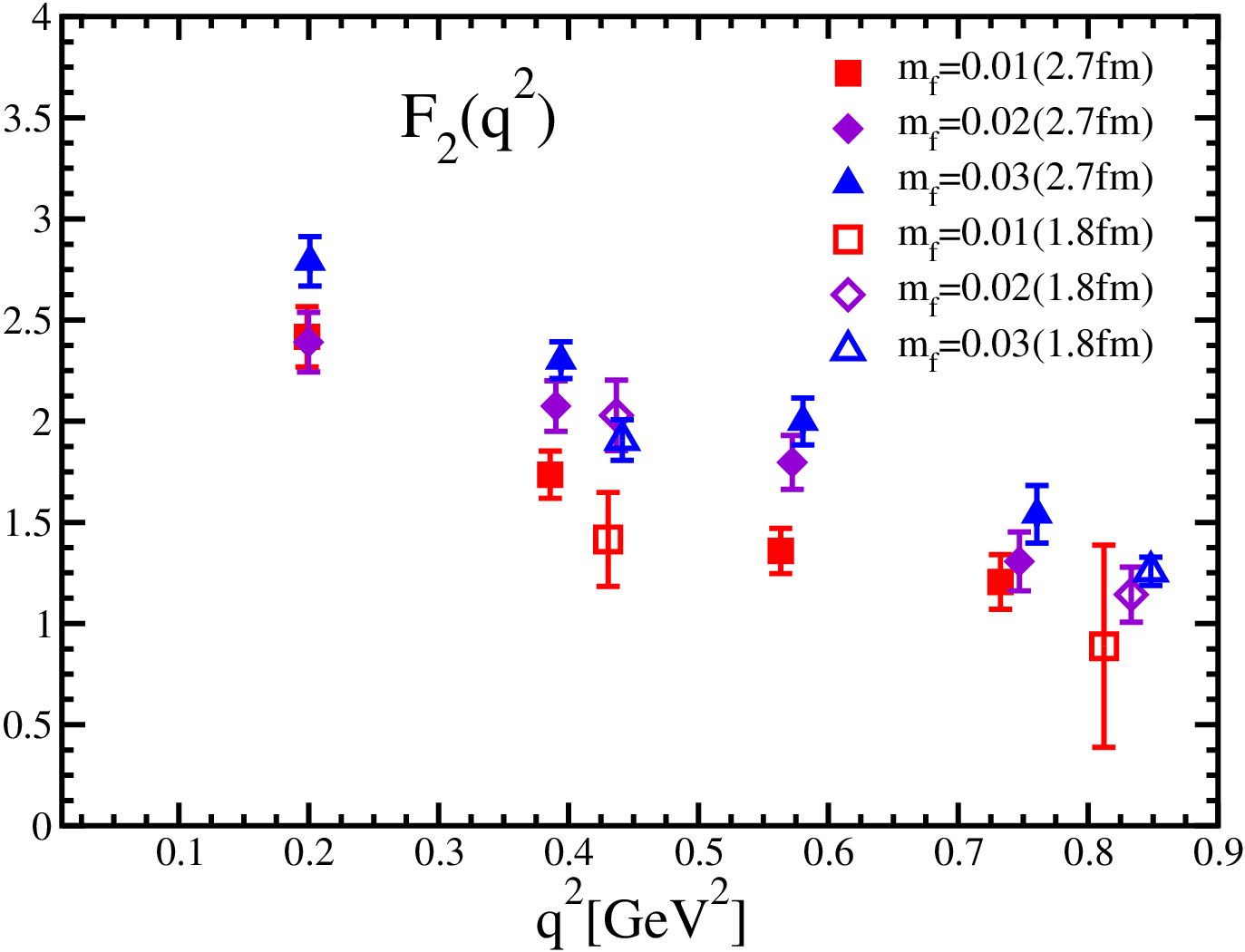}
\end{center}
\caption{
Comparison of $F_2$ with larger and smaller volumes denoted by 
closed and open symbols, respectively, at each quark mass.
}
\label{fig:f_2_c}
\end{figure}

\begin{figure}[!t]
\begin{center}
\includegraphics[width=.6\textwidth,clip]{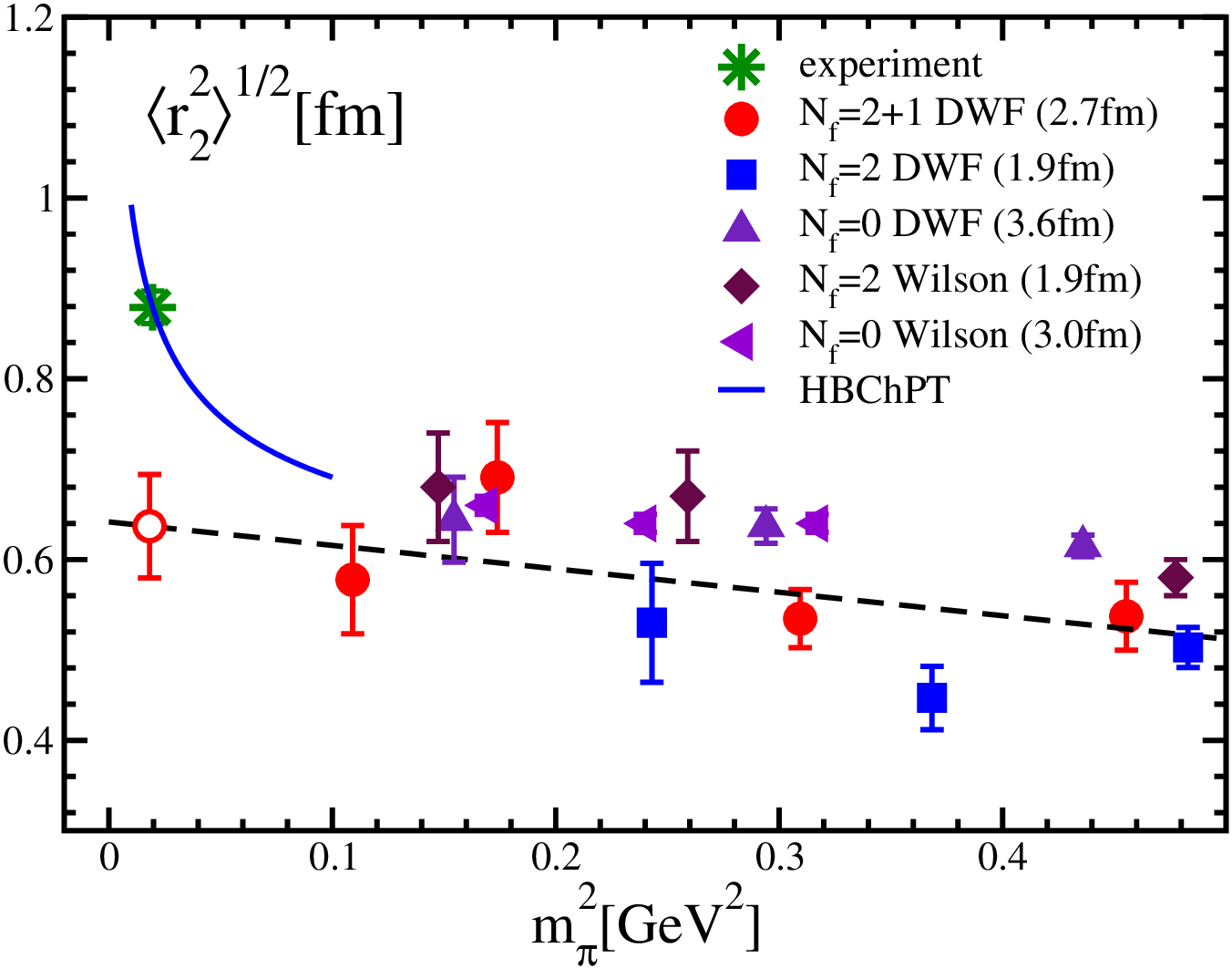}
\end{center}
\caption{
Same as Fig.~\ref{fig:r_1}
except Pauli rms radius
$\displaystyle{\langle r_2^2\rangle^{1/2}}$ determined from dipole fit.
Prediction of HBChPT with 
experimental result~\cite{Alexandrou:2006ru,Amsler:2008zz} 
is also plotted. 
}
\label{fig:r_2}
\end{figure}

We present in Fig.~\ref{fig:r_2} the result of the Pauli rms radius.
These results are obtained from a dipole fit and summarized 
in table~\ref{tab:r1_r2_ft}.
Some other lattice QCD calculations~\cite{Sasaki:2007gw,Alexandrou:2006ru} 
are also plotted in the figure for comparison.
We find the lightest point to be slightly smaller than 
the results at the other quark masses,
albeit with a large error. Thus, we consider this pion mass dependence
is due to statistics, not a finite volume effect as in 
the axial charge, and this is confirmed by our results from the
smaller volume simulations in Fig.~\ref{fig:f_2_c}.
The results are reasonably fitted by a linear function
of the pion mass squared, and we obtain 
$\langle r_2^2\rangle^{1/2} = 0.64(6)$ fm at the physical pion mass.
This result again is 27\% smaller than the
experimental value, 0.88(2) fm.

Here again the quantity is expected to diverge as $1/\sqrt{m_\pi}$ 
in the chiral limit in HBChPT~\cite{Beg:1973sc,Bernard:1998gv,Wang:2008vb},
however our results do not indicate such divergence.
In contrast to the Dirac radius case, perhaps because of the larger 
statistical errors, HBChPT can 
simultaneously fit the experiment and our data.
The fit with the prediction, however, gives larger $\chi^2/$d.o.f.
(degrees of freedom), and twice larger value at the physical pion mass
than the linear fit.
We need further light quark mass calculation with better statistics
to test the prediction in the lattice QCD calculation.

%
%
%
%
\begin{table}[!t]
\begin{tabular}{cccccc}\hline\hline
$m_f$ & 0.005 & 0.01 & 0.02 & 0.03 & $m_\pi^{\mathrm{phys}}$\\\hline
$\langle r_1^2 \rangle^{1/2}$ [fm] 
      & 0.564(23) & 0.548(24) & 0.520(31) & 0.485(16) & 0.584(23)\\
$\langle r_2^2 \rangle^{1/2}$ [fm] 
      & 0.578(60)  & 0.690(61) & 0.536(21) & 0.537(38) & 0.636(57) \\
$F_2(0)$ 
      & 2.82(26)  & 3.40(35) & 3.11(21) & 3.55(19) & 2.75(28) \\
\hline\hline
\end{tabular}
\caption{\label{tab:r1_r2_ft}
Dirac and Pauli rms radii $\langle r_1^2 \rangle^{1/2}$, 
$\langle r_2^2 \rangle^{1/2}$, and 
anomalous magnetic moment $F_2(0) = \mu_p-\mu_n-1$.
The linear fit results at $m_\pi^{\mathrm{phys}} = 135$ MeV
are also presented.
}
\end{table}

%
%
\subsection{Form factors of the axialvector current}
\label{sec:axial_vector}
\begin{figure}[!t]
\centering
\includegraphics[width=.6\textwidth,clip]{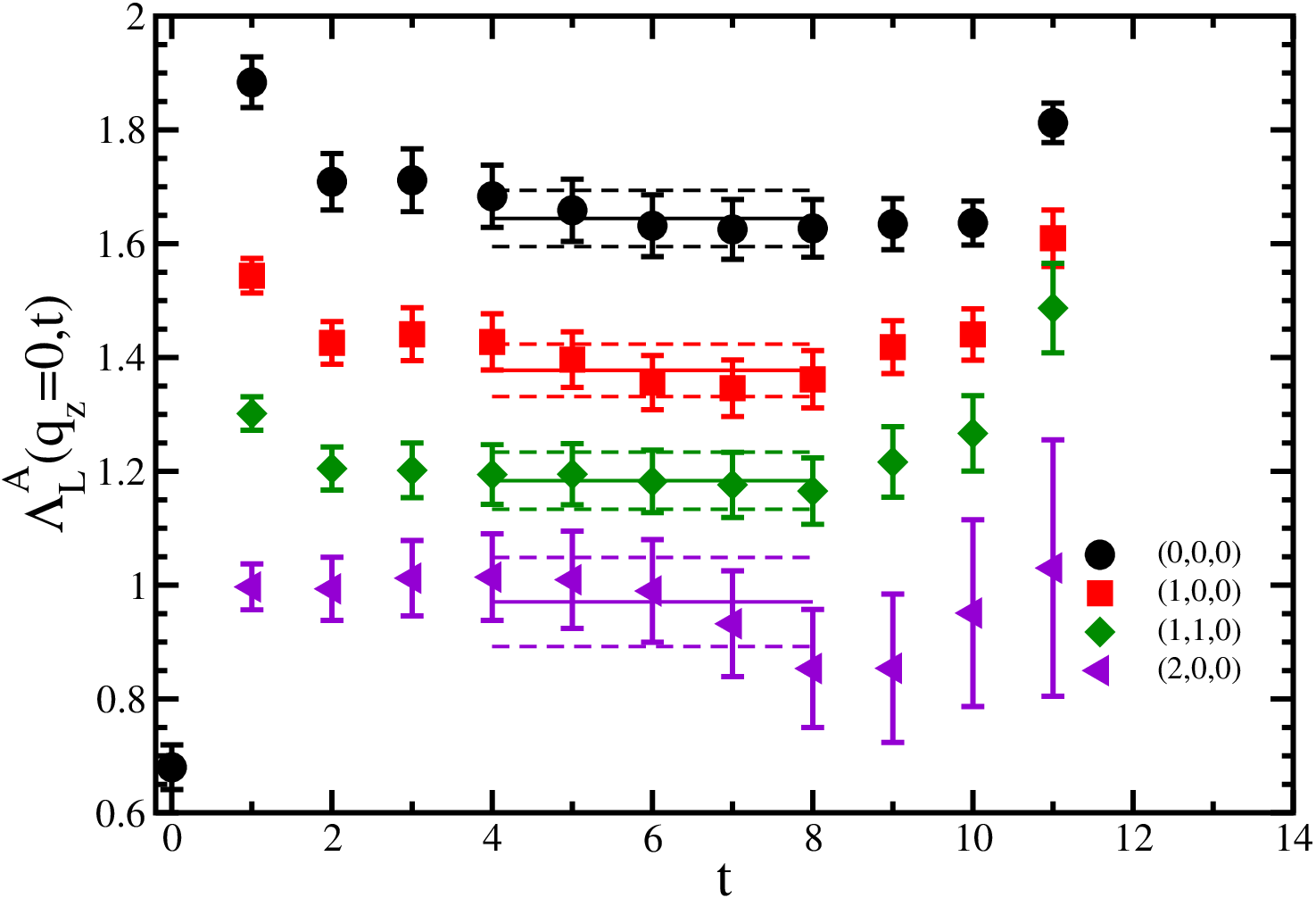}
\includegraphics[width=.6\textwidth,clip]{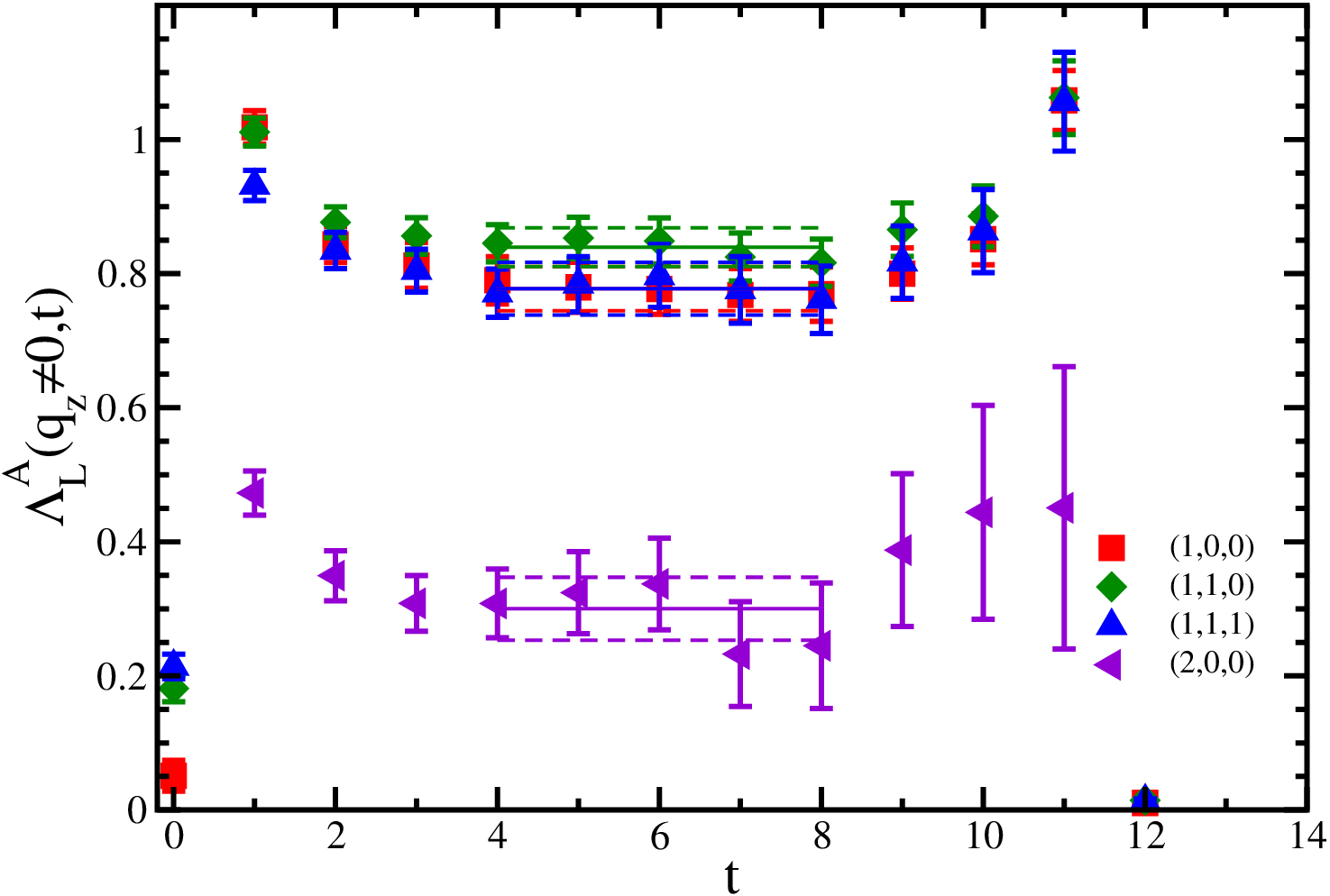}
\includegraphics[width=.6\textwidth,clip]{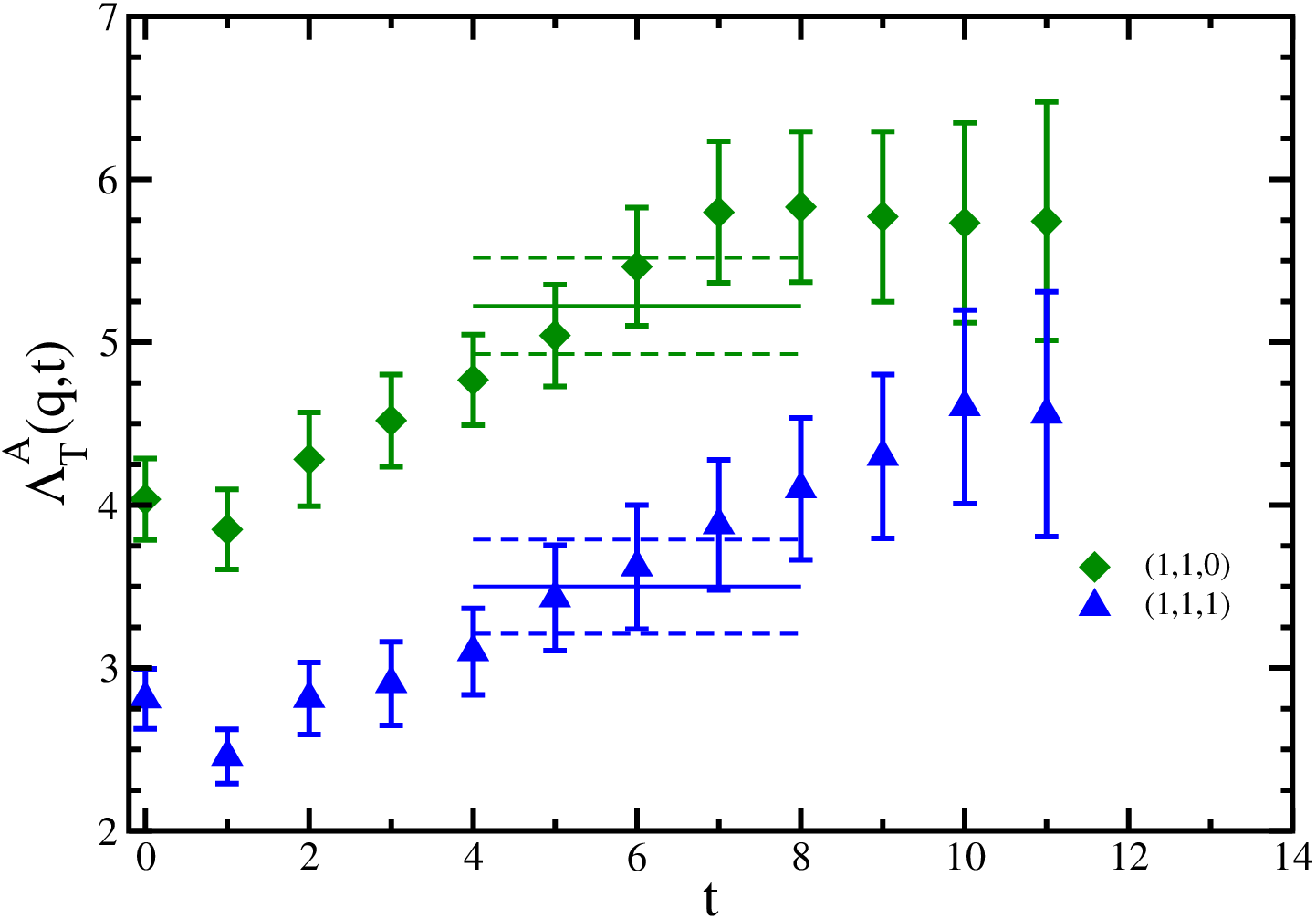}
\caption{
Ratios of 2- and 3-point functions for axialvector current,
$\Lambda_L^A(q_3=0,t), \Lambda_L^A(q_3\ne 0,t)$ and $\Lambda_T^A(q,t)$ at \(m_f=0.01\).
}
\label{fig:3pts_axa}
\end{figure}

In this subsection we show the form factors obtained from the axialvector currents,
$F_A(q^2)$ and $F_P(q^2)$.
They are extracted from the ratios of three- and two-point functions defined in eqs.(\ref{eq:lambda_al_t})
and (\ref{eq:lambda_at_t}).
Figure~\ref{fig:3pts_axa} shows that the typical plateaus of the
ratios with the $A_3$ component of the current at $m_f = 0.01$ are reasonably flat in
the middle time region between the source and sink operators.
We plot the ratios $\Lambda_L^A(q_3 = 0,t)$ and $\Lambda_L^A(q_3 \ne
0,t)$ separately, since $\Lambda_L^A(q_3 \ne
0,t)$ contains both form factors, while
$\Lambda_L^A(q_3 = 0,t)$ contains only $F_A(q^2)$.
It is worth noting that there is no 
$\Lambda_L^A(q_3 = 0,t)$ 
in the case of $\vec{q}\propto (1,1,1)$.
$\Lambda_T^A(q,t)$ has a slope in the range $t=1$--8 with large statistical errors 
as shown in the bottom panel of Fig.~\ref{fig:3pts_axa}.
We consider the slope to be caused by poor statistics
in the data.
The values of the matrix elements for all the ratios are determined by constant fits with the range of $t=4$--8.

Using the relations eqs.~(\ref{eq:lambda_al}) and (\ref{eq:lambda_at}), the two form factors are determined through the following 
equations which depend on the spatial momentum transfer in the 
three-point function,
\begin{eqnarray}
F_A(q^2) &=& \left\{ \begin{array}{cl}
\Lambda_L^A(q_3=0) & \mathrm{for}\ n = 0, 1, 2, 4\\
\Lambda_L^A(q_3\ne 0 ) + \displaystyle{\frac{q^2_3}{M_N(M_N+E(q))}}
\Lambda_T^A(q) &
 \mathrm{for}\ n = 3
\end{array}
\right.\\
F_P(q^2) &=&\left\{ \begin{array}{cl}
\Lambda_T^A(q)/M_N & \mathrm{for}\ n = 2, 3\\
\displaystyle{\frac{M_N+E(q)}{q^2_3}}
\left(\Lambda_L^A(q_3 = 0 ) - \Lambda_L^A(q_3 \ne 0)
\right) & \mathrm{for}\ n = 1, 4
\end{array}
\right.
\end{eqnarray}
where $n = \vec{q}\,^2 \cdot (L/2\pi)^2$.
The results for the two form factors are summarized in table~\ref{tab:ffs}
in the appendix.

%
%
\subsubsection{Axialvector form factor $F_A(q^2)$}

\begin{figure}[!t]
\centering
\includegraphics[width=.6\textwidth,clip]{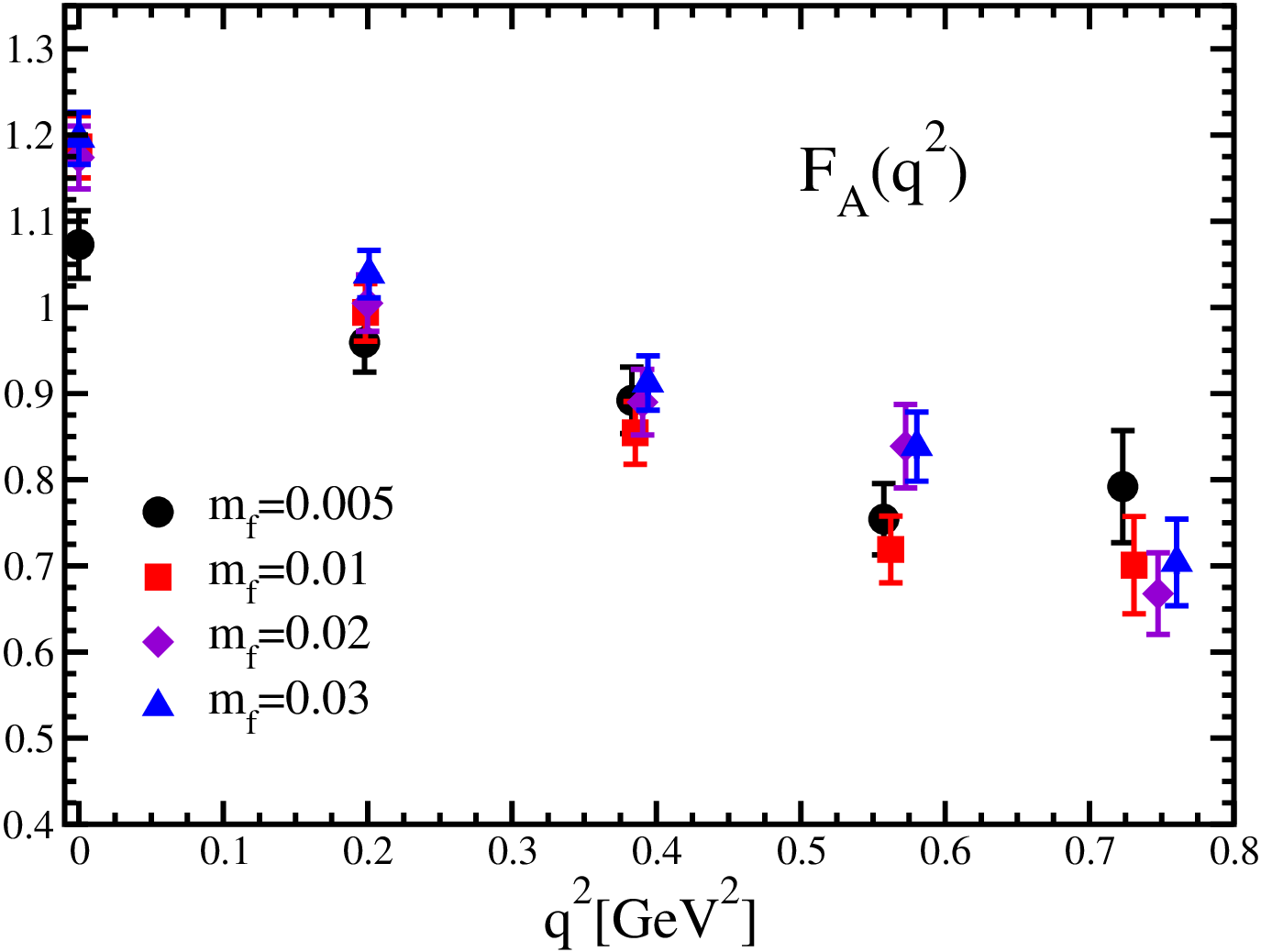}
\caption{The axialvector form factor,
$F_A(q^2)$,  renormalized by $Z_V = 1/F_1(0)$.
}
\label{fig:f_a}
\end{figure}

Figure~\ref{fig:f_a} shows the axialvector form factor at each quark mass, which is renormalized by the Dirac form factor at zero momentum transfer, $Z_V = 1/F_1(0)$.
This renormalization is valid due to the good chiral
properties of DWF.
At zero momentum transfer, the result at $m_f = 0.005$ is smaller than the other masses which 
corresponds to the bending of $g_A$ discussed in Sec.~\ref{sec:axial_charge}.
Furthermore, the $q^2$ dependence of the results at the lightest
quark mass is milder than the other masses.

\begin{figure}[!t]
\centering
\includegraphics[width=.6\textwidth,clip]{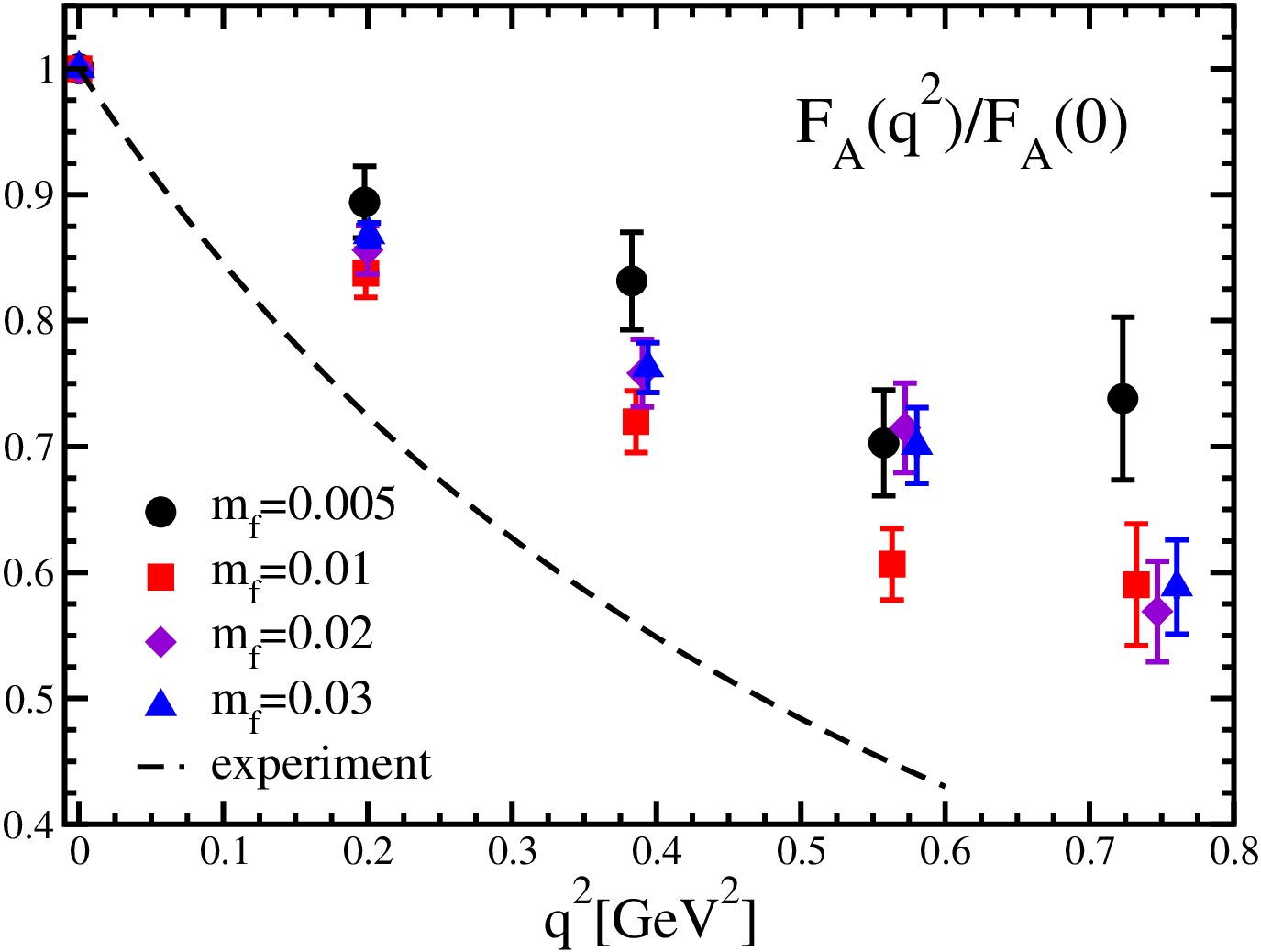}
\caption{The axialvector form factor,
$F_A(q^2)$, normalized at $q^2 = 0$. The dashed line denotes a fit to experimental data.
}
\label{fig:n_f_a}
\end{figure}

In the following we focus only on the momentum transfer dependence of the axialvector form factor:
We normalize the form factor by its value at zero momentum transfer respectively for each quark mass.
Figure~\ref{fig:n_f_a} shows the results after these normalizations, $F_A(q^2)/F_A(0)$.
For the heavier three masses, the results tend to decrease
with quark mass while the dependence is
opposite for the lightest mass. 
Similar to the vector-current form factors, the experimental axialvector form factor 
is also traditionally considered to be fitted well by the dipole form,
\begin{equation}
\frac{F_A(q^2)}{F_A(0)} = \frac{1}{(1+q^2/M_A^2)^2},
\label{eq:dipole_ga_ff}
\end{equation}
with the experimental data giving a best fit of $M_A = 1.03(2)$ GeV~\cite{Bernard:2001rs} for the axialvector dipole mass.
The experimental fit is shown by the dashed line in 
{Fig.~\ref{fig:n_f_a}}.

\begin{figure}[!t]
\centering
\includegraphics[width=.6\textwidth,clip]{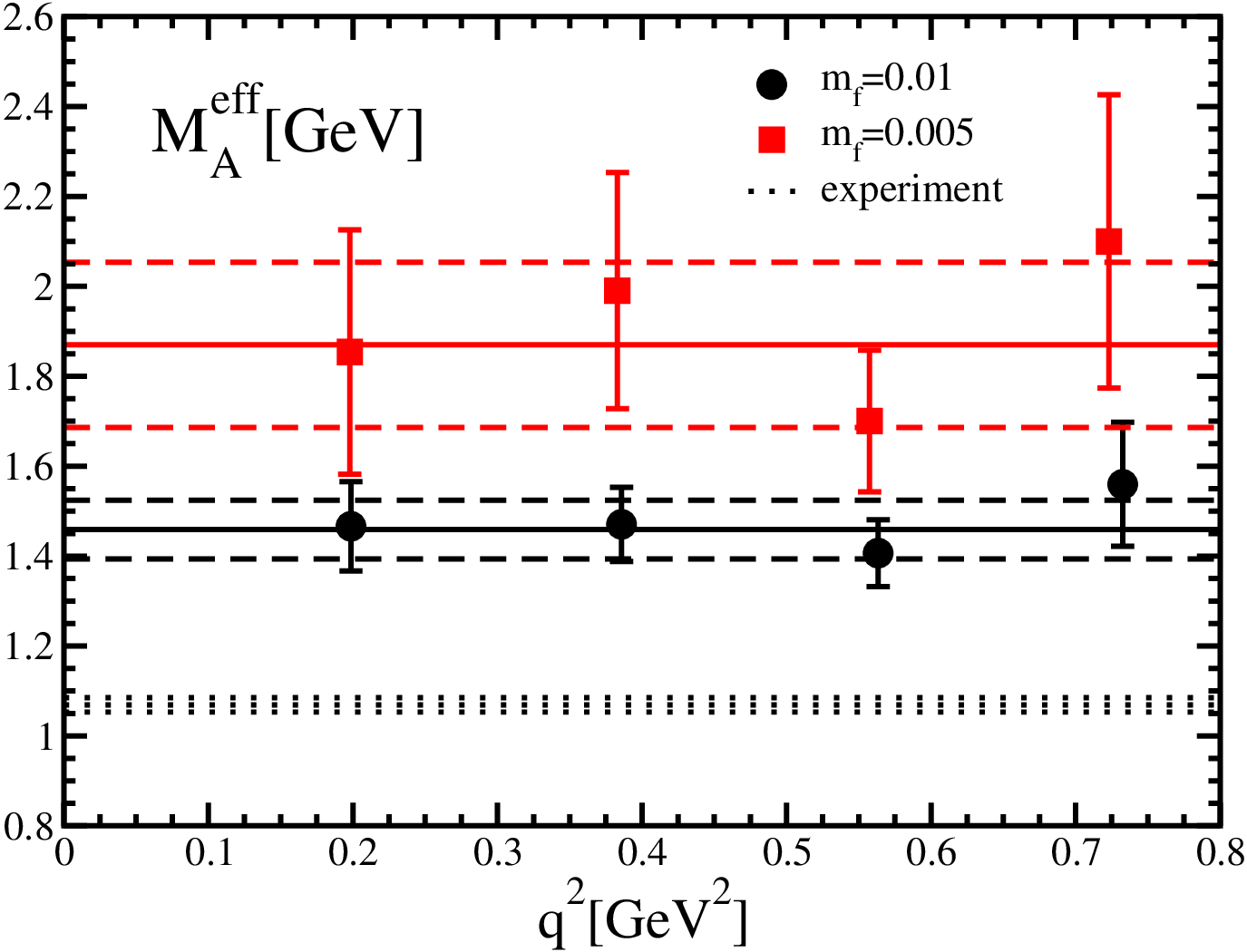}
\caption{
Effective dipole mass $M_A^{\mathrm{eff}}$ of $F_A(q^2)$ at $m_f = 0.005, 0.01$ 
together with the
experimental result~\cite{Bernard:2001rs}.
Result of Dipole fit (solid line) with one standard deviation
(dashed line) is also presented.
}
\label{fig:eff_m_a}
\end{figure}

If the dipole form is valid in the entire $q^2$ region, we can extract the effective axial dipole mass, 
\begin{equation}
M_A^{\mathrm{eff}} = \sqrt{\frac{q^2}{\sqrt{F_A(0)/F_A(q^2)}-1}}.
\end{equation}
{at each non-zero value of $q^2$.}
Figure~\ref{fig:eff_m_a} {shows} that the effective dipole mass at $m_f = 0.01$ is reasonably flat.
This means that the form factor behaves as a dipole,
as in the cases of the Dirac and Pauli form factors.
We fit the form factor with the dipole form, and the
fitted dipole mass is consistent with the
effective one, as shown in Fig.~\ref{fig:eff_m_a} 
by the solid line with the one standard deviation (dashed lines).
Figure~\ref{fig:eff_m_a} shows that the lightest quark mass data
is also well explained by the dipole form, although
the results do not approach the experimental value.

\begin{figure}[!t]
\centering
\includegraphics[width=.6\textwidth,clip]{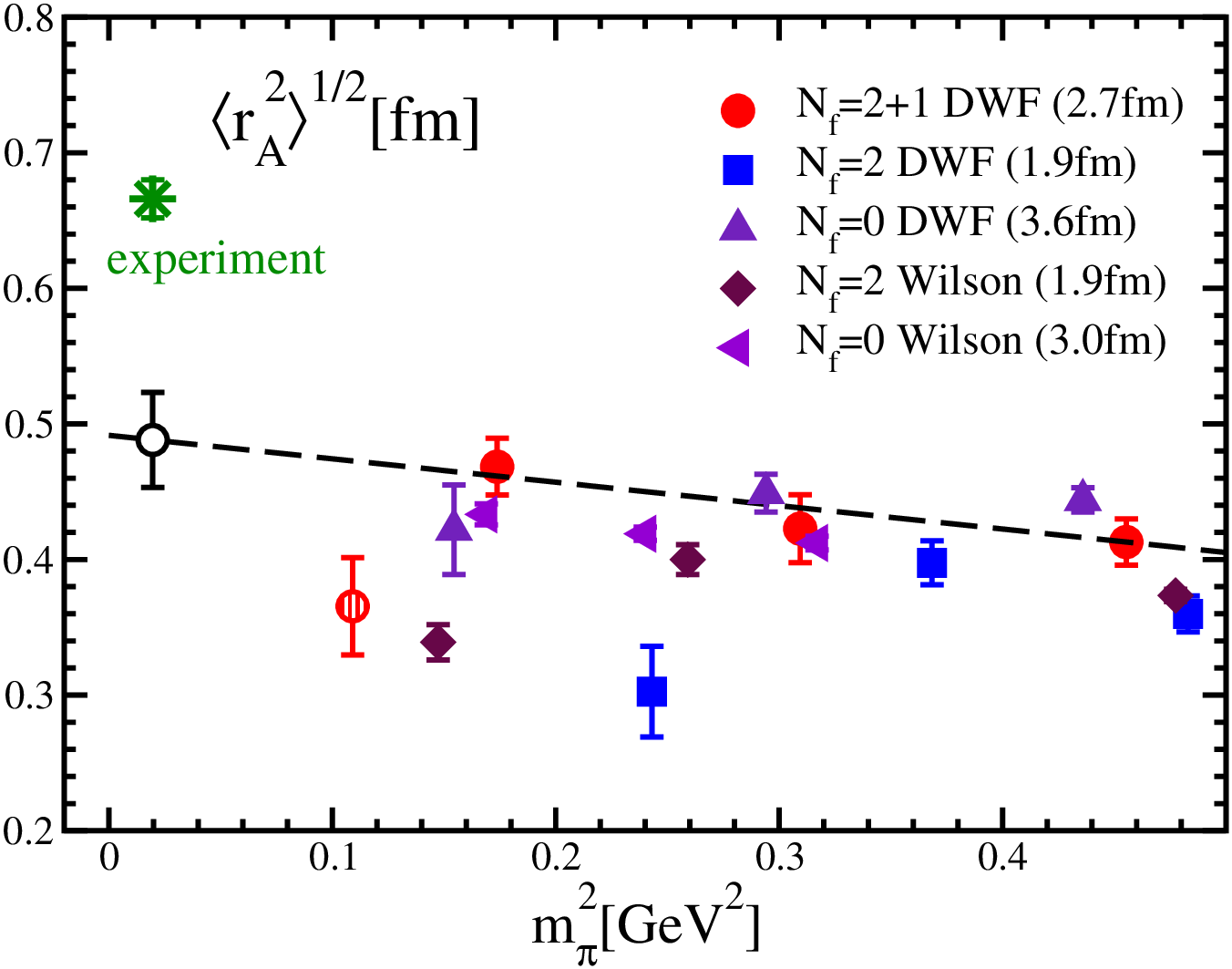}
\caption{
Axial charge rms radius 
$\displaystyle{\langle r_A^2\rangle^{1/2}}$ determined from dipole fit.
Dashed line represents a linear extrapolation of our data
excluding the lightest point(striped circle). Square, up triangle,
diamond and left triangle denote two-flavor~\cite{Lin:2008uz} and quenched DWF~\cite{Sasaki:2007gw}, and 
two-flavor and quenched Wilson~\cite{Alexandrou:2007xj} calculations, respectively.
The star denotes the experimental result~\cite{Bernard:2001rs}.
}
\label{fig:r_a}
\end{figure}

\begin{figure}[!t]
\begin{center}
\includegraphics[width=.6\textwidth,clip]{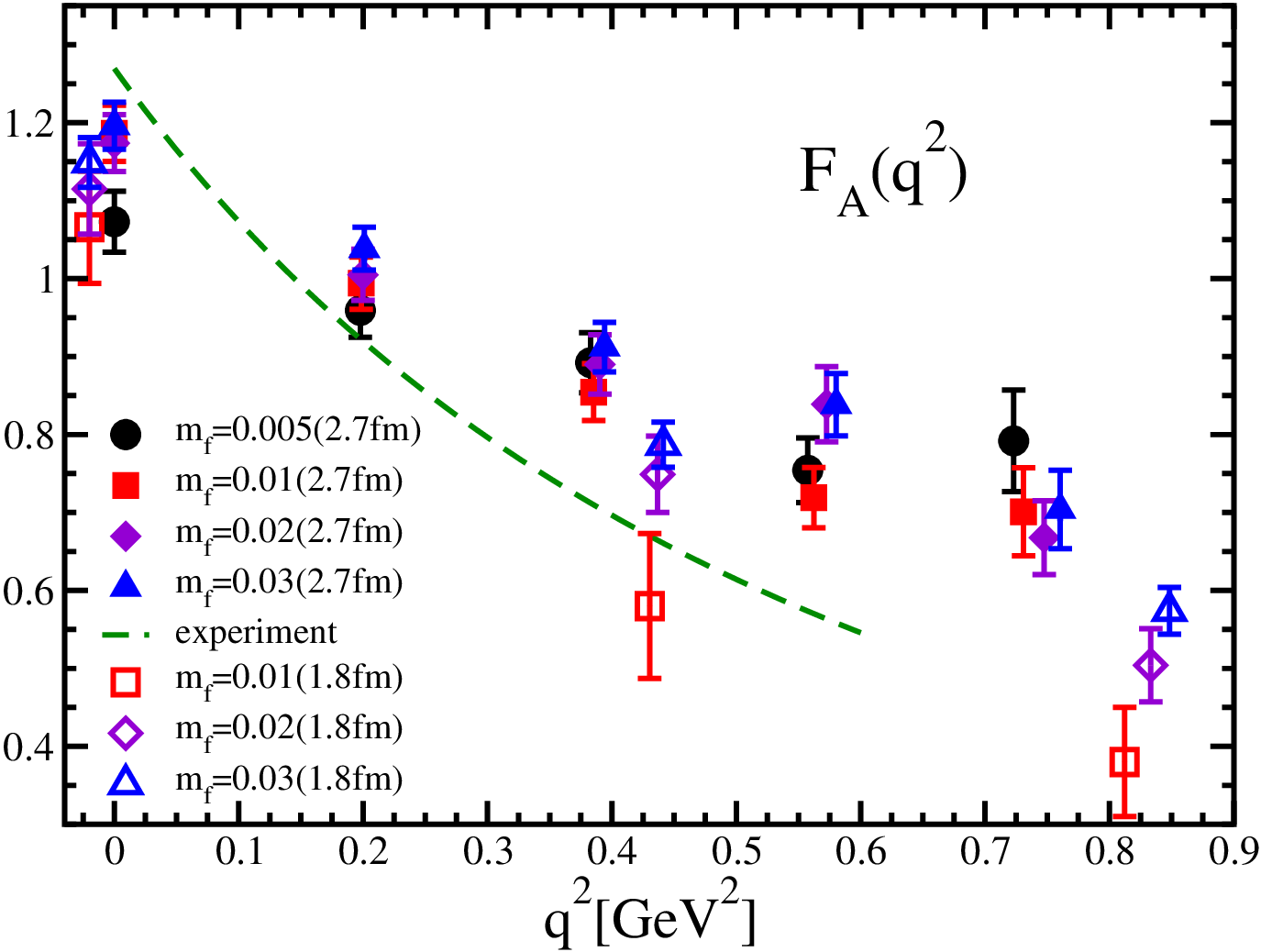}
\end{center}
\caption{
Comparison of $F_A(q^2)$ renormalized by $Z_V = 1/F_1(0)$ with larger and 
smaller volumes denoted by closed and open symbols, respectively, 
at each quark mass. Open symbols at $q^2 = 0$ are slightly shifted to
minus direction in $x$-axis.
The dashed curve is a fit to experimental data.
}
\label{fig:f_a_c}
\end{figure}

The axial rms radius is determined from the dipole mass,
\begin{equation}
\langle r^2_A \rangle^{1/2} = \sqrt{12}/M_A,
\end{equation}
and is 0.666(14) fm in the experiment.
The calculated axial rms radius from the fits is shown in
Fig.~\ref{fig:r_a} plotted as a function of the pion mass squared.
The results are summarized in table~\ref{tab:phys_axial_vector}.
\begin{table}[!t]
\begin{tabular}{cccccc}\hline\hline
$m_f$ & 0.005 & 0.01 & 0.02 & 0.03 & $m_\pi^{\mathrm{phys}}$\\\hline
$\langle r_A^2 \rangle^{1/2}$ [fm] 
      & 0.366(36) & 0.469(21) & 0.423(25) & 0.413(17) & 0.493(33)\\
$g_{\pi NN}$(def.) 
      & 8.53(82)  & 10.38(94) & 11.1(1.3) & 12.0(1.1) & 9.5(1.6) \\
$g_{\pi NN}$(GT) 
      & 11.84(45) & 13.12(47) & 12.66(78) & 13.56(57) & 12.79(79)\\
$g_P$ 
      & 6.71(60)  & 8.45(71)  & 10.31(88) & 11.93(93) & 6.6(1.2)\\
\hline\hline
\end{tabular}
\caption{\label{tab:phys_axial_vector} 
Axial charge rms radius $\langle r_A^2 \rangle^{1/2}$, 
nucleon-pion coupling $g_{\pi NN}$ and 
induced pseudoscalar coupling $g_P$. 
$g_{\pi NN}$ is calculated with the definition eq.~(\ref{eq:def_g_piNN})
and Goldberger-Treiman (GT) relation eq.~(\ref{eq:GT})
denoted as def. and GT in table, respectively.
The linear fit results at $m_\pi^{\mathrm{phys}} = 135$ MeV
obtained without the lightest quark mass,
are also presented.
}
\end{table}
While the result increases as the pion mass decreases, the lightest result significantly decreases.
This pion mass dependence is similar to that observed in the axial charge in Fig.~\ref{fig:ga_mpi}.
This, however, is not clear in $F_A(q^2)$ renormalized by $Z_V = 1/F_1(0)$
obtained on our smaller  
volume as shown in Fig.~\ref{fig:f_a_c}: the data at the lightest  
quark mass on the smaller volume shows a significant deviation from  
the larger volume result, but the statistical errors are too large to  
allow for a more quantitative comparison.

\begin{figure}[!t]
\centering
\includegraphics[width=.6\textwidth,clip]{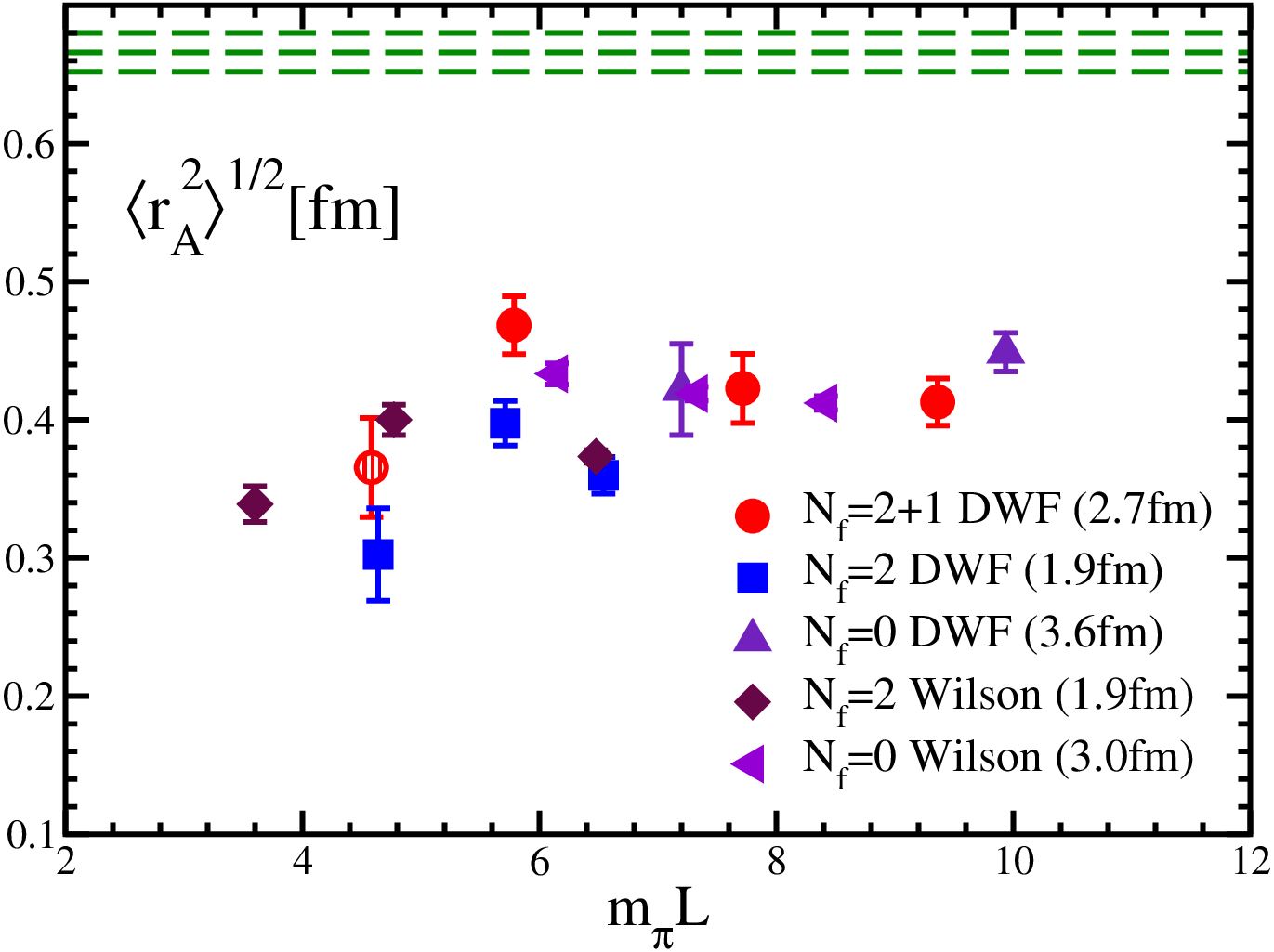}
\caption{
Same as Fig.~\ref{fig:r_a} except the horizontal axis is the scaling variable $m_\pi L$.
Dashed lines denote the experimental result~\cite{Bernard:2001rs} 
and its one standard deviation.
}
\label{fig:r_a_mpiL}
\end{figure}

Here, we similarly suspect this behavior of the larger volume
to be caused by a large finite volume effect.
A similar behavior is also seen in previous two-flavor results as 
presented in Fig.~\ref{fig:r_a}.
DWF~\cite{Lin:2008uz} and Wilson~\cite{Alexandrou:2007xj} fermion 
calculations on a smaller volume (1.9 fm)$^3$ have similar pion mass 
dependences, but the radius begins to decrease at heavier pion mass.
Once again, this behavior is quite similar to the case of the axial charge.
Moreover, previous results obtained on large volumes~\cite{Sasaki:2007gw,Alexandrou:2006ru} 
do not exhibit such strong pion mass
dependence, which is also shown in Fig.~\ref{fig:r_a}.
Figure~\ref{fig:r_a_mpiL} shows the same results of the rms radii, but
plotted as a function of $m_\pi L$.
The scaling of the rms radius with $m_\pi L$ is not 
as compelling as the axial charge case, but from the figure we
estimate that $m_\pi L > 6$ is required
to obtain the axial charge rms radius without significant finite 
volume effects. Needless to say, other systematic errors,
{\it e.g.}, due to heavier quark mass than the physical one, 
should be removed to reproduce the experimental value.

The lightest pion mass data is omitted in the following chiral
extrapolation, because we cannot rule out a large systematic error
due to the finite volume of the simulations as discussed above.
A linear fit to the heaviest three quark masses and
extrapolation to the physical pion mass yields 
$\langle r_A^2 \rangle^{1/2}=0.49(3)$ fm.
The fit result is presented in Fig.~\ref{fig:r_a}, and reproduces 73\%
of the experimental value.

%
%
\subsubsection{Induced pseudoscalar form factor $F_P(q^2)$}

The induced pseudoscalar form factor, $F_P(q^2)$, 
is expected to have a pion pole, so its momentum-transfer dependence 
should be different from the other form factors.
At the lightest quark mass this form factor is suspected
to have a large finite volume effect, since it is
obtained from the matrix element
of the axialvector current together with the
axialvector form factor, as discussed in the previous subsection.

\begin{figure}[!t]
\centering
\includegraphics[width=.6\textwidth,clip]{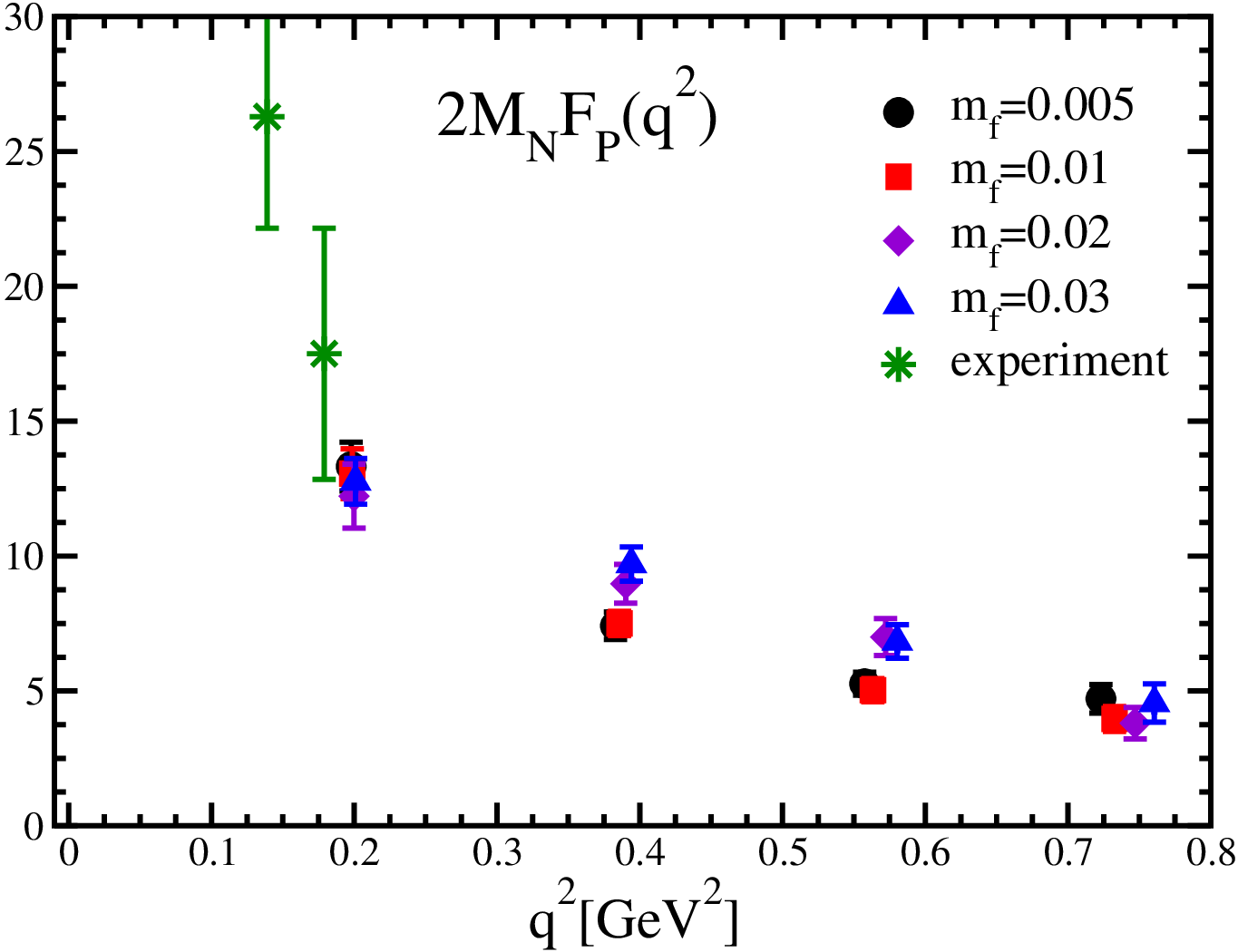}
\caption{
$2M_N F_P(q^2)$ renormalized by $Z_V = 1/F_1(0)$ with
experimental values~\cite{Choi:1993vt}.
}
\label{fig:f_p}
\end{figure}

Figure~\ref{fig:f_p} shows $2 M_N F_P(q^2)$ renormalized with $Z_V$, plotted against the momentum transfer squared at each quark mass.
We immediately notice that this form factor has a
much larger $q^2$ dependence than the other form factors.
In addition, the results from all quark
masses appear to be consistent with
the experimental data~\cite{Choi:1993vt}.
Note that our statistical error is much smaller than the experiment.

\begin{figure}[!t]
\centering
\includegraphics[width=.6\textwidth,clip]{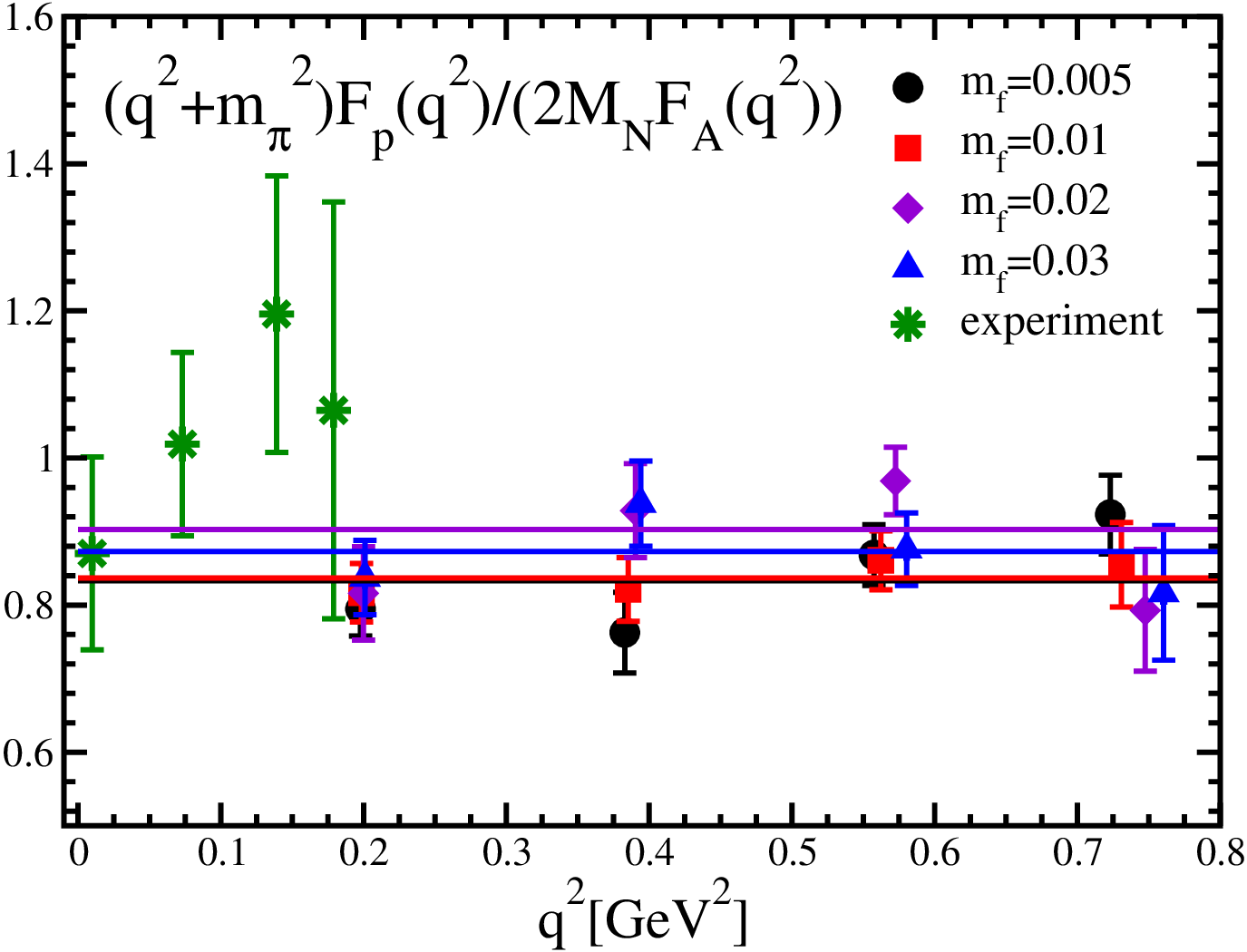}
\caption{Induced pseudoscalar form factor $F_P(q^2)$
normalized by nucleon mass, $F_A(q^2)$, and pion pole, along with
experimental values~\cite{Choi:1993vt}.
}
\label{fig:nf_p}
\end{figure}

The induced pseudoscalar form factor is related to the axial 
vector form factor through the so-called partially conserved 
axialvector current (PCAC) relation which is a manifestation 
of spontaneously broken chiral symmetry.
In the traditional PCAC current algebra with pion-pole dominance (PPD), the
PPD form,
\begin{equation}
F^{^{\mathrm{PPD}}}_P(q^2) = \frac{2 M_N F_A(q^2)}{ q^2 + m_\pi^2 },
\label{eq:pion_pole}
\end{equation}
is obtained at $m_\pi \approx 0$.  
The denominator on the right-hand side of 
this relation corresponds to the pion pole.
We investigate the validity of this relation in our results through a quantity, 
\begin{equation}
\alpha_{_\mathrm{PPD}}=
\frac{ ( q^2 + m_\pi^2 ) F_P(q^2) }{  2 M_N F_A(q^2) }.
\label{eq:a_ppd}
\end{equation}
If the relation holds we obtain unity for this quantity at all $q^2$.
Figure~\ref{fig:nf_p} shows 
$\alpha_{_\mathrm{PPD}}$ calculated using our lattice results for
$F_A$ and $F_P$. There is no significant $q^2$ dependence, 
and while the values are close to unity, they are systematically less than one.
We fit these results by a constant for each quark
mass, whose results are presented in {Fig.~\ref{fig:nf_p}} and 
table~\ref{tab:appd_mpole}.
\begin{table}[!t]
\begin{tabular}{ccccc}\hline\hline
$m_f$ & 0.005 & 0.01 & 0.02 & 0.03 \\\hline
$\alpha_{_\mathrm{PPD}}$ 
      & 0.833(25) & 0.837(29) & 0.903(32) & 0.873(31) \\
$[m^{\mathrm{pole}}_{F_P}/m_\pi]^2$ 
      & 1.044(39) & 1.009(19) & 0.940(24) & 0.977(20) \\
\hline\hline
\end{tabular}
\caption{\label{tab:appd_mpole} 
$\alpha_{_\mathrm{PPD}}$ and $m^{\mathrm{pole}}_{F_P}$. }
\end{table}
While all the fit results are consistent with the 
experimental data~\cite{Andreev:2007wg,Czarnecki:2007th,Choi:1993vt} 
within the larger error of the experiments,
they are about 10--20\% smaller than the prediction of the PPD form.

We should note that the quantity at the lightest quark mass looks
similar to the others, but
$F_A(q^2)$ at $m_f = 0.005$ is suspected to have large finite volume effect
as discussed in the last subsection.
This means that $F_P(q^2)$ at $m_f = 0.005$ is
expected to suffer from a similarly-sized effect at 
the same quark mass. Thus, it appears that
the two large finite volume effects cancel in this ratio.

\begin{figure}[!t]
\centering
\includegraphics[width=.6\textwidth,clip]{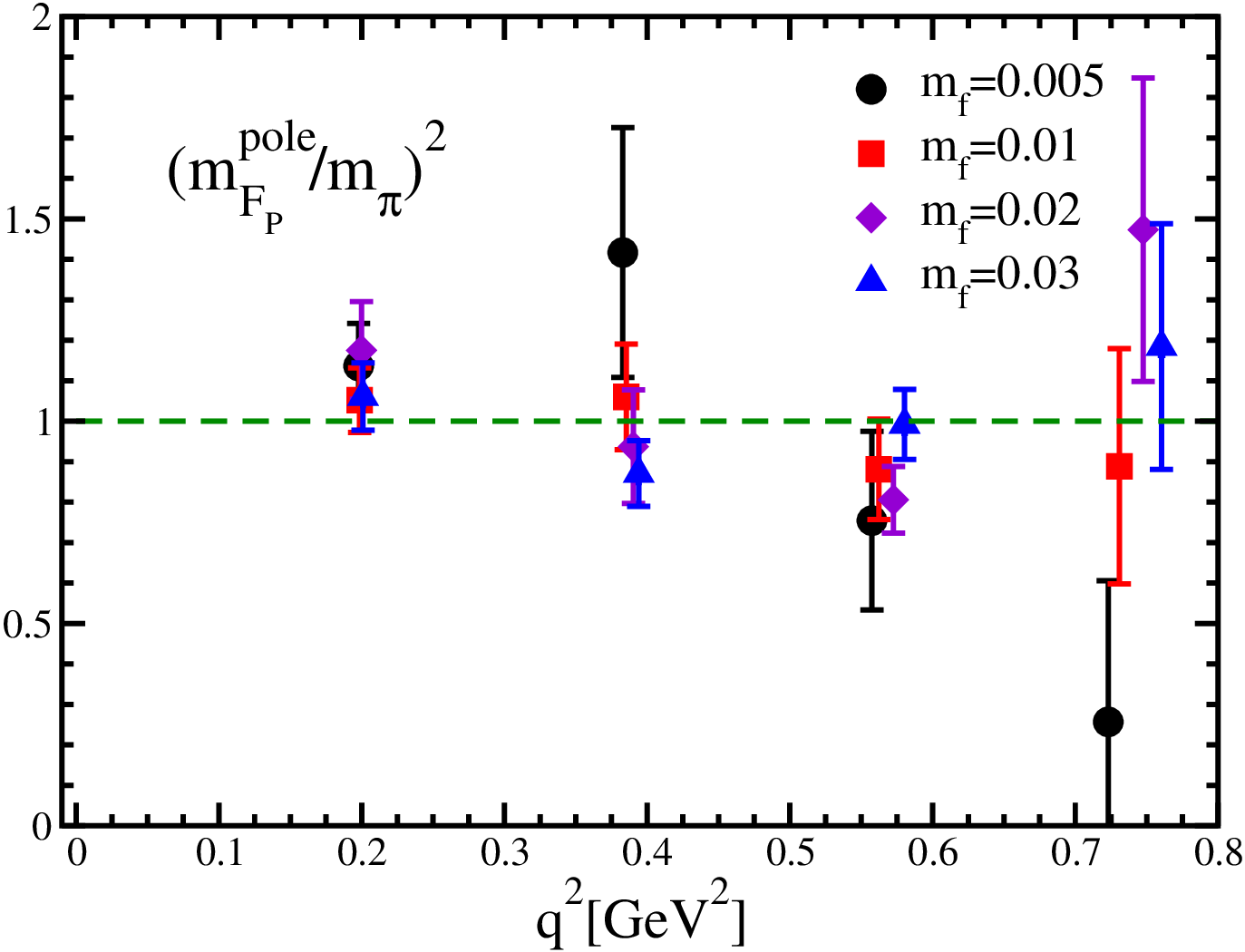}
\caption{
Ratio of the pole mass in $F_P(q^2)$ and measured pion mass.
}
\label{fig:f_p_pp}
\end{figure}

We check the consistency of the pole mass in $F_P(q^2)$ 
with the measured pion mass at each quark mass by observing that 
the pole mass is given by
\begin{equation}
\left(m_{F_P}^{\mathrm{pole}}\right)^2
=
\frac{2\alpha_{_\mathrm{PPD}}M_N F_A(q^2)}{F_P(q^2)}-q^2,
\end{equation}
where we use the fact that $\alpha_{_\mathrm{PPD}}\ne 1$ in our data.
Figure~\ref{fig:f_p_pp} shows that the ratio
$\left[m_{F_P}^{\mathrm{pole}}/m_\pi\right]^2$ is reasonably
consistent with unity, and has no large $q^2$ dependence except for
the lightest quark mass point, which has large statistical error.
The values obtained from a constant fit are presented 
in table~\ref{tab:appd_mpole}.
This consistency suggests that $F_P(q^2)$ does indeed have a 
pion pole structure,
which is consistent with the PPD form, however
$\alpha_{_{\rm PPD}} \ne 1$ in our data.
We confirmed that $m_{F_P}^{\mathrm{pole}}$ 
and $\alpha_{_\mathrm{PPD}}$ 
obtained from a monopole fit of $2 M_N F_A / F_P$
are reasonably consistent with the above results, but have larger errors.

The pion-nucleon coupling is related to the induced pseudoscalar
form factor via the relation
\begin{equation}
g_{\pi NN} = \lim_{q^2\to-m_\pi^2}\left[
\frac{(q^2 + m_\pi^2) F_P(q^2)}{2F_\pi}
\right],
\label{eq:def_g_piNN}
\end{equation}
where $F_\pi = 92.4$ MeV.
Combining the above relation with the PPD form, eq.~(\ref{eq:pion_pole}), we obtain the Goldberger-Treiman (GT) relation~\cite{Goldberger:1958tr},
\begin{equation}
g_{\pi NN} F_\pi = M_N g_A.
\label{eq:GT}
\end{equation}
In this relation we assume $F_A(0) \approx F_A(-m_\pi^2)$.
As such it suffers from a small mismatch in momentum transfer. 
Nevertheless, if we substitute the experimental values for the quantities, 
we obtain $g_{\pi NN} = 12.9$.

\begin{figure}[!t]
\centering
\includegraphics[width=.6\textwidth,clip]{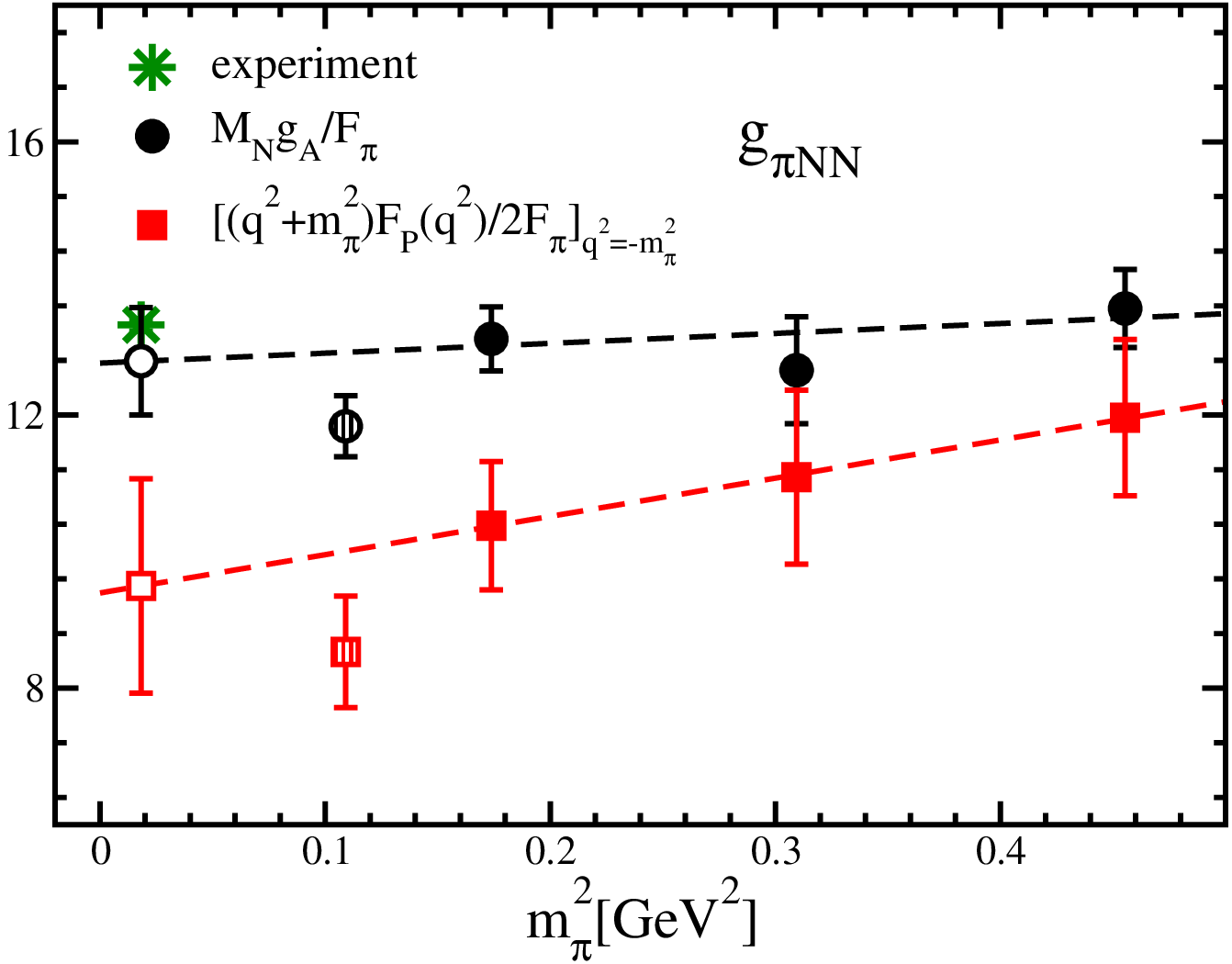}
\caption{
Two measurements of the $\pi NN$ coupling with Goldberger-Treiman relation
and definition of $g_{\pi NN}$.
The experimental value~\cite{Ericson:2000md} is indicated by the star.
Dashed lines present linear extrapolations of our data without
lightest point.
}
\label{fig:g_pinn}
\end{figure}

Figure~\ref{fig:g_pinn} shows two calculations for the \(\pi NN\) coupling,  $g_{\pi NN}$:
one uses  the definition of $g_{\pi NN}$ and another the GT relation at each quark mass, plotted against the pion mass squared.
In determining $g_{\pi NN}$, we use the measured pion decay constant at each quark mass from Ref.~\cite{Allton:2008pn}.
Results for $g_{\pi NN}$ from both methods are given in table~\ref{tab:phys_axial_vector}.
From Fig.~\ref{fig:g_pinn}, we observe that $g_{\pi NN}$ obtained
from both methods displays only a mild $m_\pi^2$ dependence, with
the exception of the lightest mass results which show a significant 
downward shift away from the trend set by the three heavier mass values.
This of course is another manifestation of the large
finite size effect observed 
in the axial charge (see Sec.~\ref{sec:axial_charge}).
Hence, for the chiral extrapolation we simply employ a linear fit form 
and exclude the lightest mass point.
We obtain the results at the physical pion mass, 
$g_{\pi NN} = 9.5(1.6)$ 
from the definition eq.(\ref{eq:def_g_piNN}), and
$g_{\pi NN} = 12.8(8)$
from the GT relation eq.(\ref{eq:GT}).
The value obtained using the GT relation agrees with a recent 
estimation of the
coupling $g_{\pi NN} = 13.3(9)$ obtained from forward $\pi N$ 
scattering data~\cite{Ericson:2000md}, and also with the previous result,
$g_{\pi NN} = 11.8(3)$,
from a quenched simulation performed using
the Wilson action~\cite{Alexandrou:2007xj} estimated by
the GT relation. The result from the definition
  eq.(\ref{eq:def_g_piNN}), on the other hand, is consistent with 
a quenched DWF determination, $g_{\pi NN} = 10.4(1.0)$~\cite{Sasaki:2007gw}, 
obtained from $F_P(-m_\pi^2)$.

Rigorously speaking, the GT relation is not valid in our data,
since our data do not satisfy the PPD form due to
$\alpha_{_\mathrm{PPD}} \approx 0.85$.
Thus, the difference between the two determinations of 
$g_{\pi NN}$ can be explained by $\alpha_{_\mathrm{PPD}}$.
Further study of the GT relation is an important future work,
since the relation should be satisfied in the chiral limit,
and at zero momentum transfer.

The induced pseudoscalar coupling for muon capture on the proton, 
$g_P = m_\mu F_P(q_c^2)$ where $q_c^2 = 0.88 m_\mu^2$, 
is defined with the muon mass $m_\mu$ and the induced pseudoscalar 
form factor $F_P$ at the specific momentum transfer
where the muon capture occurs, $p + \mu^- \to n + \nu_\mu$.

\begin{figure}[!t]
\centering
\includegraphics[width=.6\textwidth,clip]{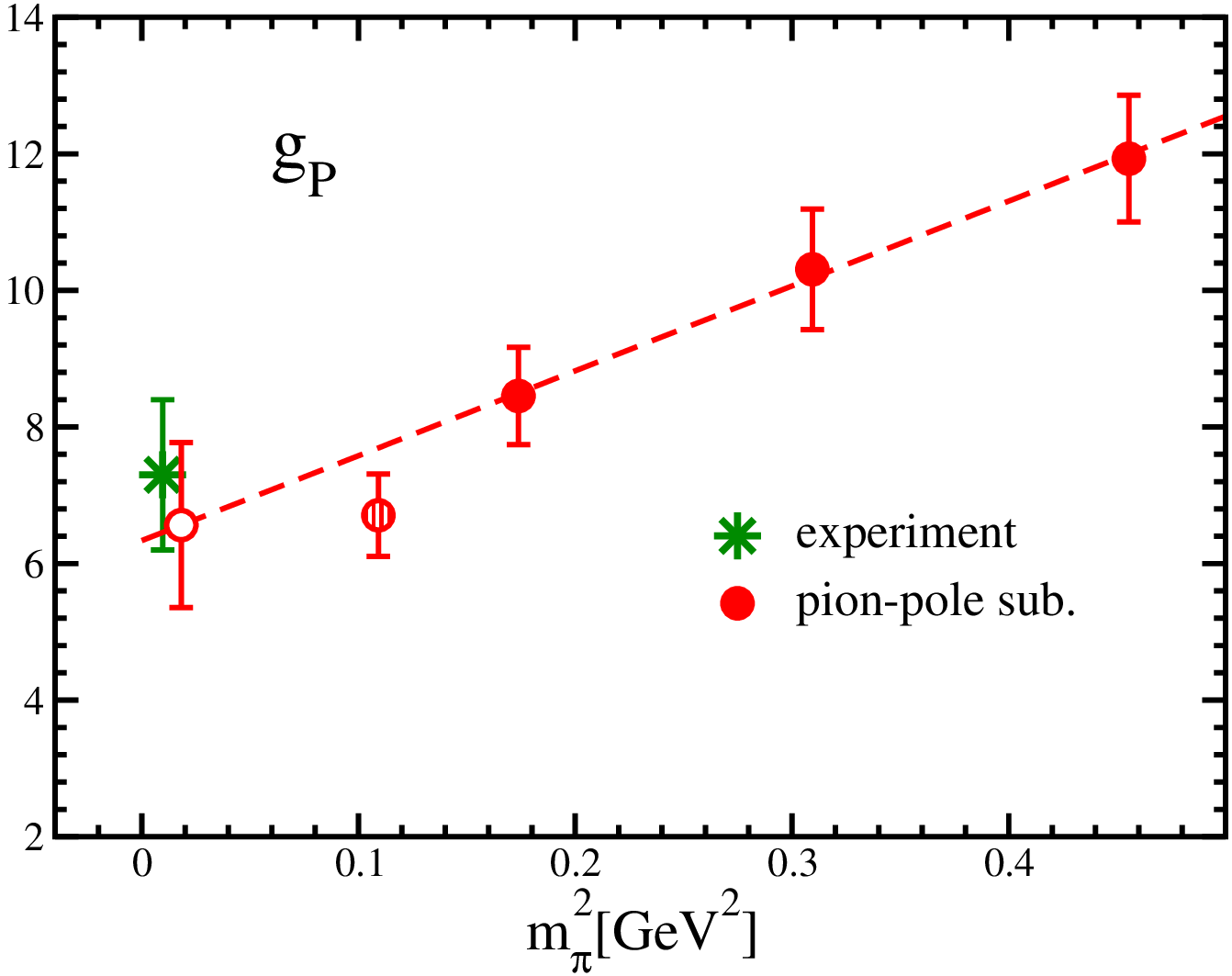}
\caption{
Induced pseudoscalar coupling for muon capture, $g_p$, estimated
with pion-pole subtraction. The experimental result~\cite{Andreev:2007wg} 
is indicated by the star.
The dashed line represents a linear extrapolation of our
data excluding the lightest point.
}
\label{fig:induced_coupling}
\end{figure}

Since $F_P(q^2)$ has significantly large pion mass and
momentum-transfer dependences due to the pion pole,
we subtract this contribution before performing the momentum transfer and
chiral extrapolations. To do this, we first define the quantity
with pion-pole subtraction by
\begin{equation}
\overline{F}_P(q^2) = ( q^2 + m_\pi^2 ) F_P( q^2 )
\end{equation}
at each $q^2$, and then extrapolate this to the required 
momentum transfer $q^2_c$.
The induced pseudoscalar coupling is estimated by the
normalization factor with the physical pion mass
\begin{equation}
g_p = m_\mu \frac{\overline{F}_P(q^2_c)}{q^2_c + (m_\pi^{phys})^2},
\end{equation}
where $m_\pi^{phys} = 135$ MeV, at each quark mass as shown in Fig.~\ref{fig:induced_coupling}.
The figure shows that the result is almost linear as 
a function of the pion mass squared and decreases toward 
the experimental result for the three heavier mass values.
Again the lightest mass result is an exception
 caused by the finite volume effect in $g_A$, as discussed 
in Sec.~\ref{sec:axial_charge}.
The result at the physical pion mass, presented in 
table~\ref{tab:phys_axial_vector}, is obtained from a linear fit
to the heaviest three pion masses, and is consistent with 
the recent experiment~\cite{Andreev:2007wg} 
and analysis~\cite{Czarnecki:2007th}.
This result also agrees with the previous quenched DWF result~\cite{Sasaki:2007gw},
while it disagrees with a quenched Wilson determination~\cite{Alexandrou:2007xj}\footnote{The value is estimated by the authors of Ref.~\cite{Sasaki:2007gw} using
the raw data in Ref.~\cite{Alexandrou:2007xj}}
which is almost half of the experimental value.

\section{Conclusions}
\label{Conclusions}
We have studied the isovector nucleon form factors with $N_f = 2 + 1$
flavors of dynamical quarks using the domain wall fermion action 
at a lattice cutoff of $a^{-1} = 1.73$ GeV.
The form factors are calculated with four light quark masses, corresponding to a lightest pion mass, $m_\pi = 0.33$ GeV, and with momentum
transfers down to $q^2 \approx 0.2$ GeV$^2$.

We have found the axial charge decreases significantly at the lightest quark mass point
on the larger volume while the effect sets in for heavier quark mass on the smaller volume.  By comparing our results with
those using different volumes, numbers of flavors, and lattice 
fermions as a function of the single variable $m_\pi L$,
we conclude that this downward trend is caused by the finite volume used in our calculation.  The fact that such an effect is absent in quenched and partially-quenched mixed-action studies on large volumes may be explained by the presence of unphysical logarithms.

We have fit the data to several forms, including 
finite volume effects, and obtain $g_A = 1.19(6)(4)$,
where the first and second errors are statistical and systematic,  respectively,
which is 7\% smaller than the experimental value.
In our estimation, a spatial volume of $V \ge ($3.5 fm)$^3$
is required to keep the finite volume effect at, or below, one percent at
$m_\pi = 0.33$ GeV. 
Hence lattice calculations should continue to push down the quark mass and increase the volume with $m_\pi L> 6$. 
Detailed analyses of the
quark mass and finite volume dependence is desirable to 
understand the systematic deviation from experiment.

Our lattice results for the form factors of the vector current are well fit by
the standard dipole form.
We have evaluated the root-mean squared radii,
and the difference of the anomalous magnetic moment between
the proton and neutron from the dipole fits.
The radii and the anomalous moment are well explained by 
a linear function of the pion mass
squared. In the radii we have not observed divergent quark mass behaviors
predicted by HBChPT.
Besides the divergent behavior, the pion mass dependences for the 
observables are quite consistent with
other lattice QCD calculations including the recent results of LHP.
Due to the linear behavior, we have concluded that the form factors
of the vector current is less sensitive to the finite volume
effect in contrast to the axial charge.
Although both Dirac and Pauli rms radii approach to the experimental values 
as the pion mass decreases, the values extrapolated by the linear
form at the physical pion mass underestimate the experiments by
about 25\%. 
Future work will involve simulating at lighter quark masses to
  search for the nonanalytic behavior predicted by HBChPT.

The axialvector form factor is also well described by the dipole form,
even at the lightest quark mass, where the axial charge, $F_A(0)$,
is suspected to have a large finite volume effect.
The axial charge radius, obtained from the dipole fit, has
a downward tendency as a function of the pion mass squared,
which drives the radius away from the experimental value.
We have considered this dependence to be caused by the finite volume
of our simulation, as in the case of the axial charge. 
{We observe that our} results seem to
scale as $m_\pi L$, as do previous calculations using several volumes.
We have concluded that the form factors of the axialvector current
are more sensitive to the finite volume than those of the vector current
from the observations of the finite volume effects.

We have checked the pion-pole structure in the induced 
pseudoscalar form factor with our simulations.
We have found that the pion-pole dominance form describes our data 
well, with the exception that $\alpha_{_{\mathrm{PPD}}} < 1$. 
Taking into account that $\alpha_{_{\mathrm{PPD}}}\ne 1$,
the pole mass of the induced pseudoscalar form factor
reasonably agrees with the measured pion mass.

For a precision test of QCD from nucleon matrix elements,
we have identified several problems that first need to be overcome, such as
finite volume systematic errors in the axial charge and the
form factors of the axialvector currents, and the underestimation 
of the radii of the form factors of the vector current.
Further lighter quark mass and larger volume calculations are essential 
to solve the problems, and such simulations are underway.
Besides the comparisons with the experimental values,
it is also important future work to study why 
$\alpha_{_{\mathrm{PPD}}}$ deviates from unity.

\section*{Acknowledgments}

We thank the members of the RIKEN-BNL-Columbia (RBC) and UKQCD Collaborations.
We also thank RIKEN, Brookhaven National Laboratory, the U.S.\ Department of Energy, Edinburgh University and the U.K. PPARC for providing the facilities essential for the completion of this work.
T.B. and T.Y. were supported 
by the U.S.\ DOE under contract
DE-FG02-92ER40716. 
T.Y. was the Yukawa Fellow supported by Yukawa Memorial Foundation.
H.L. is supported by DOE contract DE-AC05-06OR23177 under which 
JSA, LLC operates THNAF.
S.S. is supported by JSPS Grant-In-Aid
for Scientific Research (C) (No.19540265).
J.Z. is supported by STFC grant PP/F009658/1.

\bibliography{nuc_ref}

\begin{thebibliography}{66}
\expandafter\ifx\csname natexlab\endcsname\relax\def\natexlab#1{#1}\fi
\expandafter\ifx\csname bibnamefont\endcsname\relax
  \def\bibnamefont#1{#1}\fi
\expandafter\ifx\csname bibfnamefont\endcsname\relax
  \def\bibfnamefont#1{#1}\fi
\expandafter\ifx\csname citenamefont\endcsname\relax
  \def\citenamefont#1{#1}\fi
\expandafter\ifx\csname url\endcsname\relax
  \def\url#1{\texttt{#1}}\fi
\expandafter\ifx\csname urlprefix\endcsname\relax\def\urlprefix{URL }\fi
\providecommand{\bibinfo}[2]{#2}
\providecommand{\eprint}[2][]{\url{#2}}

\bibitem[{\citenamefont{Sasaki and
  Yamazaki}(2008{\natexlab{a}})}]{Sasaki:2007gw}
\bibinfo{author}{\bibfnamefont{S.}~\bibnamefont{Sasaki}} \bibnamefont{and}
  \bibinfo{author}{\bibfnamefont{T.}~\bibnamefont{Yamazaki}},
  \bibinfo{journal}{Phys. Rev.} \textbf{\bibinfo{volume}{D78}},
  \bibinfo{pages}{014510} (\bibinfo{year}{2008}{\natexlab{a}}),
  \eprint{0709.3150}.

\bibitem[{\citenamefont{Lin et~al.}(2008)\citenamefont{Lin, Blum, Ohta, Sasaki,
  and Yamazaki}}]{Lin:2008uz}
\bibinfo{author}{\bibfnamefont{H.-W.} \bibnamefont{Lin}},
  \bibinfo{author}{\bibfnamefont{T.}~\bibnamefont{Blum}},
  \bibinfo{author}{\bibfnamefont{S.}~\bibnamefont{Ohta}},
  \bibinfo{author}{\bibfnamefont{S.}~\bibnamefont{Sasaki}}, \bibnamefont{and}
  \bibinfo{author}{\bibfnamefont{T.}~\bibnamefont{Yamazaki}}
  (\bibinfo{collaboration}{RBC}), \bibinfo{journal}{Phys. Rev.}
  \textbf{\bibinfo{volume}{D78}}, \bibinfo{pages}{114505}
  (\bibinfo{year}{2008}), \eprint{0802.0863}.

\bibitem[{\citenamefont{Arrington et~al.}(2007)\citenamefont{Arrington,
  Roberts, and Zanotti}}]{Arrington:2006zm}
\bibinfo{author}{\bibfnamefont{J.}~\bibnamefont{Arrington}},
  \bibinfo{author}{\bibfnamefont{C.~D.} \bibnamefont{Roberts}},
  \bibnamefont{and} \bibinfo{author}{\bibfnamefont{J.~M.}
  \bibnamefont{Zanotti}}, \bibinfo{journal}{J. Phys.}
  \textbf{\bibinfo{volume}{G34}}, \bibinfo{pages}{S23} (\bibinfo{year}{2007}),
  \eprint{nucl-th/0611050}.

\bibitem[{\citenamefont{Hofstadter and McAllister}(1955)}]{Hofstadter:1955ae}
\bibinfo{author}{\bibfnamefont{R.}~\bibnamefont{Hofstadter}} \bibnamefont{and}
  \bibinfo{author}{\bibfnamefont{R.~W.} \bibnamefont{McAllister}},
  \bibinfo{journal}{Phys. Rev.} \textbf{\bibinfo{volume}{98}},
  \bibinfo{pages}{217} (\bibinfo{year}{1955}).

\bibitem[{\citenamefont{Bumiller et~al.}(1960)\citenamefont{Bumiller,
  Croissiaux, and Hofstadter}}]{Bumiller:1960zz}
\bibinfo{author}{\bibfnamefont{F.}~\bibnamefont{Bumiller}},
  \bibinfo{author}{\bibfnamefont{M.}~\bibnamefont{Croissiaux}},
  \bibnamefont{and}
  \bibinfo{author}{\bibfnamefont{R.}~\bibnamefont{Hofstadter}},
  \bibinfo{journal}{Phys. Rev. Lett.} \textbf{\bibinfo{volume}{5}},
  \bibinfo{pages}{261} (\bibinfo{year}{1960}).

\bibitem[{\citenamefont{Bumiller et~al.}(1961)\citenamefont{Bumiller,
  Croissiaux, Dally, and Hofstadter}}]{Bumiller:1961zz}
\bibinfo{author}{\bibfnamefont{F.}~\bibnamefont{Bumiller}},
  \bibinfo{author}{\bibfnamefont{M.}~\bibnamefont{Croissiaux}},
  \bibinfo{author}{\bibfnamefont{E.}~\bibnamefont{Dally}}, \bibnamefont{and}
  \bibinfo{author}{\bibfnamefont{R.}~\bibnamefont{Hofstadter}},
  \bibinfo{journal}{Phys. Rev.} \textbf{\bibinfo{volume}{124}},
  \bibinfo{pages}{1623} (\bibinfo{year}{1961}).

\bibitem[{\citenamefont{Janssens et~al.}(1966)\citenamefont{Janssens,
  Hofstadter, Hughes, and Yearian}}]{Janssens:1965kd}
\bibinfo{author}{\bibfnamefont{T.}~\bibnamefont{Janssens}},
  \bibinfo{author}{\bibfnamefont{R.}~\bibnamefont{Hofstadter}},
  \bibinfo{author}{\bibfnamefont{E.~B.} \bibnamefont{Hughes}},
  \bibnamefont{and} \bibinfo{author}{\bibfnamefont{M.~R.}
  \bibnamefont{Yearian}}, \bibinfo{journal}{Phys. Rev.}
  \textbf{\bibinfo{volume}{142}}, \bibinfo{pages}{922} (\bibinfo{year}{1966}).

\bibitem[{\citenamefont{Jones et~al.}(2000)}]{Jones:1999rz}
\bibinfo{author}{\bibfnamefont{M.~K.} \bibnamefont{Jones}} \bibnamefont{et~al.}
  (\bibinfo{collaboration}{Jefferson Lab Hall A}), \bibinfo{journal}{Phys. Rev.
  Lett.} \textbf{\bibinfo{volume}{84}}, \bibinfo{pages}{1398}
  (\bibinfo{year}{2000}), \eprint{nucl-ex/9910005}.

\bibitem[{\citenamefont{Gayou et~al.}(2002)}]{Gayou:2001qd}
\bibinfo{author}{\bibfnamefont{O.}~\bibnamefont{Gayou}} \bibnamefont{et~al.}
  (\bibinfo{collaboration}{Jefferson Lab Hall A}), \bibinfo{journal}{Phys. Rev.
  Lett.} \textbf{\bibinfo{volume}{88}}, \bibinfo{pages}{092301}
  (\bibinfo{year}{2002}), \eprint{nucl-ex/0111010}.

\bibitem[{\citenamefont{Nambu and
  Jona-Lasinio}(1961{\natexlab{a}})}]{Nambu:1961tp}
\bibinfo{author}{\bibfnamefont{Y.}~\bibnamefont{Nambu}} \bibnamefont{and}
  \bibinfo{author}{\bibfnamefont{G.}~\bibnamefont{Jona-Lasinio}},
  \bibinfo{journal}{Phys. Rev.} \textbf{\bibinfo{volume}{122}},
  \bibinfo{pages}{345} (\bibinfo{year}{1961}{\natexlab{a}}).

\bibitem[{\citenamefont{Nambu and
  Jona-Lasinio}(1961{\natexlab{b}})}]{Nambu:1961fr}
\bibinfo{author}{\bibfnamefont{Y.}~\bibnamefont{Nambu}} \bibnamefont{and}
  \bibinfo{author}{\bibfnamefont{G.}~\bibnamefont{Jona-Lasinio}},
  \bibinfo{journal}{Phys. Rev.} \textbf{\bibinfo{volume}{124}},
  \bibinfo{pages}{246} (\bibinfo{year}{1961}{\natexlab{b}}).

\bibitem[{\citenamefont{Amsler et~al.}(2008)}]{Amsler:2008zz}
\bibinfo{author}{\bibfnamefont{C.}~\bibnamefont{Amsler}} \bibnamefont{et~al.}
  (\bibinfo{collaboration}{Particle Data Group}), \bibinfo{journal}{Phys.
  Lett.} \textbf{\bibinfo{volume}{B667}}, \bibinfo{pages}{1}
  (\bibinfo{year}{2008}).

\bibitem[{\citenamefont{Goldberger and Treiman}(1958)}]{Goldberger:1958tr}
\bibinfo{author}{\bibfnamefont{M.~L.} \bibnamefont{Goldberger}}
  \bibnamefont{and} \bibinfo{author}{\bibfnamefont{S.~B.}
  \bibnamefont{Treiman}}, \bibinfo{journal}{Phys. Rev.}
  \textbf{\bibinfo{volume}{110}}, \bibinfo{pages}{1178} (\bibinfo{year}{1958}).

\bibitem[{\citenamefont{Bernard et~al.}(2002)\citenamefont{Bernard,
  Elouadrhiri, and Meissner}}]{Bernard:2001rs}
\bibinfo{author}{\bibfnamefont{V.}~\bibnamefont{Bernard}},
  \bibinfo{author}{\bibfnamefont{L.}~\bibnamefont{Elouadrhiri}},
  \bibnamefont{and} \bibinfo{author}{\bibfnamefont{U.~G.}
  \bibnamefont{Meissner}}, \bibinfo{journal}{J. Phys.}
  \textbf{\bibinfo{volume}{G28}}, \bibinfo{pages}{R1} (\bibinfo{year}{2002}),
  \eprint{hep-ph/0107088}.

\bibitem[{\citenamefont{Choi et~al.}(1993)}]{Choi:1993vt}
\bibinfo{author}{\bibfnamefont{S.}~\bibnamefont{Choi}} \bibnamefont{et~al.},
  \bibinfo{journal}{Phys. Rev. Lett.} \textbf{\bibinfo{volume}{71}},
  \bibinfo{pages}{3927} (\bibinfo{year}{1993}).

\bibitem[{\citenamefont{Andreev et~al.}(2007)}]{Andreev:2007wg}
\bibinfo{author}{\bibfnamefont{V.~A.} \bibnamefont{Andreev}}
  \bibnamefont{et~al.} (\bibinfo{collaboration}{MuCap}),
  \bibinfo{journal}{Phys. Rev. Lett.} \textbf{\bibinfo{volume}{99}},
  \bibinfo{pages}{032002} (\bibinfo{year}{2007}), \eprint{0704.2072}.

\bibitem[{\citenamefont{Hagler}(2007)}]{Hagler:2007hu}
\bibinfo{author}{\bibfnamefont{P.}~\bibnamefont{Hagler}},
  \bibinfo{journal}{PoS} \textbf{\bibinfo{volume}{LAT2007}},
  \bibinfo{pages}{013} (\bibinfo{year}{2007}), \eprint{0711.0819}.

\bibitem[{\citenamefont{Zanotti}(2008)}]{Zanotti:2008zm}
\bibinfo{author}{\bibfnamefont{J.~M.} \bibnamefont{Zanotti}},
  \bibinfo{journal}{PoS} \textbf{\bibinfo{volume}{LAT2008}},
  \bibinfo{pages}{007} (\bibinfo{year}{2008}), \eprint{0812.3845}.

\bibitem[{\citenamefont{Gockeler et~al.}(2005)}]{Gockeler:2003ay}
\bibinfo{author}{\bibfnamefont{M.}~\bibnamefont{Gockeler}} \bibnamefont{et~al.}
  (\bibinfo{collaboration}{QCDSF}), \bibinfo{journal}{Phys. Rev.}
  \textbf{\bibinfo{volume}{D71}}, \bibinfo{pages}{034508}
  (\bibinfo{year}{2005}), \eprint{hep-lat/0303019}.

\bibitem[{\citenamefont{Sasaki et~al.}(2003)\citenamefont{Sasaki, Orginos,
  Ohta, and Blum}}]{Sasaki:2003jh}
\bibinfo{author}{\bibfnamefont{S.}~\bibnamefont{Sasaki}},
  \bibinfo{author}{\bibfnamefont{K.}~\bibnamefont{Orginos}},
  \bibinfo{author}{\bibfnamefont{S.}~\bibnamefont{Ohta}}, \bibnamefont{and}
  \bibinfo{author}{\bibfnamefont{T.}~\bibnamefont{Blum}}
  (\bibinfo{collaboration}{RBCK}), \bibinfo{journal}{Phys. Rev.}
  \textbf{\bibinfo{volume}{D68}}, \bibinfo{pages}{054509}
  (\bibinfo{year}{2003}), \eprint{hep-lat/0306007}.

\bibitem[{\citenamefont{Tang et~al.}(2003)\citenamefont{Tang, Wilcox, and
  Lewis}}]{Tang:2003jh}
\bibinfo{author}{\bibfnamefont{A.}~\bibnamefont{Tang}},
  \bibinfo{author}{\bibfnamefont{W.}~\bibnamefont{Wilcox}}, \bibnamefont{and}
  \bibinfo{author}{\bibfnamefont{R.}~\bibnamefont{Lewis}},
  \bibinfo{journal}{Phys. Rev.} \textbf{\bibinfo{volume}{D68}},
  \bibinfo{pages}{094503} (\bibinfo{year}{2003}), \eprint{hep-lat/0307006}.

\bibitem[{\citenamefont{Boinepalli et~al.}(2006)\citenamefont{Boinepalli,
  Leinweber, Williams, Zanotti, and Zhang}}]{Boinepalli:2006xd}
\bibinfo{author}{\bibfnamefont{S.}~\bibnamefont{Boinepalli}},
  \bibinfo{author}{\bibfnamefont{D.~B.} \bibnamefont{Leinweber}},
  \bibinfo{author}{\bibfnamefont{A.~G.} \bibnamefont{Williams}},
  \bibinfo{author}{\bibfnamefont{J.~M.} \bibnamefont{Zanotti}},
  \bibnamefont{and} \bibinfo{author}{\bibfnamefont{J.~B.} \bibnamefont{Zhang}},
  \bibinfo{journal}{Phys. Rev.} \textbf{\bibinfo{volume}{D74}},
  \bibinfo{pages}{093005} (\bibinfo{year}{2006}), \eprint{hep-lat/0604022}.

\bibitem[{\citenamefont{Alexandrou et~al.}(2006)\citenamefont{Alexandrou,
  Koutsou, Negele, and Tsapalis}}]{Alexandrou:2006ru}
\bibinfo{author}{\bibfnamefont{C.}~\bibnamefont{Alexandrou}},
  \bibinfo{author}{\bibfnamefont{G.}~\bibnamefont{Koutsou}},
  \bibinfo{author}{\bibfnamefont{J.~W.} \bibnamefont{Negele}},
  \bibnamefont{and} \bibinfo{author}{\bibfnamefont{A.}~\bibnamefont{Tsapalis}},
  \bibinfo{journal}{Phys. Rev.} \textbf{\bibinfo{volume}{D74}},
  \bibinfo{pages}{034508} (\bibinfo{year}{2006}), \eprint{hep-lat/0605017}.

\bibitem[{\citenamefont{Khan et~al.}(2006)}]{Khan:2006de}
\bibinfo{author}{\bibfnamefont{A.~A.} \bibnamefont{Khan}} \bibnamefont{et~al.},
  \bibinfo{journal}{Phys. Rev.} \textbf{\bibinfo{volume}{D74}},
  \bibinfo{pages}{094508} (\bibinfo{year}{2006}), \eprint{hep-lat/0603028}.

\bibitem[{\citenamefont{Gockeler et~al.}(2007)}]{Gockeler:2007hj}
\bibinfo{author}{\bibfnamefont{M.}~\bibnamefont{Gockeler}} \bibnamefont{et~al.}
  (\bibinfo{collaboration}{QCDSF/UKQCD}), \bibinfo{journal}{PoS}
  \textbf{\bibinfo{volume}{LAT2007}}, \bibinfo{pages}{161}
  (\bibinfo{year}{2007}), \eprint{0710.2159}.

\bibitem[{\citenamefont{Edwards et~al.}(2006)}]{Edwards:2005ym}
\bibinfo{author}{\bibfnamefont{R.~G.} \bibnamefont{Edwards}}
  \bibnamefont{et~al.} (\bibinfo{collaboration}{LHPC}), \bibinfo{journal}{Phys.
  Rev. Lett.} \textbf{\bibinfo{volume}{96}}, \bibinfo{pages}{052001}
  (\bibinfo{year}{2006}).

\bibitem[{\citenamefont{Alexandrou et~al.}(2007)}]{Alexandrou:2007xj}
\bibinfo{author}{\bibfnamefont{C.}~\bibnamefont{Alexandrou}}
  \bibnamefont{et~al.}, \bibinfo{journal}{Phys. Rev.}
  \textbf{\bibinfo{volume}{D76}}, \bibinfo{pages}{094511}
  (\bibinfo{year}{2007}), \eprint{arXiv:0706.3011 [hep-lat]}.

\bibitem[{\citenamefont{Hagler et~al.}(2008)}]{Hagler:2007xi}
\bibinfo{author}{\bibfnamefont{P.}~\bibnamefont{Hagler}} \bibnamefont{et~al.}
  (\bibinfo{collaboration}{LHPC}), \bibinfo{journal}{Phys. Rev.}
  \textbf{\bibinfo{volume}{D77}}, \bibinfo{pages}{094502}
  (\bibinfo{year}{2008}), \eprint{0705.4295}.

\bibitem[{\citenamefont{Bratt et~al.}(2008)}]{Bratt:2008uf}
\bibinfo{author}{\bibfnamefont{J.~D.} \bibnamefont{Bratt}}
  \bibnamefont{et~al.}, \bibinfo{journal}{PoS}
  \textbf{\bibinfo{volume}{LATTICE2008}} (\bibinfo{year}{2008}),
  \eprint{0810.1933}.

\bibitem[{\citenamefont{Alexandrou et~al.}(2008)}]{Alexandrou:2007dt}
\bibinfo{author}{\bibfnamefont{C.}~\bibnamefont{Alexandrou}}
  \bibnamefont{et~al.}, \bibinfo{journal}{Phys. Rev.}
  \textbf{\bibinfo{volume}{D77}}, \bibinfo{pages}{085012}
  (\bibinfo{year}{2008}), \eprint{0710.4621}.

\bibitem[{\citenamefont{Guadagnoli et~al.}(2007)\citenamefont{Guadagnoli,
  Lubicz, Papinutto, and Simula}}]{Guadagnoli:2006gj}
\bibinfo{author}{\bibfnamefont{D.}~\bibnamefont{Guadagnoli}},
  \bibinfo{author}{\bibfnamefont{V.}~\bibnamefont{Lubicz}},
  \bibinfo{author}{\bibfnamefont{M.}~\bibnamefont{Papinutto}},
  \bibnamefont{and} \bibinfo{author}{\bibfnamefont{S.}~\bibnamefont{Simula}},
  \bibinfo{journal}{Nucl. Phys.} \textbf{\bibinfo{volume}{B761}},
  \bibinfo{pages}{63} (\bibinfo{year}{2007}), \eprint{hep-ph/0606181}.

\bibitem[{\citenamefont{Sasaki and
  Yamazaki}(2008{\natexlab{b}})}]{Sasaki:2008ha}
\bibinfo{author}{\bibfnamefont{S.}~\bibnamefont{Sasaki}} \bibnamefont{and}
  \bibinfo{author}{\bibfnamefont{T.}~\bibnamefont{Yamazaki}}
  (\bibinfo{year}{2008}{\natexlab{b}}), \eprint{0811.1406}.

\bibitem[{\citenamefont{Lin and Orginos}(2008)}]{Lin:2008mr}
\bibinfo{author}{\bibfnamefont{H.-W.} \bibnamefont{Lin}} \bibnamefont{and}
  \bibinfo{author}{\bibfnamefont{K.}~\bibnamefont{Orginos}}
  (\bibinfo{year}{2008}), \eprint{0812.4456}.

\bibitem[{\citenamefont{Ginsparg and Wilson}(1982)}]{Ginsparg:1981bj}
\bibinfo{author}{\bibfnamefont{P.~H.} \bibnamefont{Ginsparg}} \bibnamefont{and}
  \bibinfo{author}{\bibfnamefont{K.~G.} \bibnamefont{Wilson}},
  \bibinfo{journal}{Phys. Rev.} \textbf{\bibinfo{volume}{D25}},
  \bibinfo{pages}{2649} (\bibinfo{year}{1982}).

\bibitem[{\citenamefont{Kaplan}(1992)}]{Kaplan:1992bt}
\bibinfo{author}{\bibfnamefont{D.~B.} \bibnamefont{Kaplan}},
  \bibinfo{journal}{Phys. Lett.} \textbf{\bibinfo{volume}{B288}},
  \bibinfo{pages}{342} (\bibinfo{year}{1992}), \eprint{hep-lat/9206013}.

\bibitem[{\citenamefont{Kaplan}(1993)}]{Kaplan:1992sg}
\bibinfo{author}{\bibfnamefont{D.~B.} \bibnamefont{Kaplan}},
  \bibinfo{journal}{Nucl. Phys. Proc. Suppl.} \textbf{\bibinfo{volume}{30}},
  \bibinfo{pages}{597} (\bibinfo{year}{1993}).

\bibitem[{\citenamefont{Shamir}(1993)}]{Shamir:1993zy}
\bibinfo{author}{\bibfnamefont{Y.}~\bibnamefont{Shamir}},
  \bibinfo{journal}{Nucl. Phys.} \textbf{\bibinfo{volume}{B406}},
  \bibinfo{pages}{90} (\bibinfo{year}{1993}), \eprint{hep-lat/9303005}.

\bibitem[{\citenamefont{Furman and Shamir}(1995)}]{Furman:1994ky}
\bibinfo{author}{\bibfnamefont{V.}~\bibnamefont{Furman}} \bibnamefont{and}
  \bibinfo{author}{\bibfnamefont{Y.}~\bibnamefont{Shamir}},
  \bibinfo{journal}{Nucl. Phys.} \textbf{\bibinfo{volume}{B439}},
  \bibinfo{pages}{54} (\bibinfo{year}{1995}), \eprint{hep-lat/9405004}.

\bibitem[{\citenamefont{Yamazaki et~al.}(2008)}]{Yamazaki:2008py}
\bibinfo{author}{\bibfnamefont{T.}~\bibnamefont{Yamazaki}} \bibnamefont{et~al.}
  (\bibinfo{collaboration}{RBC and UKQCD}), \bibinfo{journal}{Phys. Rev. Lett.}
  \textbf{\bibinfo{volume}{100}}, \bibinfo{pages}{171602}
  (\bibinfo{year}{2008}), \eprint{0801.4016}.

\bibitem[{\citenamefont{Yamazaki and Ohta}(2007)}]{Yamazaki:2007mk}
\bibinfo{author}{\bibfnamefont{T.}~\bibnamefont{Yamazaki}} \bibnamefont{and}
  \bibinfo{author}{\bibfnamefont{S.}~\bibnamefont{Ohta}}
  (\bibinfo{collaboration}{RBC and UKQCD}), \bibinfo{journal}{PoS}
  \textbf{\bibinfo{volume}{LAT2007}}, \bibinfo{pages}{165}
  (\bibinfo{year}{2007}), \eprint{0710.0422}.

\bibitem[{\citenamefont{Ohta and Yamazaki}(2008)}]{Ohta:2008kd}
\bibinfo{author}{\bibfnamefont{S.}~\bibnamefont{Ohta}} \bibnamefont{and}
  \bibinfo{author}{\bibfnamefont{T.}~\bibnamefont{Yamazaki}}
  (\bibinfo{collaboration}{RBC and UKQCD}), \bibinfo{journal}{PoS}
  \textbf{\bibinfo{volume}{LAT2008}}, \bibinfo{pages}{168}
  (\bibinfo{year}{2008}), \eprint{0810.0045}.

\bibitem[{\citenamefont{Sasaki et~al.}(2002)\citenamefont{Sasaki, Blum, and
  Ohta}}]{Sasaki:2001nf}
\bibinfo{author}{\bibfnamefont{S.}~\bibnamefont{Sasaki}},
  \bibinfo{author}{\bibfnamefont{T.}~\bibnamefont{Blum}}, \bibnamefont{and}
  \bibinfo{author}{\bibfnamefont{S.}~\bibnamefont{Ohta}},
  \bibinfo{journal}{Phys. Rev.} \textbf{\bibinfo{volume}{D65}},
  \bibinfo{pages}{074503} (\bibinfo{year}{2002}), \eprint{hep-lat/0102010}.

\bibitem[{\citenamefont{Sasaki and Sasaki}(2005)}]{Sasaki:2005ug}
\bibinfo{author}{\bibfnamefont{K.}~\bibnamefont{Sasaki}} \bibnamefont{and}
  \bibinfo{author}{\bibfnamefont{S.}~\bibnamefont{Sasaki}},
  \bibinfo{journal}{Phys. Rev.} \textbf{\bibinfo{volume}{D72}},
  \bibinfo{pages}{034502} (\bibinfo{year}{2005}), \eprint{hep-lat/0503026}.

\bibitem[{\citenamefont{Alexandrou et~al.}(1994)\citenamefont{Alexandrou,
  Gusken, Jegerlehner, Schilling, and Sommer}}]{Alexandrou:1992ti}
\bibinfo{author}{\bibfnamefont{C.}~\bibnamefont{Alexandrou}},
  \bibinfo{author}{\bibfnamefont{S.}~\bibnamefont{Gusken}},
  \bibinfo{author}{\bibfnamefont{F.}~\bibnamefont{Jegerlehner}},
  \bibinfo{author}{\bibfnamefont{K.}~\bibnamefont{Schilling}},
  \bibnamefont{and} \bibinfo{author}{\bibfnamefont{R.}~\bibnamefont{Sommer}},
  \bibinfo{journal}{Nucl. Phys.} \textbf{\bibinfo{volume}{B414}},
  \bibinfo{pages}{815} (\bibinfo{year}{1994}), \eprint{hep-lat/9211042}.

\bibitem[{\citenamefont{Wilcox et~al.}(1992)\citenamefont{Wilcox, Draper, and
  Liu}}]{Wilcox:1991cq}
\bibinfo{author}{\bibfnamefont{W.}~\bibnamefont{Wilcox}},
  \bibinfo{author}{\bibfnamefont{T.}~\bibnamefont{Draper}}, \bibnamefont{and}
  \bibinfo{author}{\bibfnamefont{K.-F.} \bibnamefont{Liu}},
  \bibinfo{journal}{Phys. Rev.} \textbf{\bibinfo{volume}{D46}},
  \bibinfo{pages}{1109} (\bibinfo{year}{1992}), \eprint{hep-lat/9205015}.

\bibitem[{\citenamefont{Hagler et~al.}(2003)}]{Hagler:2003jd}
\bibinfo{author}{\bibfnamefont{P.}~\bibnamefont{Hagler}} \bibnamefont{et~al.}
  (\bibinfo{collaboration}{LHPC}), \bibinfo{journal}{Phys. Rev.}
  \textbf{\bibinfo{volume}{D68}}, \bibinfo{pages}{034505}
  (\bibinfo{year}{2003}), \eprint{hep-lat/0304018}.

\bibitem[{\citenamefont{Allton et~al.}(2008)}]{Allton:2008pn}
\bibinfo{author}{\bibfnamefont{C.}~\bibnamefont{Allton}} \bibnamefont{et~al.}
  (\bibinfo{collaboration}{RBC-UKQCD}), \bibinfo{journal}{Phys. Rev.}
  \textbf{\bibinfo{volume}{D78}}, \bibinfo{pages}{114509}
  (\bibinfo{year}{2008}), \eprint{0804.0473}.

\bibitem[{\citenamefont{Iwasaki}(1983)}]{Iwasaki:1983yi}
\bibinfo{author}{\bibfnamefont{Y.}~\bibnamefont{Iwasaki}}
  (\bibinfo{year}{1983}), \bibinfo{note}{unpublished, UTHEP-118}.

\bibitem[{\citenamefont{Berruto et~al.}(2006)\citenamefont{Berruto, Blum,
  Orginos, and Soni}}]{Berruto:2005hg}
\bibinfo{author}{\bibfnamefont{F.}~\bibnamefont{Berruto}},
  \bibinfo{author}{\bibfnamefont{T.}~\bibnamefont{Blum}},
  \bibinfo{author}{\bibfnamefont{K.}~\bibnamefont{Orginos}}, \bibnamefont{and}
  \bibinfo{author}{\bibfnamefont{A.}~\bibnamefont{Soni}},
  \bibinfo{journal}{Phys. Rev.} \textbf{\bibinfo{volume}{D73}},
  \bibinfo{pages}{054509} (\bibinfo{year}{2006}), \eprint{hep-lat/0512004}.

\bibitem[{\citenamefont{Thomas et~al.}(2005)\citenamefont{Thomas, Ashley,
  Leinweber, and Young}}]{Thomas:2005qm}
\bibinfo{author}{\bibfnamefont{A.~W.} \bibnamefont{Thomas}},
  \bibinfo{author}{\bibfnamefont{J.~D.} \bibnamefont{Ashley}},
  \bibinfo{author}{\bibfnamefont{D.~B.} \bibnamefont{Leinweber}},
  \bibnamefont{and} \bibinfo{author}{\bibfnamefont{R.~D.} \bibnamefont{Young}},
  \bibinfo{journal}{J. Phys. Conf. Ser.} \textbf{\bibinfo{volume}{9}},
  \bibinfo{pages}{321} (\bibinfo{year}{2005}), \eprint{hep-lat/0502002}.

\bibitem[{\citenamefont{Dolgov et~al.}(2002)}]{Dolgov:2002zm}
\bibinfo{author}{\bibfnamefont{D.}~\bibnamefont{Dolgov}} \bibnamefont{et~al.}
  (\bibinfo{collaboration}{LHPC}), \bibinfo{journal}{Phys. Rev.}
  \textbf{\bibinfo{volume}{D66}}, \bibinfo{pages}{034506}
  (\bibinfo{year}{2002}), \eprint[http://arXiv.org/abs]{hep-lat/0201021}.

\bibitem[{\citenamefont{Kim and Kim}(1998)}]{Kim:1996bz}
\bibinfo{author}{\bibfnamefont{M.}~\bibnamefont{Kim}} \bibnamefont{and}
  \bibinfo{author}{\bibfnamefont{S.}~\bibnamefont{Kim}},
  \bibinfo{journal}{Phys. Rev.} \textbf{\bibinfo{volume}{D58}},
  \bibinfo{pages}{074509} (\bibinfo{year}{1998}).

\bibitem[{\citenamefont{Bar et~al.}(2005)}]{Bar:2005tu}
\bibinfo{author}{\bibfnamefont{O.}~\bibnamefont{Bar}} \bibnamefont{et~al.},
  \bibinfo{journal}{Phys. Rev.} \textbf{\bibinfo{volume}{D72}},
  \bibinfo{pages}{054502} (\bibinfo{year}{2005}).

\bibitem[{\citenamefont{Prelovsek}(2006)}]{Prelovsek:2005rf}
\bibinfo{author}{\bibfnamefont{S.}~\bibnamefont{Prelovsek}},
  \bibinfo{journal}{Phys. Rev.} \textbf{\bibinfo{volume}{D73}},
  \bibinfo{pages}{014506} (\bibinfo{year}{2006}).

\bibitem[{\citenamefont{Jiang}(2007)}]{Jiang:2007sn}
\bibinfo{author}{\bibfnamefont{F.-J.} \bibnamefont{Jiang}}
  (\bibinfo{year}{2007}), \eprint{hep-lat/0703012}.

\bibitem[{\citenamefont{Chen et~al.}(2007)\citenamefont{Chen, O'Connell, and
  Walker-Loud}}]{Chen:2007ug}
\bibinfo{author}{\bibfnamefont{J.-W.} \bibnamefont{Chen}},
  \bibinfo{author}{\bibfnamefont{D.}~\bibnamefont{O'Connell}},
  \bibnamefont{and}
  \bibinfo{author}{\bibfnamefont{A.}~\bibnamefont{Walker-Loud}}
  (\bibinfo{year}{2007}), \eprint{arXiv:0706.0035 [hep-lat]}.

\bibitem[{\citenamefont{Smigielski and Wasem}(2007)}]{Smigielski:2007pe}
\bibinfo{author}{\bibfnamefont{B.}~\bibnamefont{Smigielski}} \bibnamefont{and}
  \bibinfo{author}{\bibfnamefont{J.}~\bibnamefont{Wasem}},
  \bibinfo{journal}{Phys. Rev.} \textbf{\bibinfo{volume}{D76}},
  \bibinfo{pages}{074503} (\bibinfo{year}{2007}).

\bibitem[{\citenamefont{Bernard and Meissner}(2006)}]{Bernard:2006te}
\bibinfo{author}{\bibfnamefont{V.}~\bibnamefont{Bernard}} \bibnamefont{and}
  \bibinfo{author}{\bibfnamefont{U.-G.} \bibnamefont{Meissner}},
  \bibinfo{journal}{Phys. Lett.} \textbf{\bibinfo{volume}{B639}},
  \bibinfo{pages}{278} (\bibinfo{year}{2006}), \eprint{hep-lat/0605010}.

\bibitem[{\citenamefont{Jaffe}(2002)}]{Jaffe:2001eb}
\bibinfo{author}{\bibfnamefont{R.~L.} \bibnamefont{Jaffe}},
  \bibinfo{journal}{Phys. Lett.} \textbf{\bibinfo{volume}{B529}},
  \bibinfo{pages}{105} (\bibinfo{year}{2002}).

\bibitem[{\citenamefont{Beane and Savage}(2004)}]{Beane:2004rf}
\bibinfo{author}{\bibfnamefont{S.~R.} \bibnamefont{Beane}} \bibnamefont{and}
  \bibinfo{author}{\bibfnamefont{M.~J.} \bibnamefont{Savage}},
  \bibinfo{journal}{Phys. Rev.} \textbf{\bibinfo{volume}{D70}},
  \bibinfo{pages}{074029} (\bibinfo{year}{2004}), \eprint{hep-ph/0404131}.

\bibitem[{\citenamefont{Detmold and Lin}(2005)}]{Detmold:2005pt}
\bibinfo{author}{\bibfnamefont{W.}~\bibnamefont{Detmold}} \bibnamefont{and}
  \bibinfo{author}{\bibfnamefont{C.-J.~D.} \bibnamefont{Lin}},
  \bibinfo{journal}{Phys. Rev.} \textbf{\bibinfo{volume}{D71}},
  \bibinfo{pages}{054510} (\bibinfo{year}{2005}).

\bibitem[{\citenamefont{Beg and Zepeda}(1972)}]{Beg:1973sc}
\bibinfo{author}{\bibfnamefont{M.~A.~B.} \bibnamefont{Beg}} \bibnamefont{and}
  \bibinfo{author}{\bibfnamefont{A.}~\bibnamefont{Zepeda}},
  \bibinfo{journal}{Phys. Rev.} \textbf{\bibinfo{volume}{D6}},
  \bibinfo{pages}{2912} (\bibinfo{year}{1972}).

\bibitem[{\citenamefont{Bernard et~al.}(1998)\citenamefont{Bernard, Fearing,
  Hemmert, and Meissner}}]{Bernard:1998gv}
\bibinfo{author}{\bibfnamefont{V.}~\bibnamefont{Bernard}},
  \bibinfo{author}{\bibfnamefont{H.~W.} \bibnamefont{Fearing}},
  \bibinfo{author}{\bibfnamefont{T.~R.} \bibnamefont{Hemmert}},
  \bibnamefont{and} \bibinfo{author}{\bibfnamefont{U.~G.}
  \bibnamefont{Meissner}}, \bibinfo{journal}{Nucl. Phys.}
  \textbf{\bibinfo{volume}{A635}}, \bibinfo{pages}{121} (\bibinfo{year}{1998}),
  \eprint{hep-ph/9801297}.

\bibitem[{\citenamefont{Wang et~al.}(2008)\citenamefont{Wang, Leinweber,
  Thomas, and Young}}]{Wang:2008vb}
\bibinfo{author}{\bibfnamefont{P.}~\bibnamefont{Wang}},
  \bibinfo{author}{\bibfnamefont{D.~B.} \bibnamefont{Leinweber}},
  \bibinfo{author}{\bibfnamefont{A.~W.} \bibnamefont{Thomas}},
  \bibnamefont{and} \bibinfo{author}{\bibfnamefont{R.~D.} \bibnamefont{Young}}
  (\bibinfo{year}{2008}), \eprint{0810.1021}.

\bibitem[{\citenamefont{Czarnecki et~al.}(2007)\citenamefont{Czarnecki,
  Marciano, and Sirlin}}]{Czarnecki:2007th}
\bibinfo{author}{\bibfnamefont{A.}~\bibnamefont{Czarnecki}},
  \bibinfo{author}{\bibfnamefont{W.~J.} \bibnamefont{Marciano}},
  \bibnamefont{and} \bibinfo{author}{\bibfnamefont{A.}~\bibnamefont{Sirlin}}
  (\bibinfo{year}{2007}), \eprint{arXiv:0704.3968 [hep-ph]}.

\bibitem[{\citenamefont{Ericson et~al.}(2002)\citenamefont{Ericson, Loiseau,
  and Thomas}}]{Ericson:2000md}
\bibinfo{author}{\bibfnamefont{T.~E.~O.} \bibnamefont{Ericson}},
  \bibinfo{author}{\bibfnamefont{B.}~\bibnamefont{Loiseau}}, \bibnamefont{and}
  \bibinfo{author}{\bibfnamefont{A.~W.} \bibnamefont{Thomas}},
  \bibinfo{journal}{Phys. Rev.} \textbf{\bibinfo{volume}{C66}},
  \bibinfo{pages}{014005} (\bibinfo{year}{2002}), \eprint{hep-ph/0009312}.

\end{thebibliography}

\appendix

\section{Results of form factors}
%
%
%
%
\begin{table}[!h]
\begin{tabular}{cccccc}\hline\hline
$m_f$ & $q^2$ [GeV$^2$] & 
$F_1(q^2)$ & $F_2(q^2)$ & $F_A(q^2)$ & $2 M_N F_P(q^2)$ \\\hline
0.005 & 0.0  & 1.0000(11)& N/A      & 1.073(39)& N/A       \\
      & 0.198& 0.785(19) & 2.20(13) & 0.959(34)& 13.32(91) \\
      & 0.383& 0.622(22) & 1.716(97)& 0.892(39)&  7.42(51) \\
      & 0.557& 0.505(28) & 1.40(10) & 0.754(41)&  5.27(42) \\
      & 0.723& 0.516(53) & 1.36(16) & 0.792(65)&  4.71(53) \\
0.01  & 0.0  & 1.0000(10)& N/A      & 1.186(33)& N/A       \\
      & 0.199& 0.787(17) & 2.38(15) & 0.994(37)& 13.00(90) \\
      & 0.385& 0.641(22) & 1.71(12) & 0.854(37)&  7.48(44) \\
      & 0.562& 0.524(31) & 1.34(11) & 0.719(39)&  5.01(42) \\
      & 0.731& 0.506(49) & 1.19(13) & 0.701(57)&  3.95(46) \\
0.02  & 0.0  & 1.0000(15)& N/A      & 1.174(37)& N/A       \\
      & 0.200& 0.805(20) & 2.40(15) & 1.005(33)& 12.3(1.2) \\
      & 0.390& 0.686(32) & 2.08(13) & 0.890(38)&  9.04(73) \\
      & 0.573& 0.599(49) & 1.80(13) & 0.839(48)&  7.06(70) \\
      & 0.748& 0.443(37) & 1.31(14) & 0.668(47)&  3.84(59) \\
0.03  & 0.0  & 1.0000(11)& N/A      & 1.196(30)& N/A       \\
      & 0.201& 0.8302(99)& 2.79(12) & 1.038(28)& 12.74(88) \\
      & 0.394& 0.700(15) & 2.302(92)& 0.912(32)&  9.68(62) \\
      & 0.580& 0.595(22) & 2.00(12) & 0.838(40)&  6.81(61) \\
      & 0.760& 0.500(31) & 1.54(15) & 0.704(50)&  4.55(73) \\
\hline\hline
\end{tabular}
\caption{\label{tab:ffs} Form factors of vector, axialvector currents
on (2.7 fm)$^3$.
All form factors are renormalized.
}
\end{table}
\begin{table}[!h]
\begin{tabular}{cccccc}\hline\hline
$m_f$ & $q^2$ [GeV$^2$] & 
$F_1(q^2)$ & $F_2(q^2)$ & $F_A(q^2)$ & $2 M_N F_P(q^2)$ \\\hline
0.01  & 0.0  & 1.000(27) & N/A      & 1.066(72)& N/A       \\
      & 0.430& 0.621(52) & 1.42(23) & 0.580(93)&  3.2(1.4) \\
      & 0.812& 0.46(14)  & 0.89(50) & 0.380(70)&  1.83(68) \\
0.02  & 0.0  & 1.000(14) & N/A      & 1.115(58)& N/A       \\
      & 0.437& 0.705(38) & 2.03(17) & 0.749(49)&  6.16(97) \\
      & 0.833& 0.501(36) & 1.14(14) & 0.504(47)&  2.70(38) \\
0.03  & 0.0  & 1.0000(6) & N/A      & 1.149(32)& N/A       \\
      & 0.441& 0.686(18) & 1.91(10) & 0.787(29)&  6.56(70) \\
      & 0.848& 0.522(23) & 1.258(70)& 0.574(30)&  3.48(30) \\
\hline\hline
\end{tabular}
\caption{\label{tab:ffs_s} Form factors of vector, axialvector currents
on (1.8 fm)$^3$.
All form factors are renormalized. }
\end{table}

\end{document}